# Molecular gas in the *Herschel*⋆-selected strongly lensed submillimeter galaxies at $z \sim 2$–4 as probed by multi-$J$ CO lines ⋆⋆

C. Yang(杨辰涛)[1,2,3,4,5], A. Omont[4,5], A. Beelen[2], Y. Gao(高煜)[1], P. van der Werf[6], R. Gavazzi[4,5], Z.-Y. Zhang(张智昱)[7,8], R. Ivison[7,8], M. Lehnert[4,5], D. Liu(刘岱钟)[9], I. Oteo[7,8], E. González-Alfonso[10], H. Dannerbauer[11,12], P. Cox[13], M. Krips[14], R. Neri[14], D. Riechers[15], A. J. Baker[16], M.J. Michałowski[17,7], A. Cooray[18], and I. Smail[19]

*(Affiliations can be found after the references)*

**ABSTRACT**

We present the IRAM-30 m observations of multiple-$J$ CO ($J_{\rm up}$ mostly from 3 up to 8) and [C I]($^3P_2 \rightarrow {}^3P_1$) ([C I](2–1) hereafter) line emission in a sample of redshift ∼ 2–4 submillimeter galaxies (SMGs). These SMGs are selected among the brightest-lensed galaxies discovered in the *Herschel*-Astrophysical Terahertz Large Area Survey (*H*-ATLAS). Forty-seven CO lines and 7 [C I](2–1) lines have been detected in 15 lensed SMGs. A non-negligible effect of differential lensing is found for the CO emission lines, which could have caused significant underestimations of the linewidths, and hence of the dynamical masses. The CO spectral line energy distributions (SLEDs), peaking around $J_{\rm up} \sim 5$–7, are found to be similar to those of the local starburst-dominated ultra-luminous infrared galaxies and of the previously studied SMGs. After correcting for lensing amplification, we derived the global properties of the bulk of molecular gas in the SMGs using non-LTE radiative transfer modelling, such as the molecular gas density $n_{\rm H_2} \sim 10^{2.5}$–$10^{4.1}$ cm$^{-3}$ and the kinetic temperature $T_{\rm k} \sim 20$–750 K. The gas thermal pressure $P_{\rm th}$ ranging from $\sim 10^5$ K cm$^{-3}$ to $10^6$ K cm$^{-3}$ is found to be correlated with star formation efficiency. Further decomposing the CO SLEDs into two excitation components, we find a low-excitation component with $n_{\rm H_2} \sim 10^{2.8}$–$10^{4.6}$ cm$^{-3}$ and $T_{\rm k} \sim 20$–30 K, which is less correlated with star formation, and a high-excitation one ($n_{\rm H_2} \sim 10^{2.7}$–$10^{4.2}$ cm$^{-3}$, $T_{\rm k} \sim 60$–400 K) which is tightly related to the on-going star-forming activity. Additionally, tight linear correlations between the far-infrared and CO line luminosities have been confirmed for the $J_{\rm up} \geq 5$ CO lines of these SMGs, implying that these CO lines are good tracers of star formation. The [C I](2–1) lines follow the tight linear correlation between the luminosities of the [C I](2–1) and the CO(1–0) line found in local starbursts, indicating that [C I] lines could serve as good total molecular gas mass tracers for high-redshift SMGs as well. The total mass of the molecular gas reservoir, (1–30) × $10^{10}$ $M_\odot$, derived based on the CO(3–2) fluxes and $\alpha_{\rm CO(1-0)} = 0.8 M_\odot$ ( K km s$^{-1}$ pc$^2$)$^{-1}$, suggests a typical molecular gas depletion time $t_{\rm dep} \sim 20$–100 Myr and a gas to dust mass ratio $\delta_{\rm GDR} \sim 30$–100 with ∼ 20%–60% uncertainty for the SMGs. The ratio between CO line luminosity and the dust mass $L'_{\rm CO}/M_{\rm dust}$ appears to be slowly increasing with redshift for high-redshift SMGs, which need to be further confirmed by a more complete SMG sample at various redshifts. Finally, through comparing the linewidth of CO and $H_2O$ lines, we find that they agree well in almost all our SMGs, confirming that the emitting regions of the CO and $H_2O$ lines are co-spatially located.

## 1. Introduction

The strongest starbursts throughout the star formation history of our Universe are the high-redshift hyper- and ultra-luminous infrared galaxies (HyLIRGs and ULIRGs). With infrared luminosities integrated over 8–1000 $\mu$m $L_{\rm IR} \geq 10^{13}$ $L_\odot$ and $10^{13}$ $L_\odot > L_{\rm IR} \geq 10^{12}$ $L_\odot$, respectively, and star formation rate (SFR) around 1000 $M_\odot$ yr$^{-1}$, they approach the limit of maximum starbursts (Barger et al. 2014). Despite having comparable or slightly higher luminosities than the local ULIRGs (Tacconi et al. 2010), these submillimeter (submm) bright galaxies (SMGs, see reviews of Blain et al. 2002; Casey et al. 2014) are different, being more extended and unlike nuclear starbursts of local ULIRGS. This population of dusty starburst galaxies was first discovered in the submm band using Submillimeter Common-User Bolometer Array (SCUBA, Holland et al. 1999) on the James Clerk Maxwell Telescope (Barger et al. 1998; Hughes et al. 1998; Smail et al. 1997), and later the spectroscopy observations revealed a median redshift of ∼ 2.5 (Chapman et al. 2005; Danielson et al. 2017). Their extremely intense star formation activity indicates that these "vigorous monsters" generating enormous energy at far-infrared (FIR) are in the critical phase of rapid stellar mass assembly. They are believed to be the progenitors of the most massive galaxies today (e.g. Simpson et al. 2014). Nevertheless, theoretical models of galaxy evolution have been challenged by the observed large number of high-redshift SMGs (e.g. Casey et al. 2014).

Since the initial discovery of SMGs at 850 $\mu$m with SCUBA at the end of the last century, Chapman et al. (2005) carefully studied the properties of this 850 $\mu$m-selected SMG population and concluded that those with $S_{850\mu{\rm m}} > 1$ mJy contribute a significant fraction to the cosmic star formation around $z$ = 2–3, that is ≳ 10%. Several other works have also confirmed that SMGs play a key role in the cosmic star formation at high-redshift (e.g. Murphy et al. 2011; Magnelli et al. 2013; Swinbank et al. 2014; Michałowski et al. 2017). For the ULIRGs studied with a median redshift of 2.2, it can be > 65% according to Le Floc'h et al. 2005 and Dunlop et al. 2017 (see ALMA counts by Karim et al. 2013; Oteo et al. 2016; Aravena et al.

⋆ *Herschel* is an ESA space observatory with science instruments provided by European-led Principal Investigator consortia and with important participation from NASA.
⋆⋆ Based on observations carried out under project number 076-16, 196-15 and 079-15 (PI: C. Yang); 252-11 and 124-11 (PI: P. van de Werf) with the IRAM-30m Telescope. IRAM is supported by INSU/CNRS (France), MPG (Germany) and IGN (Spain).

2016a; Dunlop et al. 2017 and references therein for updated SMG counts; and Casey et al. 2014 for redshift distributions of SMGs selected at 850–870 $\mu$m by SCUBA/LABOCA and at 1.1 mm by AzTEC).

It is important to understand the extreme star-forming activity within SMGs through studying their molecular gas content which serves as the basic ingredient for star formation, especially those at the peak of the star formation history (i.e. $z \sim 2$–3, Madau & Dickinson 2014). Nevertheless, due to their great distances, the number of well-studied high-redshift SMGs with several CO transitions at different energy levels is limited (see reviews of Solomon & Vanden Bout 2005; Carilli & Walter 2013) and this is mostly achieved through strong gravitation lensing and/or in quasi-stellar objects (QSOs), including IRAS F10214+4724 (Ao et al. 2008), APM 08279+5455 (Weiß et al. 2007), Cloverleaf (Bradford et al. 2009), SMM J2135-0102 (Danielson et al. 2011), G15v2.779 (Cox et al. 2011) and in the weakly lensed SMG, HFLS3 at $z = 6.34$ (Riechers et al. 2013; Cooray et al. 2014). Our knowledge of the detailed physical and chemical properties and processes related to star formation within these high-redshift Hy/ULIRGs is still limited.

Tacconi et al. (2008) found that high-redshift SMGs have large reservoirs of molecular gas about $10^{10-11}\,M_\odot$ (see also Ivison et al. 2011; Riechers et al. 2011c). CO rotational lines are contributing a significant amount of cooling of the molecular gas. By measuring the multiple-$J$ CO lines, we can constrain the kinetic temperature and the gas density of the emitting regions (e.g. Rangwala et al. 2011) using non-local thermodynamic equilibrium (non-LTE) models. From the observations of the aforementioned individual high-redshift galaxies, the variety of CO spectral line energy distribution (SLED) shows that multiple molecular gas components in terms of their different gas densities and kinetic temperatures are required to explain the entire CO SLEDs. The mid/high-$J$ CO emission can be explained by a warm component with molecular gas volume density of $10^3$–$10^4$ cm$^{-3}$ which is more closely related to the ongoing star formation, while there is also an extended cool component dominating the low-$J$ CO (e.g. Ivison et al. 2010; Danielson et al. 2011). Recent works with *Herschel* SPIRE/FTS spectra of 167 local galaxies by Liu et al. (2015) and 121 local LIRGs by Lu et al. (2017) also favour the presence of multiple CO excitation components. Daddi et al. (2015) reached similar conclusions for $z \sim 1.5$ normal star-forming galaxies. The differences in the $J_{\rm up} > 6$ part of the CO SLEDs reveal different excitation processes (e.g. Lu et al. 2017): in most cases, the CO emission is insignificant for $J_{\rm up} > 7$ CO lines; in the few cases ($\lesssim 10\%$) where $L_{\rm IR}$ is dominated by an active galactic nucleus (AGN) there is a substantial excess of CO emission in the $J_{\rm up} > 10$ CO lines (van der Werf et al. 2010), likely associated with AGN heating of molecular gas; there could also be a small number of exceptional cases, like NGC 6240, where shock excitation dominates (Meijerink et al. 2013).

Thanks to the extra-galactic surveys at FIR and submm bands like the *Herschel*-Astrophysical Terahertz Large Area Survey (*H*-ATLAS, Eales et al. 2010), the *Herschel* Multi-tiered Extragalactic Survey (*Her*MES, Oliver et al. 2012) and South Pole Telescope (SPT) survey (Vieira et al. 2013), large and statistically significant samples of SMGs have been built. It was found that with a criterium of source flux at 500 $\mu$m, namely $S_{500\,\mu m} > 100$ mJy, (galaxy-galaxy) strongly lensed high-redshift SMGs can be efficiently selected (e.g. Negrello et al. 2007, 2010, 2017; Vieira et al. 2010; Wardlow et al. 2013; Nayyeri et al. 2016). The strong lensing effect not only boosts the sensitivity of observations but also improves the spatial resolution so that we can study the high-redshift galaxies in unprecedented detail (see e.g. Swinbank et al. 2010).

The spectroscopy redshifts (mostly determined from CO lines) have now been determined in more than 24 *Herschel*-selected, lensed *H*-ATLAS SMGs thanks to the combined use of various telescopes; for example, *Herschel* itself, using the SPIRE/FTS (George et al. 2013; Zhang et al., in prep), CSO with Z-Spec (Scott et al. 2011; Lupu et al. 2012), APEX (Ivison et al., in prep), IRAM/PdBI (Cox et al. 2011; Krips et al. in prep.), LMT, ALMA (Asboth et al. 2016) and especially the Zpectrometer on the GBT (Frayer et al. 2011; Harris et al. 2012, and in prep.) and CARMA (Riechers et al. 2011b, and in prep.).

In their parallel work on strongly lensed SMGs (Vieira et al. 2010, 2013; Hezaveh et al. 2013; Spilker et al. 2016, and the references therein), the SPT group used a selection based on the 1.4 mm continuum flux density. The ALMA blind redshift survey of these 1.4 mm-selected SMGs shows a flat redshift distribution in the range $z = 2$–4, with a mean value of $\langle z \rangle = 3.5$, being in contrast to the 850–870 $\mu$m SCUBA/LABOCA-selected sample (Weiß et al. 2013; Spilker et al. 2016). This can be explained by the different flux limits of the two samples, namely, the SPT-selected sources are intrinsically brighter than the classic 850–870 $\mu$m SCUBA/LABOCA-selected SMGs (Koprowski et al. 2014).

Efficient CO detection in lensed SMGs has significantly enlarged the sample size of multi-$J$ CO detections, with the aim of allowing statistical studies. Thus, we present here our observations of multi-$J$ CO emission lines in 16 *H*-ATLAS lensed SMGs at $z \sim 2$–4, for a better understanding of the physical conditions of the ISM in high-redshift SMGs on a statistical basis.

Although there is a large number of CO observations in high-redshift sources, only a few high-density tracers with high dipole, for example, HCN, have so far been detected, most of which in QSOs (e.g. Gao et al. 2007; Riechers et al. 2010), and even fewer detections in SMGs (e.g. Oteo et al. 2017). Submm $H_2O$ lines, another dense gas tracer, have been reported in 12 *H*-ATLAS lensed SMGs (Omont et al. 2011; Omont et al. 2013, O13 hereafter and Yang et al. 2016, Y16 hereafter) using IRAM NOrthern Extended Millimeter Array (NOEMA), and also in other galaxies (see the review by van Dishoeck et al. 2013). An open question is whether or not the submm $H_2O$ emission lines trace similar regions as traced by mid/high-$J$ CO and HCN. The difficulty of the comparison is coming from the currently limited high-resolution mapping of the submm $H_2O$ lines. However, by comparing line profiles of unresolved observations of lensed SMGs, Y16 argue that the mid-$J$ CO lines originate in similar conditions to the submm $H_2O$ lines. This can be further tested by a larger sample from this work, and more directly, the high angular-resolution mapping of the emissions: see, for example, the cases of SDP81 as probed by ALMA (ALMA Partnership, Vlahakis et al. 2015), NCv1.143 observed by NOEMA and of G09v1.97 through ALMA observations (Yang et al. in prep.).

In this paper, we study the physical properties of the molecular gas in a sample of 16 lensed SMGs at $z \sim 2$–4 by analysing their multiple-$J$ CO emission lines. This paper is organised as follows: we describe our sample, the observations and data reduction in Section 2. The observed properties of the multi-$J$ CO emission lines are presented in Section 3. The global properties of the SMGs together with the differential lensing effect is discussed in Section 4. A detailed discussion of the CO excitation is given in Section 5. Section 5.3 describes the discussion of molecular gas mass and star formation. We compare the emission lines of CO and submm $H_2O$ in Section 5.4. Finally, we summarise our results in Section 6. A spatially-flat $\Lambda$CDM cosmology with $H_0 = 67.8 \pm 0.9\,\mathrm{km\,s^{-1}\,Mpc^{-1}}$, $\Omega_M = 0.308 \pm 0.012$ (Planck Collaboration et al. 2016) and Salpeter's (1955) initial mass function (IMF) has been adopted throughout this paper.

**Table 1:** Basic information on the CO rotational lines and [C I] $^3$P fine structure lines used in this paper.

| Molecule | Transition $J_U \rightarrow J_L$ | $\nu_{rest}$ (GHz) | $E_{up}/k$ (K) | $A_{UL}$ ($s^{-1}$) | $n_{crit}$ ($cm^{-3}$) |
|---|---|---|---|---|---|
| CO | $1 \rightarrow 0$ | 115.271 | 5.5 | $7.20 \times 10^{-8}$ | $2.4 \times 10^{2}$ |
| | $2 \rightarrow 1$ | 230.538 | 16.6 | $6.91 \times 10^{-7}$ | $2.1 \times 10^{3}$ |
| | $3 \rightarrow 2$ | 345.796 | 33.2 | $2.50 \times 10^{-6}$ | $7.6 \times 10^{3}$ |
| | $4 \rightarrow 3$ | 461.041 | 55.3 | $6.12 \times 10^{-6}$ | $1.8 \times 10^{4}$ |
| | $5 \rightarrow 4$ | 576.268 | 83.0 | $1.22 \times 10^{-5}$ | $3.6 \times 10^{4}$ |
| | $6 \rightarrow 5$ | 691.473 | 116.2 | $2.14 \times 10^{-5}$ | $6.3 \times 10^{4}$ |
| | $7 \rightarrow 6$ | 806.652 | 154.9 | $3.42 \times 10^{-5}$ | $1.0 \times 10^{5}$ |
| | $8 \rightarrow 7$ | 921.800 | 199.1 | $5.13 \times 10^{-5}$ | $1.5 \times 10^{5}$ |
| | $9 \rightarrow 8$ | 1036.912 | 248.9 | $7.33 \times 10^{-5}$ | $2.1 \times 10^{5}$ |
| | $10 \rightarrow 9$ | 1151.985 | 304.2 | $1.00 \times 10^{-4}$ | $2.9 \times 10^{5}$ |
| | $11 \rightarrow 10$ | 1267.014 | 365.0 | $1.34 \times 10^{-4}$ | $3.9 \times 10^{5}$ |
| [C I] | $^3P_1 \rightarrow {}^3P_0$ | 492.161 | 23.6 | $7.88 \times 10^{-8}$ | $4.9 \times 10^{2}$ |
| | $^3P_2 \rightarrow {}^3P_1$ | 809.342 | 62.4 | $2.65 \times 10^{-7}$ | $9.3 \times 10^{2}$ |

**Notes.** Critical density $n_{crit,UL} \equiv A_{UL}/\Sigma_{i\neq U}\gamma_{Ui}$ (e.g. Tielens 2005). $A_{UL}$ is the Einstein coefficient for spontaneous emission from level U to L, and $\gamma_{Ui}$ is the collision rate coefficient. The critical densities ($n_{crit}$) are calculated by assuming a gas temperature $T_k$ = 100 K, and an ortho-$H_2$ to para-$H_2$ ratio of 3 and an optically thin regime. The rest-frame frequencies ($\nu_{rest}$), upper-level energies ($E_{up}/k$) and Einstein $A$ coefficients are taken from the LAMDA database (Schöier et al. 2005). The collision rate coefficients are from Yang et al. (2010). Throughout this paper, we refer to [C I]($^3P_1 \rightarrow {}^3P_0$) as [C I](1–0) and [C I]($^3P_2 \rightarrow {}^3P_1$) as [C I](2–1).

## 2. Sample, observations, and data reduction

### 2.1. Selection of the lensed SMGs

Unlike the previously studied SMGs, our sample is drawn from shorter wavelengths using *Herschel* SPIRE photometric data at 250, 350, and 500 $\mu$m. In order to find the strongly lensed SMGs, all of our targets were selected from the *H*-ATLAS catalogue (Valiante et al. 2016) with a criterion of $S_{100\,\mu m} > 100$ mJy based on the theoretical models of the submm source number counts (e.g. Negrello et al. 2010, 2017). Then, a Submillimeter Array (SMA) subsample was constructed based on the availability of previously spectroscopically confirmed redshifts obtained by CO observations (Bussmann et al. 2013, hereafter Bu13); it includes all high-redshift *H*-ATLAS sources with $F_{500\,\mu m} > 200$ mJy in the GAMA and NGP fields (300 deg$^2$). From SMA 880 $\mu$m images and the identification of the lens deflectors and their redshifts, Bu13 built lensing models for most of them.

Our sample was thus extracted from Bu13's *H*-ATLAS-SMA sources with the initial goal of studying their $H_2O$ emission lines (see Table 6 of Y16). It consists of 17 lensed SMGs with redshift from 1.6 to 4.2. We have detected submm $H_2O$ emission lines in 16 sources observed with only one non-detection from the AGN-dominated source, G09v1.124 (O13; Y16, Table 2). However, for this CO follow-up observation, we dropped three sources among the $H_2O$-detected 16: SDP 11 due to its low redshift z < 2, NCv1.268 because of its broad linewidth that brings difficulties for line detection in a reasonable observing time, and G15v2.779 because it has already been well observed by Cox et al. (2011). Nevertheless, we included G15v2.779 in discussing the main results to have a better view of CO properties for the whole sample. Our CO sample of 14 sources (13 observed with the IRAM's Eight Mixer Receiver, for example, EMIR, in this work plus G15v2.779 studied by Cox et al. 2011) is thus a good representative for the brightest high-redshift *H*-ATLAS lensed sources with $F_{500\,\mu m} > 200$ mJy and at $z > 2$ (except SDP 81 with $F_{500\,\mu m} \sim 174$ mJy). Besides these 14 sources, we also include two slightly less bright sources, G12v2.890 and G12v2.257, down to $F_{500\,\mu m} > 100$ mJy. In the end, as listed in Table 3, the entire sample includes 16 lensed SMGs from redshift 2.2 to 4.2.

The lensing models for twelve of the SMGs are provided by Bu13 through SMA 880 $\mu$m continuum observations. Table 3 lists the magnification factors ($\mu_{880}$) and inferred intrinsic properties of these galaxies together with their CO redshifts from previous blind CO redshift observations. After correcting for the magnification, their intrinsic infrared luminosities are $\sim$4–20 $\times$ $10^{12}\,L_\odot$. Since the lensed nature of these SMGs and their submm selection may bias the sample, we will compare their properties with other SMG samples later from Section 3 to Section 5.3.

In this work, in order to explore the physical properties of the bulk of the molecular gas, we targeted the rotational emission lines of CO, mostly from $J_{up}$ = 3 to 8 and up to 11 in a few cases. [C I](2–1) line is also observed "for free" together with CO(7–6). Basic information such as the frequencies, upper-level energies, Einstein $A$ coefficients and critical densities of the CO and [C I] lines are listed in Table 1. The targeted CO lines are selected based on their redshifted frequencies so that they could be observed in a reasonably good atmospheric window in EMIR bands. In total, we observed 55 CO lines, with 8 [C I](2–1) lines acquired simultaneously with CO(7–6) in 15 sources (Table 2).

### 2.2. Observation and data reduction

The observations were carried out from 2011 June 30th to 2012 March 13th, and from 2015 May 26th to 2016 February 22nd using the multi-band heterodyne receiver EMIR (Carter et al. 2012) on the IRAM-30m telescope. Bands at 3 mm, 2 mm, 1.3 mm and 0.8 mm (corresponding to E090, E150, E230 and E330 receivers, respectively) were used for detecting multiple CO transitions. Each bandwidth covers a frequency range of 8 GHz. We selected the wide-band line multiple auto-correlator (WILMA) with a 2 MHz spectral resolution and the fast Fourier Transform Spectrometer with a 200 kHz resolution (FTS200) as back ends simultaneously during the observations. Given that the angular sizes of our sources are all less than 8″, observations were performed in wobbler switching mode with a throw of 30″. Bright planet/quasar calibrators including Mars, 0316+413, 0851+202, 1226+023, 1253-055, 1308+326 and 1354+195 were used for pointing and focusing. The pointing model was checked every two hours for each source using the pointing calibrators, while the focus was checked after sunrise and sunset. The data were calibrated using the standard dual method. The observations were performed in average weather conditions with $\tau_{225\,GHz} \lesssim 0.5$ during 80% of the observing time.

Data reduction was performed using the `GILDAS`[1] packages `CLASS` and `GREG`. Each scan of the spectrum was inspected by eye and the bad data (up to 10%) were discarded. The baseline-removed spectra were co-added according to the weights derived from the noise level of each. We also note that due to the upgrade of the optical system of the IRAM-30m telescope in November 2015, the telescope efficiency has been changed by small factors for lower band receivers (see the EMIR commissioning report by

[1] See `http://www.iram.fr/IRAMFR/GILDAS` for more information about the GILDAS softwares.

**Table 2:** Observation log.

| Source | IAU name | RA (J2000) | DEC (J2000) | Observed Lines | $H_2O$ observation | $H_2O$ ref |
|---|---|---|---|---|---|---|
| G09v1.97 | J083051.0+013224 | 08:30:51.156 | +01:32:24.35 | CO(3–2), (5–4), (6–5), (7–6), [C I](2–1) | $2_{11}$–$2_{02}$, $3_{21}$–$3_{12}$ | 1 |
| G09v1.40 | J085358.9+015537 | 08:53:58.862 | +01:55:37.70 | CO(2–1), CO(4–3), (6–5), (7–6), [C I](2–1) | $2_{11}$–$2_{02}$ | 1 |
| SDP17b | J090302.9−014127 | 09:03:03.031 | −01:41:27.11 | CO(3–2), (4–3), (7–6), (8–7), [C I](2–1) | $2_{02}$–$1_{11}$ | 2, 3 |
| SDP81 | J090311.6+003906 | 09:03:11.568 | +00:39:06.43 | CO(3–2), (5–4), (6–5), (10-9) | $2_{02}$–$1_{11}$ | 2 |
| G12v2.43 | J113526.3−014605 | 11:35:26.273 | −01:46:06.55 | CO(3–2), (4–3), (5–4), (6–5), (8–7), (10-9) | $2_{02}$–$1_{11}$, $3_{21}$–$3_{12}$ | 1 |
| G12v2.30 | J114637.9−001132 | 11:46:37.980 | −00:11:31.80 | CO(4–3), (5–4), (6–5), (8–7), (11-10) | $2_{02}$–$1_{11}$ | 2 |
| NCv1.143 | J125632.7+233625 | 12:56:32.544 | +23:36:27.63 | CO(3–2), (5–4), (6–5), (7–6), (10-9)[a], [C I](2–1) | $2_{11}$–$2_{02}$, $3_{21}$–$3_{12}$ | 1 |
| NAv1.195 | J132630.1+334410 | 13:26:30.216 | +33:44:07.60 | CO(5–4) | $2_{02}$–$1_{11}$, $3_{21}$–$3_{12}$[b] | 1 |
| NAv1.177 | J132859.3+292317 | 13:28:59.246 | +29:23:26.13 | CO(3–2), (5–4), (7–6), (8–7) , [C I](2–1) | $2_{02}$–$1_{11}$, $3_{21}$–$3_{12}$ | 1 |
| NBv1.78 | J133008.4+245900 | 13:30:08.520 | +24:58:59.17 | CO(5–4), (6–5) | $2_{02}$–$1_{11}$, $3_{21}$–$3_{12}$ | 1 |
| NAv1.144 | J133649.9+291801 | 13:36:49.900 | +29:18:01.00 | CO(3–2), (4–3), (7–6), (8–7), [C I](2–1) | $2_{11}$–$2_{02}$ | 2 |
| NAv1.56 | J134429.4+303036 | 13:44:29.518 | +30:30:34.05 | CO(5–4) | $2_{11}$–$2_{02}$ | 1 |
| G15v2.235 | J141351.9−000026 | 14:13:51.900 | −00:00:26.00 | CO(3–2), (4–3), (5–4), (7–6), (9–8), [C I](2–1) | – | – |
| G12v2.890 | J113243.1−005108 | 11:32:42.970 | −00:51:08.90 | CO(3–2), (5–4), (9–8) | – | – |
| G12v2.257 | J115820.2−013753 | 11:58:20.190 | −01:37:55.20 | CO(3–2), (4–3), (7–6), (8–7), [C I](2–1) | – | – |

**Notes.** RA and DEC are the J2000 coordinates of the SMA 880 $\mu$m images from Bu13 (except for G12v2.890 and G12v2.257 which were not observed by SMA, *Herschel* SPIRE image coordinates in Valiante et al. 2016 are then used instead). These coordinates were used for observations. See in Appendix Table B.1 for the observing frequencies. The $H_2O$ observations for each source are reported in (1) Yang et al. (2016); (2) Omont et al. (2013); (3) Omont et al. (2011). The sources have been divided into two groups as in the table, see Section 2.1 for more details. [(a)] This CO(10-9) data is taken from NOEMA/IRAM project S15CV (Yang et al. in prep.). [(b)] Except for this line, the rest $H_2O$ lines are detected.

**Table 3:** Previously observed properties of the entire sample.

| Source | ID | $z_{\rm spec}$ | ref$_{z_{\rm spec}}$ | $F_{250}$ (mJy) | $F_{350}$ (mJy) | $F_{500}$ (mJy) | $F_{880}$ (mJy) | $f_{1.4\rm GHz}$ (mJy) | $T_{\rm d}$ (K) | $\mu L_{\rm IR}$ ($10^{13}$ $L_\odot$) | $\mu_{880}$ | $L_{\rm IR}$ ($10^{12}$ $L_\odot$) | $r_{\rm half}$ (kpc) | $\Sigma_{SFR}$ ($M_\odot$ yr$^{-1}$ kpc$^{-2}$) |
|---|---|---|---|---|---|---|---|---|---|---|---|---|---|---|
| G09v1.97 | 1 | 3.634 | 1 | 260 ± 7 | 321 ± 8 | 269 ± 9 | 85.5 ± 4.0 | < 0.45 | 44 ± 1 | 15.5 ± 4.3 | 6.9 ± 0.6 | 22.5 ± 6.5 | 0.9 | 910 ± 147 |
| G09v1.40 | 2 | 2.0923 | 1 | 389 ± 7 | 381 ± 8 | 241 ± 9 | 61.4 ± 2.9 | 0.75 ± 0.15 | 36 ± 1 | 6.6 ± 2.5 | 15.3 ± 3.5 | 4.3 ± 1.9 | 0.4 | 775 ± 303 |
| SDP17b | 3 | 2.3051 | 2 | 347 ± 7 | 339 ± 8 | 219 ± 9 | 54.7 ± 3.1 | < 0.51 | 38 ± 1 | 7.1 ± 2.6 | 4.9 ± 0.7 | 14.5 ± 5.7 | 3.1 | 52 ± 36 |
| SDP81 | 4 | 3.042 | 3 | 138 ± 7 | 199 ± 8 | 174 ± 9 | 78.4 ± 8.2 | 0.61 ± 0.16 | 34 ± 1 | 5.9 ± 1.5 | 11.1 ± 1.1 | 5.3 ± 1.5 | 3.3 | 14 ± 6 |
| G12v2.43 | 5 | 3.1276 | 4 | 290 ± 7 | 295 ± 8 | 216 ± 9 | 48.6 ± 2.3 | < 0.45 | 39 ± 2[a] | 9.0 ± 0.2[a] | – | – | – | – |
| G12v2.30 | 6 | 3.2592 | 4 | 290 ± 6 | 356 ± 7 | 295 ± 8 | 86.0 ± 4.9 | < 0.42 | 41 ± 1 | 15.6 ± 4.1 | 9.5 ± 0.6 | 16.4 ± 4.4 | 1.6 | 166 ± 27 |
| NCv1.143 | 7 | 3.565 | 1 | 214 ± 7 | 291 ± 8 | 261 ± 9 | 97.2 ± 6.5 | 0.61 ± 0.16 | 40 ± 1 | 13.0 ± 4.0 | 11.3 ± 1.7 | 11.4 ± 3.9 | 0.8[b] | 1043 ± 384[b] |
| NAv1.195 | 8 | 2.951 | 5 | 179 ± 7 | 279 ± 8 | 265 ± 9 | 65.2 ± 2.3 | < 0.42 | 36 ± 1 | 7.5 ± 2.0 | 4.1 ± 0.3 | 18.3 ± 5.1 | 1.6 | 213 ± 44 |
| NAv1.177 | 9 | 2.778 | 6 | 264 ± 9 | 310 ± 10 | 261 ± 10 | 50.1 ± 2.1 | < 0.45 | 32 ± 1[a] | 6.2 ± 0.2[a] | – | – | – | – |
| NBv1.78 | 10 | 3.1112 | 1 | 273 ± 7 | 282 ± 8 | 214 ± 9 | 59.2 ± 4.3 | 0.67 ± 0.20 | 43 ± 1 | 10.8 ± 3.9 | 13.0 ± 1.5 | 8.4 ± 3.1 | 0.6 | 1094 ± 1411 |
| NAv1.144 | 11 | 2.2024 | 4 | 295 ± 8 | 294 ± 9 | 191 ± 10 | 36.8 ± 2.9 | < 0.42 | 39 ± 1 | 6.0 ± 3.5 | 4.4 ± 0.8 | 13.6 ± 8.3 | 0.9 | 615 ± 581 |
| NAv1.56 | 12 | 2.3010 | 4 | 481 ± 9 | 484 ± 13 | 344 ± 11 | 73.1 ± 2.4 | 1.12 ± 0.27 | 38 ± 1 | 11.5 ± 3.1 | 11.7 ± 0.9 | 9.8 ± 2.8 | 1.5 | 138 ± 82 |
| G15v2.235 | 13 | 2.4782 | 4 | 190 ± 7 | 240 ± 8 | 200 ± 9 | 33.3 ± 2.6 | < 0.59 | 32 ± 2 | 2.8 ± 0.7 | 1.8 ± 0.3 | 15.6 ± 4.7 | 1.7 | 275 ± 101 |
| G12v2.890 | 14 | 2.5778 | 4 | 74 ± 13 | 118 ± 19 | 106 ± 18 | – | < 0.45 | 30 ± 2 | 2.5 ± 0.3[c] | – | – | – | – |
| G12v2.257 | 15 | 2.1911 | 4 | 132 ± 21 | 152 ± 24 | 107 ± 18 | – | < 0.82 | 32 ± 2 | 2.6 ± 0.3[c] | – | – | – | – |
| G15v2.779 | 16 | 4.243 | 7 | 115 ± 19 | 308 ± 47 | 220 ± 34 | 90.0 ± 5.0 | < 0.46 | 41 ± 1 | 10.1 ± 3.0 | 4.6 ± 0.5 | 22.0 ± 7.0 | 3.8 | 53 ± 11 |

**Notes.** $z_{\rm spec}$ is the redshift inferred from previous CO detection as reported by: (1) Riechers et al. (in prep.); (2) Lupu et al. (2012); (3) Fu et al. (2012); (4) Harris et al. (2012); (5) Harris et al. (in prep.); (6) Krips et al. (in prep.); (7) Cox et al. (2011). $F_{250}$, $F_{350}$ and $F_{500}$ are the *Herschel* SPIRE flux densities at 250, 350 and 500 $\mu$m, respectively (Valiante et al. 2016); $F_{880}$ is the 880 $\mu$m SMA flux density (Bu13); $f_{1.4\rm GHz}$ is the 1.4 GHz band flux density from the VLA FIRST survey (Becker et al. 1995), and we use 3 $\sigma$ as upper limits for non-detections; $T_{\rm d}$ is the cold-dust temperature taken from Bu13 (note that the errors quoted here are underestimated since the uncertainties from differential lensing and single-temperature dust SED assumption were not fully considered). $\mu L_{\rm IR}$ is the apparent total infrared luminosity (8–1000 $\mu$m) mostly inferred from Bu13. $\mu_{880}$ is the lensing magnification factor for the 880 $\mu$m images (Bu13); $r_{\rm half}$ and $\Sigma_{\rm SFR}$ are the intrinsic half-light radius at 880 $\mu$m and the lensing-corrected surface *SFR* density (*SFR* is derived from $L_{\rm IR}$ using the calibration of Kennicutt 1998a, $SFR = 1.73 \times 10^{-10} L_{\rm IR}$ $M_\odot$ yr$^{-1}$, by assuming a Salpeter IMF); Since G12v2.890 and G12v2.257 are significantly weaker in submm fluxes compared with other sources, and also they lack SMA 880 $\mu$m observation, we put them into a separate group. G15v2.779 is also included in the table for comparison. [(a)] These values of $T_{\rm d}$ and $\mu L_{\rm IR}$ are not given in Bu13, thus we infer them from modified black-body dust SED fitting using the submm/mm photometry data listed in this table. [(b)] This $r_{\rm half}$ is obtained based on the A-configuration NOEMA observation (Yang et al., in prep.), with a better spatial resolution and image quality comparing to the SMA one. [(c)] The values are from Harris et al. (2012).

Marka & Kramer in 2015[2] for details). All our sources are a factor of 3–7 smaller compared with the beamsize of IRAM-30m at the observing frequencies, so that they can be treated as point sources. Accordingly, we apply the different point source conversion factors (in the range of 5.4–9.7 Jy/K depending on the optics and the frequency) that convert $T_{\rm A}^*$ in units of K into flux density in units of Jy for the spectra. A typical absolute flux calibration uncertainty of $\sim 10\%$ is also taken into account. We then fit the co-added spectra with Gaussian profiles using the Levenberg-Marquardt least-square minimisation code `MPFIT` (Markwardt 2009) for obtaining the velocity integrated line fluxes, linewidths (FWHM), and the line centroid positions.

[2] Report is available on the IRAM-30m wiki page: `https://www.iram.es/IRAMES/mainWiki/Iram30mEfficiencies`.

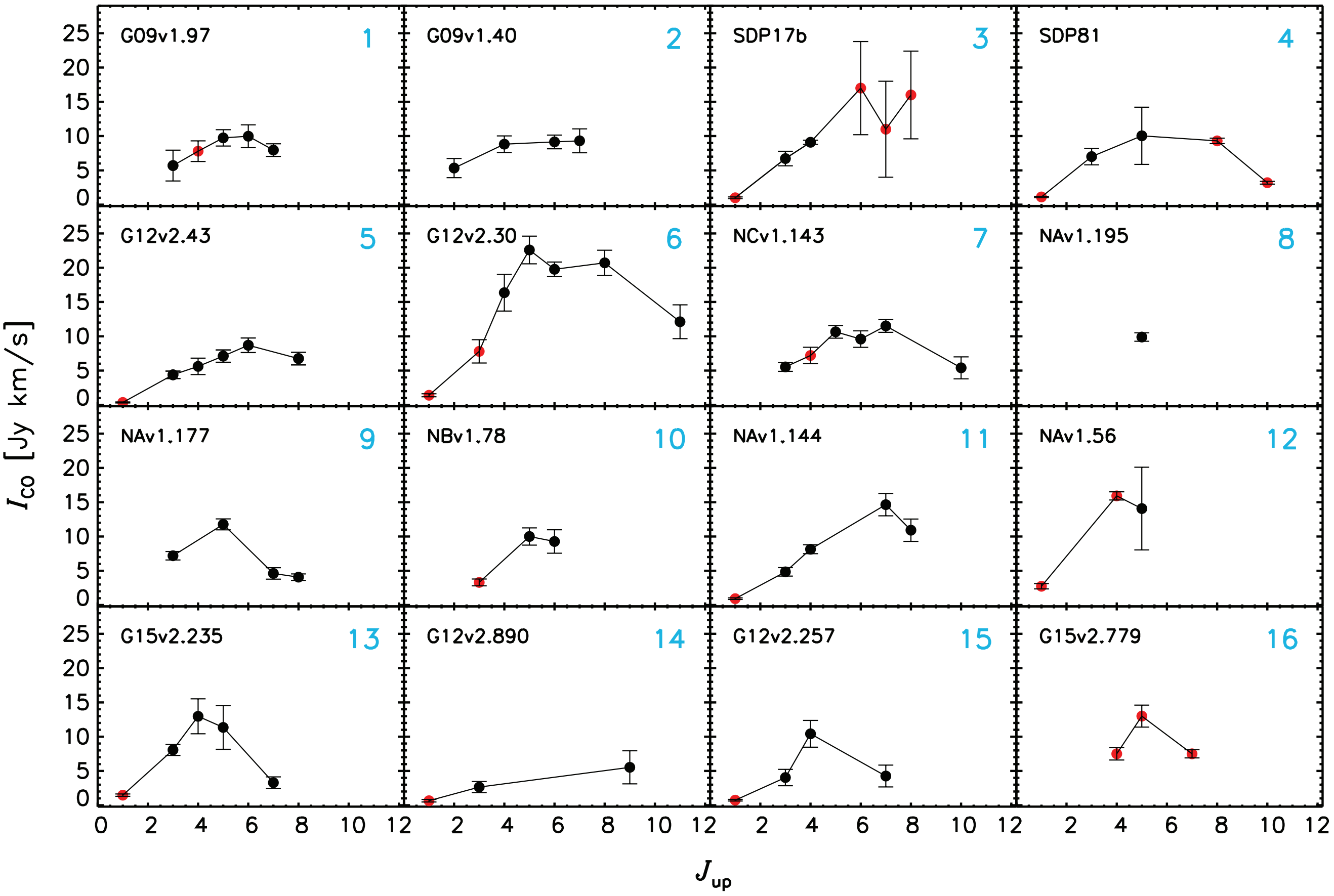

**Fig. 1:** The distribution of the observed velocity-integrated CO line flux density versus the rotational quantum number $J_{\rm up}$ for each transition, i.e. CO SLEDs. Black dots with error bars are the velocity integrated flux densities from this work. Red dots are the data from other works: all CO(1–0) data are from Harris et al. (2012); CO(4–3) in G09v1.97 is from Riechers et al. (in prep.); CO(6–5), CO(7–6) and CO(8–7) in SDP 17b are from Lupu et al. (2012); CO(8–7) and CO(10-9) in SDP 81 are from ALMA Partnership, Vlahakis et al. (2015); CO(3–2) in G12v2.30, CO(4–3) in NCv1.143 and CO(3–2) in NBv1.78 are from O13; CO(4–3) in NAv1.56 is from Oteo et al. (in prep.). For a comparison, we also plot the CO SLED of G15v2.779 (Cox et al. 2011). We mark an index number for each source in turquoise following Table 3 for the convenience of discussion.

## 3. Observation results

### 3.1. Observed CO line properties

We have detected 47 out of 55 $J \geq 2$ CO and 7 out of 8 [C I](2–1) observed emission lines in 15 *H*-ATLAS lensed SMGs (signal to noise ratio S/N $\gtrsim 3$, see Table B.1). The observed spectra are displayed in Fig. A.1 and the fluxes are also shown in the form of CO SLEDs in Fig. 1, indicated by black data points. Detected multi-*J* CO lines are bright with velocity-integrated flux densities ranging from 2 to 22 Jy km s$^{-1}$. To further compare the CO SLEDs, the CO(3–2) normalised CO SLEDs are plotted in Fig. 2 for all the *H*-ATLAS sources with CO(3–2) detections, overlaid with those of the Milky Way (Fixsen et al. 1999) and the Antennae Galaxy (Zhu et al. 2003). The CO SLEDs are mostly peaking from $J_{\rm up} = 5$ to $J_{\rm up} = 8$. The histogram of the flux ratio between CO(1–0) and CO(3–2) shows that the average $I_{\rm CO(1-0)}/I_{\rm CO(3-2)}$ ratio is $0.17 \pm 0.05$, which is $1.3 \pm 0.4$ times smaller than that of the unlensed SMGs (Bothwell et al. 2013, hereafter Bo13). This is likely to be related to differential lensing, in that the magnification factor of CO(3–2) is larger than that of CO(1–0) due to the differences in their emitting sizes. The resulting ratio of $I_{\rm CO(3-2)}/I_{\rm CO(1-0)}$ is thus larger in our lensed sources compared to the unlensed SMGs. We further discuss this in Section 4.3. Here we define the ratio between the lensing magnification factor of CO(3–2) (assumed to be equal to the magnification factor $\mu_{880}$ derived from SMA 880 $\mu$m images) and CO(1–0) to be

$$\frac{\mu_{\rm CO(3-2)}}{\mu_{\rm CO(1-0)}} = \frac{\mu_{880}}{\mu_{\rm CO(1-0)}} = 1.3 \pm 0.4. \quad (1)$$

We correct for differential lensing for CO(1–0) data using this factor as described in Section 3.3.

One of the most important characteristics of the CO lines is its linewidth. The CO linewidth (FHWM) distribution of our lensed SMGs is displayed by the black solid line in the upper panel of Fig. 3 with the corresponding cumulative fraction shown in the lower panel. This curve shows that the linewidths are distributed between 208 and 830 km s$^{-1}$ (see Table 4 for the weighted average values of the linewidth). Around 50% of the sources have linewidths close to or smaller than 300 km s$^{-1}$. The median of the whole distribution is 333 km s$^{-1}$ and its average value $418 \pm 216$ km s$^{-1}$. Figure 3 also displays the linewidth distributions and the cumulative curves of two other samples of unlensed SMGs (orange dash-dotted lines) and lensed SPT-selected SMGs (green dashed lines) for comparison as discussed in Section 3.2.

Among our 16 sources, 12 of them show a single Gaussian CO line profile. SDP 81, NBv1.78 and G15v2.235 have double Gaussian CO line profiles. Although G09v1.97 might show a single Gaussian line profile, it is likely that there is a weak component in the blue wing, that we have confirmed by a higher sensitivity ALMA observation (Yang et al. in prep.). The high S/N PdBI spectrum of CO(4–3) line in NAv1.56 (Oteo in prep.) also shows a line profile consisting of a narrow blue velocity component and a broad red component. However, due to the limited S/N, we can only identify the CO(5–4) line observed by EMIR with a single Gaussian profile.

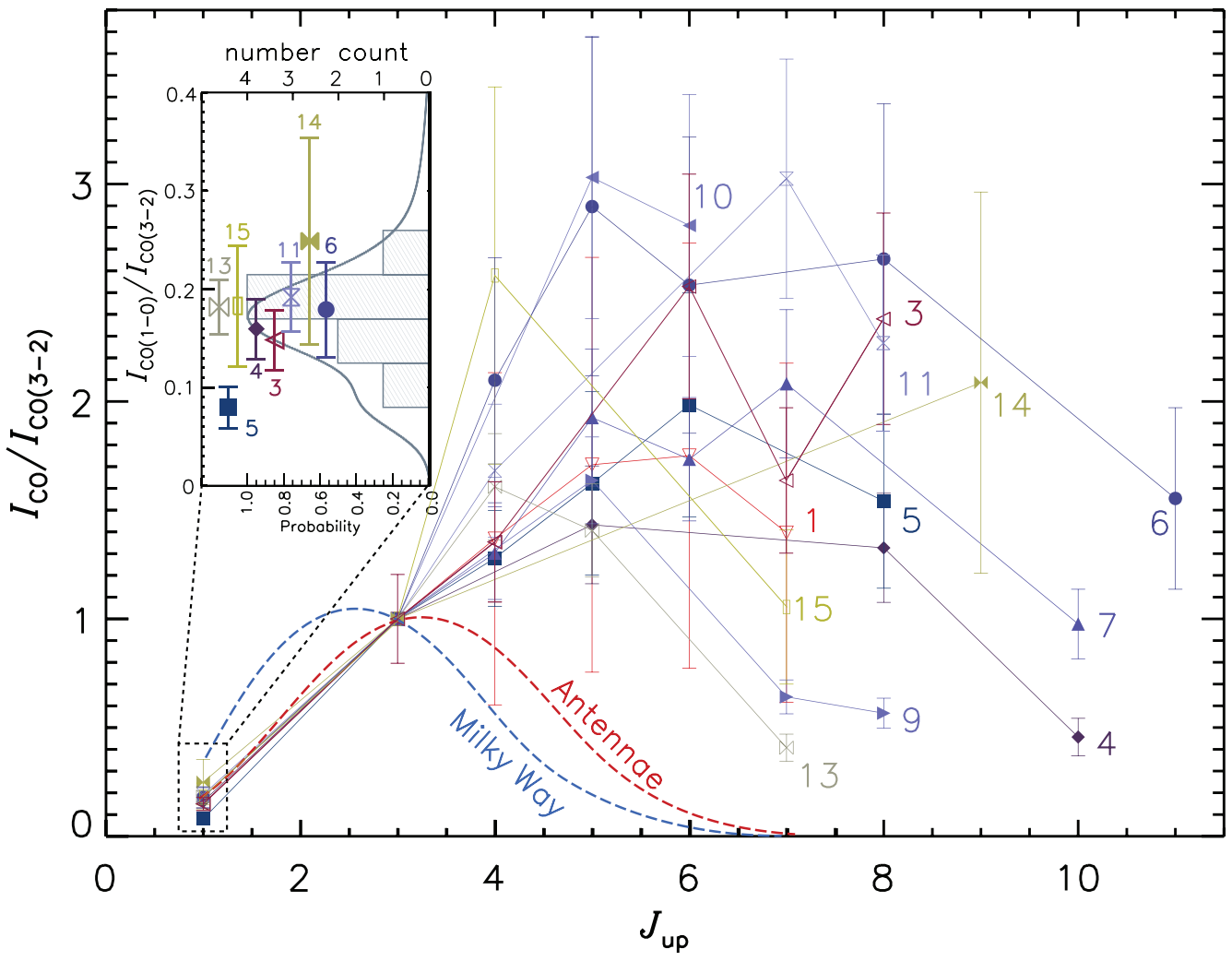

**Fig. 2:** The observed CO(3–2)-normalised CO SLED (without lensing correction) of the *H*-ATLAS SMGs, in which both $J_{up} = 1$ and $J_{up} = 3$ CO data are available. The inset shows a zoom-in plot of the flux ratio of CO(1–0)/CO(3–2). The grey histogram shows the ratio distribution, while the grey line shows the probability density plot of the line ratio (considering the error). A mean ratio of $I_{CO(1-0)}/I_{CO(3-2)} = 0.17 \pm 0.05$ has been found for our lensed SMGs. This is $1.3 \pm 0.4$ times smaller than that of the unlensed SMGs of Bo13. For comparison, we also plot the SLED of the Milky Way and the Antennae Galaxy.

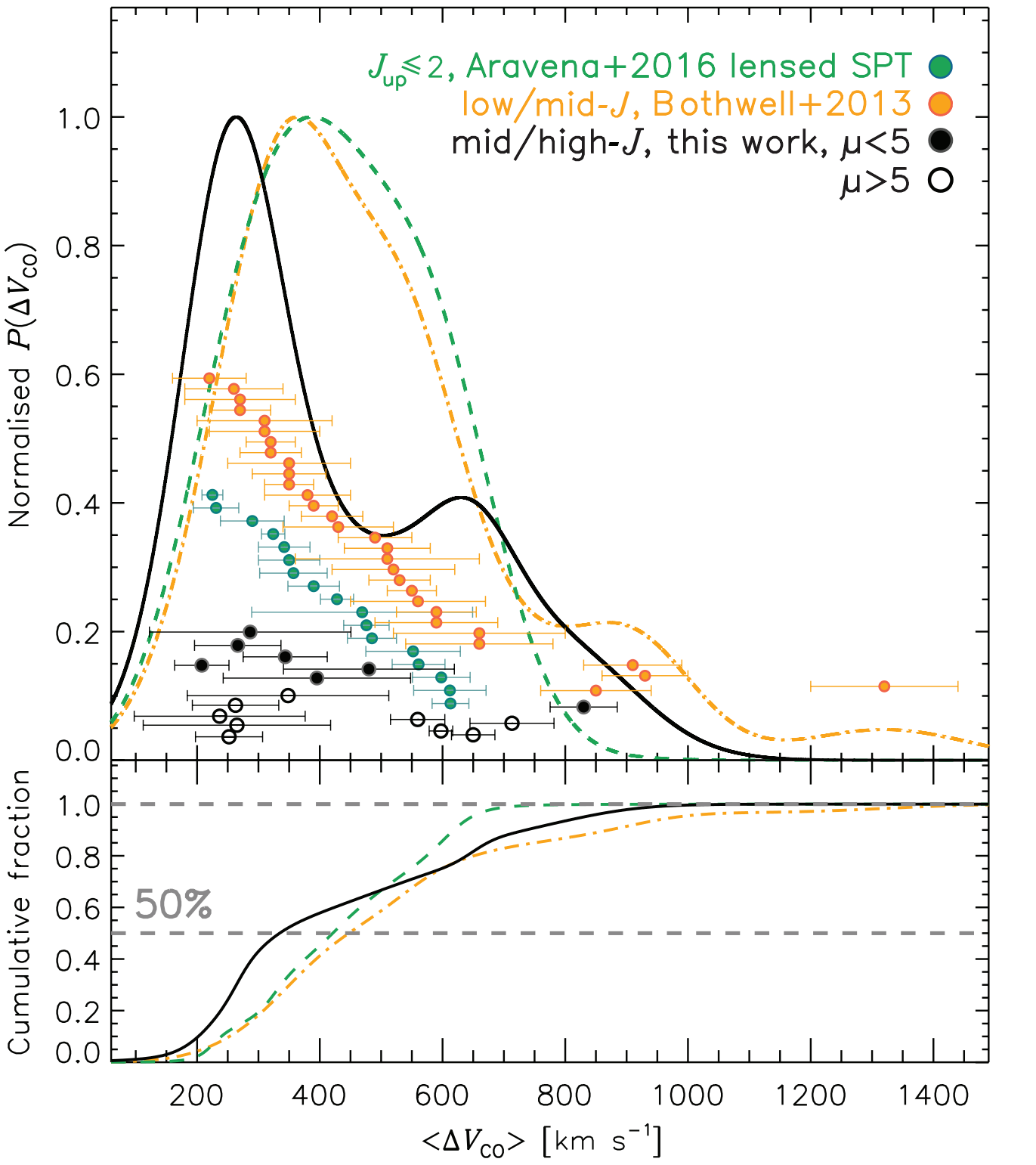

**Fig. 3:** *Upper panel*: Linewidths with errors from three different samples, with probability distributions obtained by adaptive kernel density estimate (Silverman 1986): black symbols and line are from this work, orange symbols and dashed-dotted line are the $J_{up} \geq 2$ CO linewidth distribution in unlensed SMGs (Bo13) and the green symbols and dashed line represent the linewidth from the $J_{up} \leq 2$ CO lines of the lensed SPT sources (Aravena et al. 2016b). Our lensed sources with $\mu > 5$ are indicated with open circles while the other sources are shown in filled circles. We note that although there is no lensing model for G12v2.43 and NAv1.144, it is suggested that their $\mu$ are likely to be $\sim 10$ (see Section 4.2 and Fig. 6). Thus, they are also marked with open circles. *Lower panel*: Cumulative distribution of $\langle \Delta V_{CO} \rangle$ for the three samples with the same colour code.

The CO line profiles between different $J_{up}$ levels within each source may vary, since their critical density and excitation temperature are different. However, by checking our CO spectral data as displayed in Fig. A.1, we find the differences between the line profiles (mostly by checking the linewidth) are insignificant given the current S/N. Their linewidths generally agree with each other within their uncertainties.

### 3.2. Comparing our sample to the general SMG population

If we wish to use our sample of lensed sources and the increased sensitivity allowed by magnification to infer general properties of the SMG population, it is important to investigate whether or not it is representative of this population and to recognise the possible biases introduced by lensing selection. For this purpose, we may compare it, especially for CO emission, with the sample of unlensed SMGs of the comprehensive CO study by Bo13. Thanks to early redshift determination, this sample of 32 SMGs initially detected at 850 $\mu$m was the object of a large program at IRAM/NOEMA detecting multiple low/mid-*J* CO lines. As discussed by the authors, although not completely free from possible biases, the sample appears to be a good representative of the whole SMG population. Compared to ours, its redshift distribution is similarly concentrated in the redshift range 2 to 3, with a similar extension up to $\sim 3.5$, but it also extends below 2 down to z~1 in contrast to our sample. Both samples have very comparable distributions of their FIR luminosity $L_{FIR}$ (a typical ratio between $L_{IR}$ and $L_{FIR}$ is 1.9; e.g. Dale et al. 2001), from a few $10^{12}\,L_\odot$ to just above $10^{13}\,L_\odot$, with a mean value of $6.0 \times 10^{12}\,L_\odot$ for the Bo13 sample and $8.3 \times 10^{12}\,L_\odot$ for ours. As expected from the *Herschel* selection of our sample, its dust temperature $T_d$ (Bu13) is slightly higher ($\langle T_d \rangle = 37$ K) than for typical samples of 850 $\mu$m-selected SMGs such as that of Bo13; but there is no obvious evidence of any bias in our lensed sample with respect to the whole *Herschel* SPIRE SMG population.

An important parameter is the extension radii of the dust emission at submm, which is believed to be comparable to that of high-*J* CO emission as discussed by Bo13 (note that the CO(1–0) line is expected to be more extended, see below). Values of this radius for our sources are reported in Table 3 as computed in Bu13 lens models. All values remain <~3 kpc, with a mean value of ~1.5 kpc. A similar distribution was found by Spilker et al. (2016) for a larger sample of similar strongly lensed sources found in the SPT survey. These authors have compared the intrinsic size distribution of the strongly lensed sources (in-cluding Bu13 ones) to a similar number of unlensed SMGs and found no significant differences.

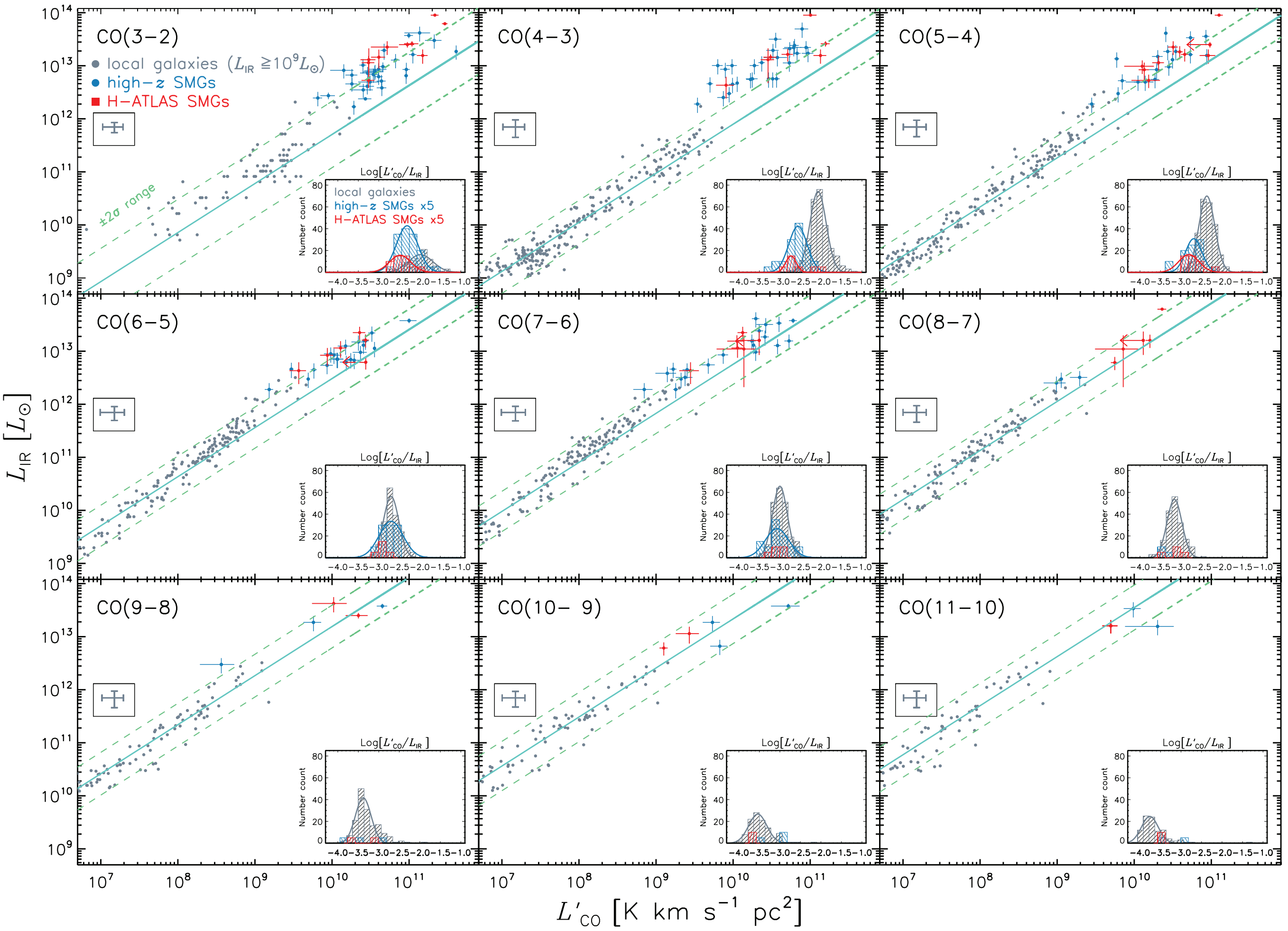

**Fig. 4:** $L_{IR}$ vs. $L'_{CO}$ from local star-forming galaxies to high-redshift SMGs. The low-$z$ data shown in grey including galaxies with $10^9 \leq L_{IR} \leq 10^{12}\ L_\odot$ (only < 20% of the local sources are ULIRGs) are from Liu et al. (2015) and Kamenetzky et al. (2016), with the typical error shown by the grey error-bars. The high-redshift SMG data in blue are from Carilli & Walter (2013) (including HFLS3 from Riechers et al. 2013). The red data points represent the $H$-ATLAS SMGs from this work. Solid light blue lines are linear fits to the local galaxies, showing the average ratios of $L'_{CO}/L_{IR}$, with the $\pm 2\,\sigma$ limits indicated by the dashed green lines. The insets show the histograms of the distribution of the ratio between $L'_{CO}$ and $L_{IR}$ for the three samples. It is clear from the correlation plots and the histograms that the high-redshift SMGs are above the low-redshift correlation for $J_{up} = 3$ and $J_{up} = 4$, with a significant smaller ratio of $L'_{CO}/L_{IR}$. Our $H$-ATLAS SMGs are located in the same region as other SMGs.

In contrast with these similarities of lensed and unlensed SMG samples, the CO linewidths of our lensed flux-limited sample appear anomalously low on average as quoted above. This is obvious from the comparison with the Bo13 sample: see Fig. 3 and the comparison of the distribution of the linewidth, the mean values ($\pm 1\,\sigma$) are 418 ± 216 km s$^{-1}$ for our $H$-ATLAS flux-limited sample, 502 ± 249 km s$^{-1}$ for Bo13 sources with $z \geq 2$, and 430±140 km s$^{-1}$ for the SPT lensed SMG based on CO(1–0) and CO(2–1) observations by Aravena et al. (2016b). The median values of linewidth for the three samples are 333 km s$^{-1}$, 445 km s$^{-1}$ and 420 km s$^{-1}$, respectively, while the mode values are 264 km s$^{-1}$, 346 km s$^{-1}$ and 328 km s$^{-1}$, respectively. The range of the CO linewidths of our lensed SMGs are similar to those of the unlensed Bo13 sample, although the former has a concentration towards a narrower linewidth; more precisely, 50% of them have linewidths $\lesssim 333$ km s$^{-1}$. In order to further compare these three samples, KS-tests were performed. The value of KS probability $P_{KS}$ will be small if the two comparing data sets are significantly different. For the linewidth of our sample and the unlensed SMG sample, $P_{KS} = 0.23$ with a maximum deviation of 0.3; while for comparing our sample with the SPT lensed SMG sample, $P_{KS} = 0.30$ and the maximum deviation equals 0.3. These values of $P_{KS}$ show that the differences among the samples are not statistically significant, indicating that they could arise from similar distributions. Nevertheless, the shapes of the probability distributions and the accumulative distributions of the linewidth for the three samples show some differences as displayed in Fig. 3. The difference between our lensed sample and the SPT one might be expected since the lensed SPT linewidths come from CO(1–0) and CO(2–1) observations which likely trace a larger velocity range of the gas, and thus tend to have larger linewidths compared with mid/high-$J$ CO lines. However, linewidths of the Bo13 SMG sample are also from low/mid-$J$ CO observations. The difference between this unlensed sample and our $H$-ATLAS flux-limited sample is rather likely coming from differential lensing, as discussed in the subsequent subsection. We note, nevertheless, that the per-centage of double-peak CO profiles appears consistent (∼25%) for our sources and those of Bo13.

### 3.3. Intrinsic CO emission properties

We derive the apparent line luminosities, for example, $\mu L_{\rm line}$ (in units of $L_\odot$) and $\mu L'_{\rm line}$ (in units of $\rm K\,km\,s^{-1}\,pc^2$), from the observed line flux densities using the classical formulae as given by Solomon et al. (1992): $L_{\rm line} = 1.04 \times 10^{-3} I_{\rm line} \nu_{\rm rest} (1+z)^{-1} D_{\rm L}^2$ and $L'_{\rm line} = 3.25 \times 10^7 I_{\rm line} \nu_{\rm obs}^{-2} (1+z)^{-3} D_{\rm L}^2$. The resulting line luminosities are listed in Appendix Table B.1. The range of the apparent line luminosities is $\mu L'_{\rm line} \sim 2$–$48 \times 10^{10}\ \rm K\,km\,s^{-1}\,pc^2$. After correcting the lensing magnification, the range of the intrinsic CO line luminosities is $\sim (1$–$60) \times 10^7\,L_\odot$ or $\sim (2$–$170) \times 10^9\ \rm K\,km\,s^{-1}\,pc^2$. As usual, the value of $L'_{\rm CO}$ decreases with increasing $J_{\rm up}$ of the CO lines. Besides CO, we have also derived the instrinsic luminosities of the [C I](2–1) line, observed together with CO(7–6), to be $\sim (3$–$23) \times 10^7\,L_\odot$ or $\sim (2$–$13) \times 10^9\ \rm K\,km\,s^{-1}\,pc^2$.

In the following analysis, we have included multi-$J$ CO data found in the literature for our sources, especially CO(1–0) from Harris et al. (2012), compensating for the absence of this line in our observations (see Caption of Fig. 1). However, due to the differential lensing effect on the CO(1–0) data as discussed in Section 4.3, we only use these CO(1–0) fluxes for the CO line excitation modelling, after applying a factor of $1.3 \pm 0.4$ to correct the differences between the magnification factors of mid/high-$J$ CO and that of CO(1–0) following Eq. (1) (as argued in Section 4.3, we assumed the magnification of mid/high-$J$ CO lines is equal to $\mu_{880}$, and we use $\mu$ as $\mu_{880}$ hereafter if not specified).

After correcting for the lensing magnification, Fig. 4 shows the correlation between the intrinsic values of $L_{\rm IR}$ and $L'_{\rm CO}$ lines from $J_{\rm up} = 3$ to $J_{\rm up} = 11$, over-plotted on the local correlations (Liu et al. 2015, see also Greve et al. 2014; Kamenetzky et al. 2016; Lu et al. 2017). One should note that > 80% of the local sources in Liu et al. (2015) are galaxies with $L_{\rm IR} \leq 10^{12}\,L_\odot$, that is, luminous infrared galaxies (LIRGs) and normal star-forming galaxies. As found previously, most of these local sources can be found well within a tight linear correlation between $L_{\rm IR}$ and $L'_{\rm CO}$ for the mid-$J$ and high-$J$ CO lines, although for the low-$J$ CO lines, the local ULIRGs seem to be lying above the correlation at a $\gtrsim 2\,\sigma$ level, having larger $L_{\rm IR}/L'_{\rm CO}$ ratios (e.g. Arp 220). As shown by the histograms of the $L_{\rm IR}/L'_{\rm CO}$ ratios in Fig. 4, comparing with local galaxies (mostly populated by galaxies with $L_{\rm IR} = 10^9$–$10^{12}\,L_\odot$), both our $H$-ATLAS SMGs and the previously studied SMGs are slightly above the correlation with larger $L_{\rm IR}/L'_{\rm CO}$ ratios for $J_{\rm up} = 3$ to $J_{\rm up} = 5$ CO lines. In contrast, for the $J_{\rm up} \geq 6$ CO lines, both the local galaxies and the high-redshift SMGs with $L_{\rm IR}$ from $10^9\,L_\odot$ to a few $10^{13}\,L_\odot$ can be found within tight linear correlations. The $H$-ATLAS SMGs show no difference with other previously studied SMGs. Among the CO transitions, CO(7–6) has the tightest correlation across different galaxy populations (∼ 0.17 dex), which agrees well with Lu et al. (2015). This again indicates that the dense warm gas traced by the $J_{\rm up} \geq 6$ CO lines is more tightly correlated with on-going active star formation (without considering AGN contamination to the excitation of CO), and CO(7–6) may be the most reliable star formation tracer among the CO lines.

We have also compared the CO line ratios in local ULIRGs with those in our lensed SMGs, by taking CO(5–4) and CO(6–5) for example. The ratios of $L'_{\rm CO(5-4)}/L'_{\rm CO(6-5)}$ from the two subsamples turn out to be similar within the uncertainties. Their mean values are 1.6 and 1.4 with the standard deviations of 0.35 and 0.37 for local ULIRGs and high-redshift lensed SMGs, respectively. This suggests that the differential lensing is unlikely introducing a large bias of choosing molecular gas with very different gas conditions.

## 4. Galactic properties and differential lensing

### 4.1. Molecular gas mass

One of the most commonly used methods to derive the mass of molecular gas in galaxies is to assume that it is proportional to the luminosity $L'_{\rm CO(1-0)}$ through a conversion factor $\alpha_{\rm CO}$ such as $M_{\rm H_2} = \alpha_{\rm CO} L'_{\rm CO(1-0)}$, where $M_{\rm H_2}$ is the mass of molecular hydrogen and $\alpha_{\rm CO}$ is the conversion factor to convert observed CO line luminosity to the molecular gas mass without helium correction (see Bolatto et al. 2013 for a review). Here we adopt a typical value of $\alpha_{\rm CO} = 0.8\,M_\odot\,(\rm K\,km\,s^{-1}\,pc^2)^{-1}$ which is usually found in starbursts as observed in local ULIRGs (Downes & Solomon 1998). The total mass of molecular gas $M_{\rm gas}$ is then inferred by multiplying $M_{\rm H_2}$ by the factor 1.36 to include helium. One should also note that at $z = 2.1$–$4.2$, the cosmic microwave background (CMB) temperature reaches ∼ 8.5–14.2 K, which is non-negligible to the low-$J$ CO lines. A typical underestimation of the CO(1–0) luminosity could be around 10%–25% if $T_{\rm k} = 50$ K, and for the bulk of the molecular gas, which is normally colder than 50 K and is only bright in the low-$J$ transitions, the CMB effect may be even more severe as pointed out by Zhang et al. (2016) (see also da Cunha et al. 2013). Although far from being settled, recent observations of high-redshift SMGs favour $\alpha_{\rm CO}$ being close to the value of local ULIRGs with large uncertainties (Ivison et al. 2011; Magdis et al. 2011; Messias et al. 2014; Spilker et al. 2015; Aravena et al. 2016b).

Half of our sources were observed in their CO(1–0) line with the Green Bank Telescope (GBT) by Harris et al. (2012). The corresponding apparent luminosities $\mu L'_{\rm CO(1-0)}$ (not corrected for lensing) are reported in Table 5. However, it is impossible to infer the total mass of molecular gas in the absence of a detailed lensing model including the extended part of CO(1–0) emission. We may nevertheless directly compare the CO(3–2) and CO(1–0) apparent luminosities $\mu L'_{\rm CO}$ for the seven Harris' sources for which we observed the CO(3–2) line (Table 5). The error-weighted mean ratio of the luminosity of CO(3–2) to CO(1–0) is 0.65±0.19. This is marginally larger at about the 1$\sigma$ level by a factor 1.3 ± 0.4 than the median brightness temperature ratios $r_{32}/r_{10}$ of 0.52±0.09 reported for unlensed SMGs by Bo13, and 0.55±0.05 reported by Ivison et al. (2011) (as described in Eq. (1)). This difference seems to suggest an effect of differential lensing, the more compact CO(3–2) emission being more magnified than the extended CO(1–0) emission (see Section 4.3 for a detail discussion).

However, the mass of molecular gas $M_{\rm gas}$ can be directly inferred from higher $J_{\rm up}$ CO lines, mostly CO(3–2), as for cases of other high-redshift SMGs where CO(1–0) observations are lacking. Moreover, comparing with the CO(1–0) line, the CO(3–2) line tends to be less affected by differential lensing because its spatial distribution is closer to that of the submm dust emission upon which the lensing models are built. Therefore, by assuming that our lensed SMGs are similar to the unlensed high-redshift SMGs, the brightness temperature ratio $r_{32}/r_{10} = 0.52 \pm 0.09$ from Bo13 yields $\beta_{\rm CO32} = 1.36 \times 0.8/0.52 = 2.09\,M_\odot\,(\rm K\,km\,s^{-1}\,pc^2)^{-1}$ for the conversion factor defined as

$$M_{\rm gas} = \beta_{\rm CO32} L'_{\rm CO(3-2)}. \qquad (2)$$

The masses of molecular gas (including He) are thus derived and reported in Table 5. [3] These values for $M_{\rm gas}$ are in the same range, $10^{10}$–$10^{11}\,M_\odot$, as those derived for unlensed SMGs by Bo13. This is confirmed by the direct comparison of the distributions of $L'_{\rm CO}$ after lensing correction (Fig.6). But one should keep in mind the accumulation of uncertainties about our $M_{\rm gas}$ estimates: to the usual uncertainty on $\alpha_{\rm CO}$ or $\beta_{\rm CO32}$, one should add that of the lensing model, especially in the absence of high-resolution CO imaging. The derived gas mass appears exceptionally high for G15v2.235, about three times larger than for any other source and twice more massive than for any unlensed SMG of Bo13. Either the magnification factor is larger than the low value, 1.8±0.3, derived by Bu13, or this source is an exceptional galaxy.

These masses of gas may be compared with the mass of dust derived, for example, through the gas to dust mass ratio

$$\delta_{\rm GDR} = M_{\rm gas}/M_{\rm dust} \quad (3)$$

The dust masses were taken from Bu13. We recall that they are derived by performing a single component modified black body model with the *Herschel* SPIRE and SMA photometric fluxes, with mass absorption coefficient $\kappa_{\rm dust}$ interpolated from Draine (2003). The values of $\delta_{\rm GDR}$ for our sample are given in Table 5. They range from $31 \pm 14$ to $100 \pm 33$ with a mean of $56 \pm 28$. Our value is generally in agreement with the mean value of $\delta_{\rm GDR} = 75 \pm 10$ for ALESS high-redshift SMGs (Simpson et al. 2014; Swinbank et al. 2014) within $1\,\sigma$ level. This range is also similar to that of the local ULIRGs (Solomon et al. 1997).

**Table 4:** Dynamical masses of the sample

| Source | $\langle\Delta V_{\rm CO}\rangle$ (km s$^{-1}$) | $r_{\rm half}$ (kpc) | $i$ (deg) | $M_{\rm dyn,vir}$ ($10^{10}\,M_\odot$) | $M_{\rm dyn,rot}$ ($10^{10}\,M_\odot$) |
|---|---|---|---|---|---|
| G09v1.97 | 348 ± 164 | 0.9 ± 0.1 | 47 ± 5 | 2.9 ± 2.7 | 2.5 ± 2.4 |
| G09v1.40 | 263 ± 70 | 0.4 ± 0.1 | 48 ± 11 | 0.8 ± 0.5 | 0.7 ± 0.5 |
| SDP17b | 286 ± 44 | 3.1 ± 0.9 | 39 ± 11 | 7.0 ± 3.0 | 8.3 ± 5.4 |
| SDP81 | 560 ± 139 | 3.3 ± 0.7 | 50 ± 7 | 29.1 ± 15.6 | 23.3 ± 13.6 |
| G12v2.43 | 237 ± 68 | 3 ± 1 | 55 | 4.7 ± 3.2$^a$ | 3.3 ± 2.2$^a$ |
| G12v2.30 | 713 ± 153 | 1.6 ± 0.1 | 77 ± 2 | 22.8 ± 9.9 | 11.1 ± 4.9 |
| NCv1.143 | 265 ± 55 | 0.8 ± 0.2 | 51 ± 10 | 1.6 ± 0.8 | 1.2 ± 0.7 |
| NAv1.195 | 266 ± 19 | 1.6 ± 0.2 | 35 ± 9 | 3.1 ± 0.6 | 4.4 ± 2.2 |
| NAv1.177 | 252 ± 35 | 3 ± 1 | 55 | 5.4 ± 2.3$^a$ | 3.7 ± 1.6$^a$ |
| NBv1.78 | 597 ± 121 | 0.6 ± 0.2 | 51 ± 7 | 5.5 ± 3.3 | 4.3 ± 2.7 |
| NAv1.144 | 208 ± 35 | 0.9 ± 0.3 | 55 ± 8 | 1.1 ± 0.5 | 0.7 ± 0.4 |
| NAv1.56 | 650 ± 28$^b$ | 1.5 ± 0.4 | 52 ± 5 | 17.8 ± 5.0 | 13.2 ± 4.1 |
| G15v2.235 | 480 ± 111 | 1.7 ± 0.3 | 59 ± 8 | 11.2 ± 5.6 | 7.1 ± 3.8 |
| G12v2.890 | 344 ± 92 | 3 ± 1 | 55 | 10.0 ± 6.3$^a$ | 6.9 ± 4.4$^a$ |
| G12v2.257 | 395 ± 206 | 3 ± 1 | 55 | 13.2 ± 14.4$^a$ | 9.2 ± 10.0$^a$ |
| G15v2.779$^c$ | 830 ± 86 | 3.8 ± 0.4 | 43 ± 7 | 73.5 ± 17.0 | 72.8 ± 22.6 |

**Notes.** $\langle\Delta V_{\rm CO}\rangle$ is the average value of the CO linewidths, the errors are from standard deviations. We recall the values of half-light radius $r_{\rm half}$ in this table. $i$ is the inclination angle derived from the major and minor axis ratio from lensing models in Bu13. $M_{\rm dyn,vir}$ and $M_{\rm dyn,rot}$ are the dynamical masses enclosed in $r_{\rm half}$. $^{(a)}$ Due to lacking $r_{\rm half}$ and $b/a$ ratio of the rotating disk from lensing models, we use a typical value of $r_{\rm half} = 3 \pm 1$ kpc (by assigning a 30% uncertainty) and $i = 55°$ (see text) for the estimation of these dynamical masses. $^{(b)}$ Because of the limited data quantity for this source, we take the CO(4–3) data of NAv1.56 from the NOEMA observation (Oteo et al. in prep.), which offers better accuracy. $^{(c)}$ The physical properties of G15v2.779 are taken from or computed according to Cox et al. (2011) and O13.

### 4.2. Dynamical mass

In high-redshift SMGs, an important fraction of the baryonic mass is in the form of molecular gas, and the CO linewidth can serve as a good dynamical mass indicator with an assumption about the dynamical structure and extent of the system (e.g. Tacconi et al. 2006; Bouché et al. 2007). From the measured linewidth of the CO lines, we can in principle derive the dynamical mass within the half-light radius ($r_{\rm half}$) by assuming that the lensed SMG can be treated as either a virialised system or a rotating disk with an inclination angle $i$.

If the system is virialised, the dynamical mass can be calculated following the approach of Bo13 as

$$M_{\rm dyn,vir} = 1.56 \times 10^6 \left(\frac{\sigma}{\rm km\,s^{-1}}\right)^2 \left(\frac{r}{\rm kpc}\right) M_\odot, \quad (4)$$

in which the velocity dispersion $\sigma = \Delta V_{\rm CO}/(2\sqrt{2\ln 2})$, $r$ is the radius of the enclosed region for calculating the dynamical mass and $\Delta V_{\rm CO}$ is the CO linewidth. If the system is a rotating disk, the dynamical mass can be derived from

$$M_{\rm dyn,rot} = 2.32 \times 10^5 \left(\frac{v_{\rm cir}}{\rm km\,s^{-1}}\right)^2 \left(\frac{r}{\rm kpc}\right) M_\odot, \quad (5)$$

following Wang et al. (2013) and Venemans et al. (2016), in which $v_{\rm cir}$ is the circular velocity equal to $0.75\Delta V_{\rm CO}/\sin i$, $r$ is the disk radius and $i$ is the inclination angle of the disk on the sky in a range from 0° to 90°. By assuming an inclined thin-disk geometry, we can derive the inclination angle from the minor to major axis ratio $b/a$ of the rotating disk as $i = \cos^{-1}(b/a)$. When possible, we use the minor to major axis ratio of the source image which was derived in the lensing models of (Bu13). This yields the values of $i$ reported in Table 4. Otherwise, we assume an average inclination angle of 55° as suggested by Wang et al. (2013). However, we also note that $1/\sin i$ can take large values for galactic disks seen close to face on. We can thus calculate the two estimates of the dynamical mass enclosed in the half-light radius $r_{\rm half}$, for example, $M_{\rm dyn,vir}$ and $M_{\rm dyn,rot}$ from the CO linewidth and/or the minor to major axis ratios following Eqs. (4) and (5).

However, it is important to note that there is certainly a significant fraction of the SMG mass distributed outside $r_{\rm half}$. Accordingly, the value of $M_{\rm dyn,vir}$ or $M_{\rm dyn,rot}$ should serve as a lower limit of the dynamical mass of the entire region where molecular gas resides. This is a fortiori true for the total mass of the galaxy including the extended diffuse, cool component beyond ~3 kpc. As suggested by the CO(1–0) observations of a sample of SMGs at $z \sim 2.4$, Ivison et al. (2011) find a typical size of ~7 kpc for the CO(1–0), with a linewidth of ~563 km s$^{-1}$. Using CO(2–1) data, Hodge et al. (2012) also suggest a rotating disk of molecular gas with a radius of ~7 kpc in GN20 and a CO linewidth equal to $575 \pm 100$ km s$^{-1}$. JVLA and ATCA observations of CO(1–0) emission at high-redshift reported by several other works (e.g. Greve et al. 2003; Riechers et al. 2011c; Deane et al. 2013; Emonts et al. 2016; Dannerbauer et al. 2017) also support the existence of such an extended cold gas component. Both their linewidth and the size of the emitting region are larger than those of most of our sources, suggesting an underestimation for the dynamical mass for our sample. But even the mass of the starburst, FIR-emitting core is likely to be underestimated by these formulas for $M_{\rm dyn}$. It is challenging to estimate the ratio between the total dynamical mass within the entire CO-emitting

[3] As stated in the caption of Table 5, when CO(3–2) observation is absent we have used another line with similar factors, $\beta_{\rm CO43} = 2.65\,M_\odot\,({\rm K\,km\,s^{-1}\,pc^2})^{-1}$ for CO(4–3) and $\beta_{\rm CO21} = 1.30\,M_\odot\,({\rm K\,km\,s^{-1}\,pc^2})^{-1}$ for CO(2–1) based on Bo13.

**Table 5:** Observationally derived physical properties of the *H*-ATLAS SMGs.

| Source | $z_{\rm CO}$ | $D_{\rm L}$ (Mpc) | $L_{\rm IR}$ ($10^{12}\,L_\odot$) | $\frac{\mu L'_{\rm CO(1-0),\,Ha}}{10^{11}}$ ($M_\odot$) | $\frac{\mu L'_{\rm CO(1-0)}}{10^{11}}$ (K km s$^{-1}$ pc$^2$) | $\mu M_{\rm H_2}$ ($11^{11}\,M_\odot$) | $\frac{L'_{\rm CO(1-0)}}{10^{10}}$ (K km s$^{-1}$ pc$^2$) | $M_{\rm gas}$ ($10^{10}\,M_\odot$) | $\frac{M_{\rm gas}}{M_{\rm dyn,vir}}$ | $\delta_{\rm GDR}$ | $t_{\rm dep}$ (Myr) |
|---|---|---|---|---|---|---|---|---|---|---|---|
| G09v1.97 | 3.6345 ± 0.0001 | 32751 ± 588 | 22.5 ± 6.5 | – | 6.9 ± 3.0 | 5.5 ± 2.4 | 10.0 ± 4.4 | **1.0 ± 0.5** | 3.8 ± 3.9 | 75 ± 35 | 28 ± 15 |
| G09v1.40 | 2.0924 ± 0.0001 | 16835 ± 283 | 4.3 ± 1.9 | – | 3.6 ± 1.1 | 2.8 ± 0.9 | 2.3 ± 0.9 | 2.5 ± 1.0 | 3.2 ± 2.2 | 31 ± 14 | 34 ± 20 |
| SDP17b | 2.3053 ± 0.0001 | 18942 ± 322 | 14.5 ± 5.7 | 2.7 ± 0.4 | 3.8 ± 0.9 | 3.0 ± 0.7 | 7.8 ± 2.1 | 8.5 ± 2.3 | 1.2 ± 0.6 | 43 ± 14 | 34 ± 16 |
| SDP81 | 3.0413 ± 0.0005 | 26469 ± 466 | 5.3 ± 1.5 | 4.8 ± 0.4 | 6.4 ± 1.6 | 5.1 ± 1.2 | 5.7 ± 1.5 | 6.2 ± 1.6 | 0.2 ± 0.1 | 41 ± 12 | 68 ± 26 |
| G12v2.43 | 3.1271 ± 0.0001 | 27367 ± 484 | (90 ± 2)/$\mu$ | 1.6 ± 0.4 | 4.2 ± 0.9 | 3.3 ± 0.7 | (41 ± 9)/$\mu$ | (45 ± 10)/$\mu$ | – | – | – |
| G12v2.30 | 3.2596 ± 0.0002 | 28761 ± 511 | 16.4 ± 4.4 | 4.7 ± 0.8 | 7.9 ± 2.2 | 6.3 ± 1.8 | 8.3 ± 2.4 | 9.1 ± 2.6 | 0.4 ± 0.2 | 69 ± 22 | 32 ± 13 |
| NCv1.143 | 3.5650 ± 0.0004 | 32007 ± 574 | 11.4 ± 3.9 | – | 6.5 ± 1.4 | 5.2 ± 1.1 | 5.7 ± 1.5 | 6.2 ± 1.6 | 4.0 ± 2.2 | 50 ± 15 | 31 ± 14 |
| NAv1.195 | 2.9510 ± 0.0001 | 25528 ± 448 | 18.3 ± 5.1 | – | – | – | – | – | – | – | – |
| NAv1.177 | 2.7778 ± 0.0001 | 23736 ± 414 | (62 ± 2)/$\mu$ | – | 5.6 ± 1.1 | 4.5 ± 0.9 | (56 ± 11)/$\mu$ | (61 ± 12)/$\mu$ | – | – | – |
| NBv1.78 | 3.1080 ± 0.0003 | 27167 ± 480 | 8.4 ± 3.1 | – | 3.1 ± 0.7 | 2.5 ± 0.6 | 2.4 ± 0.6 | 2.6 ± 0.7 | 0.5 ± 0.3 | 43 ± 13 | 18 ± 8 |
| NAv1.144 | 2.2023 ± 0.0001 | 17918 ± 303 | 13.6 ± 8.3 | 2.3 ± 0.3 | 2.5 ± 0.6 | 2.0 ± 0.4 | 5.8 ± 1.6 | 6.3 ± 1.8 | 5.9 ± 3.5 | 44 ± 16 | 27 ± 18 |
| NAv1.56 | 2.3001 ± 0.0009 | 18890 ± 321 | 9.8 ± 2.8 | 7.3 ± 1.1 | 6.4 ± 1.1[a] | 5.1 ± 0.9 | 5.5 ± 1.1 | 6.0 ± 1.2 | 0.3 ± 0.1 | 52 ± 11 | 35 ± 12 |
| G15v2.235 | 2.4789 ± 0.0001 | 20686 ± 355 | 15.6 ± 4.7 | 4.4 ± 0.5 | 5.2 ± 1.0 | 4.2 ± 0.8 | 28.8 ± 7.6 | 31.4 ± 8.2 | 2.8 ± 1.6 | 99 ± 33 | 117 ± 46 |
| G12v2.890 | 2.5783 ± 0.0003 | 21694 ± 375 | (25 ± 3)/$\mu$ | 2.1 ± 0.6 | 1.8 ± 0.5 | 1.5 ± 0.4 | (18 ± 5)/$\mu$ | (19 ± 5)/$\mu$ | – | – | – |
| G12v2.257 | 2.1914 ± 0.0001 | 17810 ± 301 | (26 ± 3)/$\mu$ | 1.8 ± 0.3 | 2.1 ± 0.6 | 1.7 ± 0.5 | (21 ± 6)/$\mu$ | (23 ± 7)/$\mu$ | – | – | – |
| G15v2.779[b] | 4.243 ± 0.001 | 39349 ± 718 | 22.0 ± 7.0 | – | 8.3 ± 1.8 | 6.6 ± 1.4 | 18.0 ± 4.3 | 19.5 ± 4.6 | 0.3 ± 0.1 | 85 ± 23 | 51 ± 20 |

**Notes.** $z_{\rm CO}$ is derived from the error-weighted mean of the multi-$J$ CO spectral redshifts from this work. For the double-peak sources, we take an average redshift of the two components. The luminosity distance $D_{\rm L}$ is calculated using `Cosmology.jl` with the of `Julia` language (Bezanson et al. 2012) and the errors are propagated using `Measurements.jl` (Giordano 2016). We also recall values of $L_{\rm IR}$ in this table. For $\mu L'_{\rm CO(1-0)}$, most of the values are converted from CO(3–2) fluxes as described in the text. For G09v1.40, we use CO(2–1)/CO(1–0) ratio to infer the flux of CO(1–0). And for NAv1.56 and G15v2.779, we use CO(4–3)/CO(1–0) ratio (flux ratios are from Bo13). The calculation of apparent molecular gas mass $\mu M_{\rm H_2}$ takes a conversion factor $\alpha_{\rm CO} = 0.8$ (see text). $\mu M_{\rm H_2,Ha}$ is the molecular gas mass calculated from CO(1–0) fluxes reported in Harris et al. (2012). Gas mass $M_{\rm gas}$ is calculated by considering a 36% Helium contribution, e.g. $M_{\rm gas} = 1.36 M_{\rm H_2}$. $\delta_{\rm GDR}$ and $t_{\rm dep}$ are gas to dust mass ratio and molecular gas depletion time, respectively (see the detail definitions in Section 4 and Section 5.3). [a] Because of the limited data quantity, we take the NOEMA CO(4–3) data of NAv1.56, which offers better accuracy (Oteo et al. in prep.). [b] The physical properties of G15v2.779 are taken from or computed according to Cox et al. (2011) and O13.

region and the one we calculated from $r_{\rm half}$, although Tacconi et al. (2006) suggest a value about 5 for such a ratio.

As seen in Table 4, the range of $M_{\rm dyn,vir}$, given by Eqs. (4) and (5), varies from $\sim 10^{10}\,M_\odot$ to $3\times10^{11}\,M_\odot$ while values of $M_{\rm dyn,rot}$ are comparable but slightly smaller by a factor of $\sim 1.4$ on average. This is again $\sim$3–10 times smaller than the value of $M_{\rm dyn,rot}$ given in Ivison et al. (2011) and Hodge et al. (2012). We compare these values with the derived masses of gas and discuss them in the following subsection.

### 4.3. Possible lensing biases

It is well known that differential lensing may be a serious problem for galaxy-galaxy strong-lensing studies of extended objects, especially for multi-line and continuum comparison (e.g. Serjeant 2012; Hezaveh et al. 2012). Although the problem should be dimmer for the compact cores ($r \lesssim 1$–3 kpc) emitting the continuum and high-$J$ CO lines in our sources, it needs consideration in case of complex caustics at this scale. It may become worse for low-$J$ CO studies since they may involve more extended SMG components (Ivison et al. 2011). In addition, the flux-limited selection of our lensed sources may bias our sample towards the most compact objects.

Because different excitation levels of CO trace predominantly regions with different gas density and different temperature (see the critical densities and the energy levels of the CO lines in Table 1), the sizes of the emitting regions of each $J$ CO line are expected to be somewhat different. This variation of the emitting region of each CO transition will certainly bring differences in the resulting parameters, such as the total magnification factor, and derived quantities such as the molecular gas mass and line ratios with respect to the intrinsic ratios of the unlensed galaxy. The differential lensing effect could arise in a complex way from the specific spatial configuration of the caustic line with respect to the background emission. However, a detailed modelling of complex effects of differential lensing is beyond the scope of the present study for two main reasons: The low resolution of our single-dish CO data and their limited range of $J_{\rm up}$ values, mostly from 3 to 8. It is expected that the regions emitting such lines will not differ very much with $J_{\rm up}$ for most sources and remain close to that of the observed 880 $\mu$m dust continuum. This is also consistent with the similarity that we find for the linewidths of the different CO transitions within each source. We will therefore neglect the effects of differential lensing in estimating the ratios of these different mid/high-$J$ emissions. Of course, the validity of this assumption should be verified, taking into account the particularities of each source and its lensing, when high-angular-resolution images are available. However, we can perform a first verification in the only case of our sources, SDP 81, for which ALMA high-resolution CO images have been published, noting that it is one of our most extended sources. These images from the ALMA long baseline campaign observation (ALMA Partnership, Vlahakis et al. 2015) show the resolved structure of the dust and CO. From lens modelling, the studies of Dye et al. (2015) and Rybak et al. (2015a,b) show that the differences between the magnification factors of the CO(5–4) and CO(8–7) lines are within 1 $\sigma$ and close to that of dust emission.

On the other hand, the effects of differential lensing might be much more severe and non-negligible when comparing CO(1–0) with high-$J$ CO lines. Previous high-angular-resolution imaging studies of SMGs show evidence that the cooler, low-density emitting regions of low-$J$ CO lines are more extended than the dust continuum emission and that of the high-$J$ CO lines (e.g. Tacconi et al. 2008; Bo13; Spilker et al. 2015; Casey et al. 2014). Especially for CO(1–0), JVLA images of high-redshift SMGs reveal a significant extension, usually several times larger on average, compared to high-$J$ CO (Ivison et al. 2011, see also Engel et al. 2010; Riechers et al. 2011a). This is probably also true for the [C I](1–0) line because its spatial distribution agrees well with that of CO(1–0) (e.g. Ikeda et al. 2002; Glover et al. 2015).

With such an extension, typically ∼7 kpc, substantial differential lensing seems unavoidable for most strong lensing configurations, yielding a lower magnification for CO(1–0) compared with more compact mid/high-$J$ CO emission. Indeed, such an effect has already been directly observed in at least two strongly lensed SMGs: SDP 81 (Rybak et al. 2015b) and SPT0538-50 (Spilker et al. 2015). As suggested in Hezaveh et al. (2012) such a difference will be moderate in the low-magnified system with small $\mu$ but can be non-negligible for the highly magnified systems where $\mu > 10$, which is likely the case of G09v1.40, SDP 81, NCv1.143, NAv1.144 and NAv1.56 in our sample (Table 3). By assuming that $\mu$ for CO(3–2) is similar to the dust continuum, the ratio of the velocity integrated flux density between CO(1–0) (from Harris et al. 2012) and CO(3–2) for our sources has a mean value of $0.17 \pm 0.05$ (Fig. 2), that is, a factor $1.3 \pm 0.4$ lower than in other high-redshift unlensed SMGs (Eq. (1)). Such a difference in the ratio of CO(1–0) over CO(3–2) can be explained by differential lensing. Nevertheless, one should note that the difference is not at a very significant level. Because we have no high-resolution maps of the CO lines, it is beyond our ability to reconstruct the exact magnification factor for each emission line. Here we assumed that the magnification factors are the same for all the $J_{\rm up} \geq 3$ CO lines as that of dust emission, and we applied the magnification factor $\mu_{880}$ derived from the SMA 880 $\mu$m images (Table 3). For the CO(1–0) line, we thus applied the factor $\mu_{880}/1.3$, as derived above for the multi-$J$ CO line excitation modelling as described in Section 5.

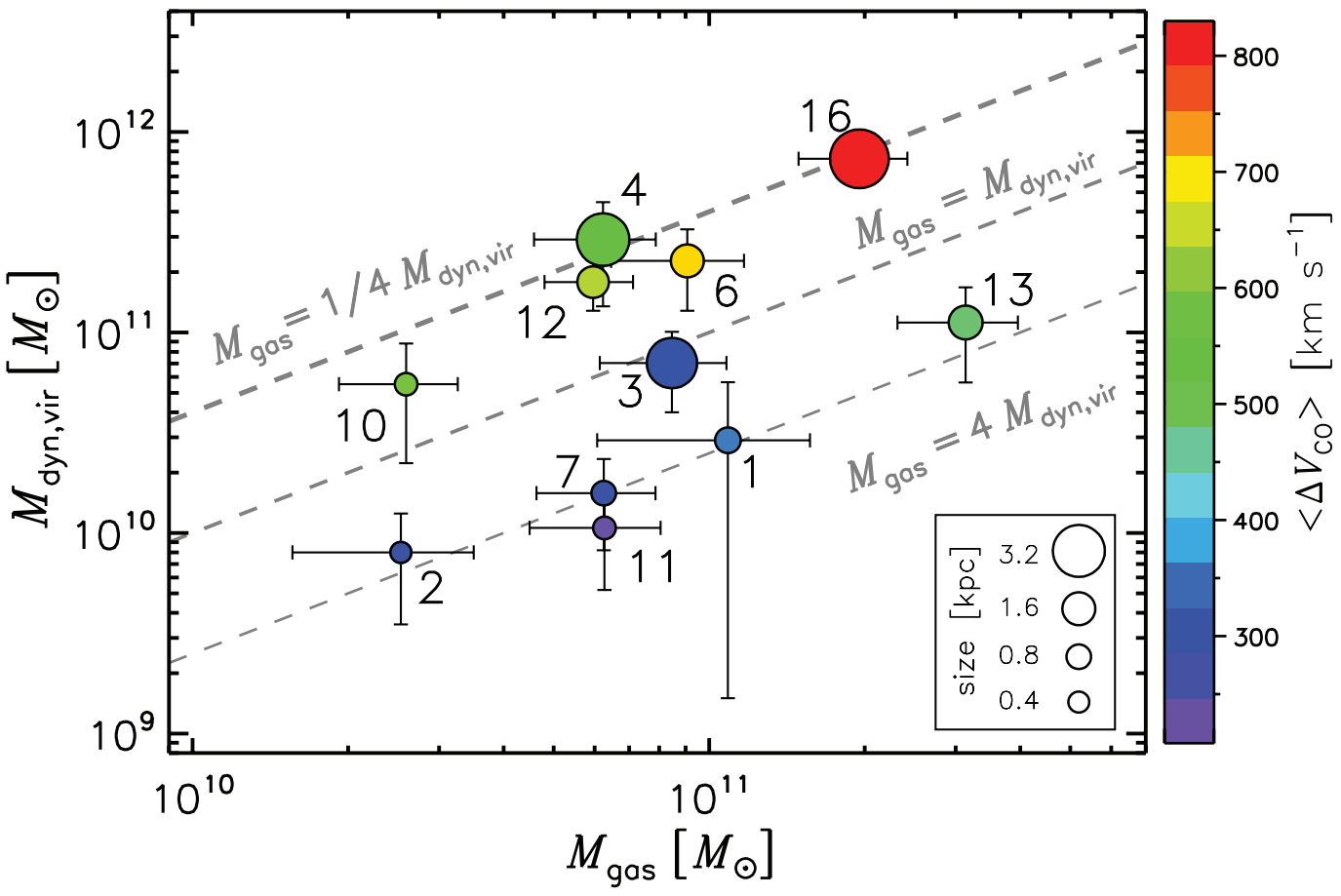

**Fig. 5:** A comparison between molecular gas mass and dynamical mass of our $H$-ATLAS lensed sample. The grey dashed lines indicate the ratio of $M_{\rm gas}/M_{\rm dyn,vir}$. Colours are coded according to the average CO linewidth. The size of the symbol represents the value of $r_{\rm half}$ of each source. There is a clear trend that sources with smaller linewidths have large ratios of $M_{\rm gas}/M_{\rm dyn,vir}$. The source index can be found in Fig. 1 and Table B.1.

Another important aspect of differential lensing is the possible distortion of the line profile. As shown in the case of the high-resolution and high-sensitivity CO spectrum of SDP 81, the line profiles show asymmetry features with a prominent red component accompanied by a weaker blue component (ALMA Partnership, Vlahakis et al. 2015). By reconstructing the source in the image plane, Swinbank et al. (2015) show that SDP 81 is a clumpy rotating disk and the red part of the disk is more magnified than the blue part, which causes the line-profile asymmetry. This might also happen to our other sources, especially for the case in which the caustic lines cross only part of the galaxy. It is not impossible that such effects might lead to underestimate the wings of some lines and thus explain at least part of the excess of narrow linewidths that we observed (see the case of G09v1.97 in Sect. 5.4 and Fig. A.1). Another cause of this excess could be a possible bias between the magnification and intrinsic source size (Spilker et al. 2016) which could perhaps bias against composite broad profiles of slightly extended sources in an early merger state. However, it seems that further observation and modelling of high-resolution CO images is needed to progress in completely explaining if this excess is real.

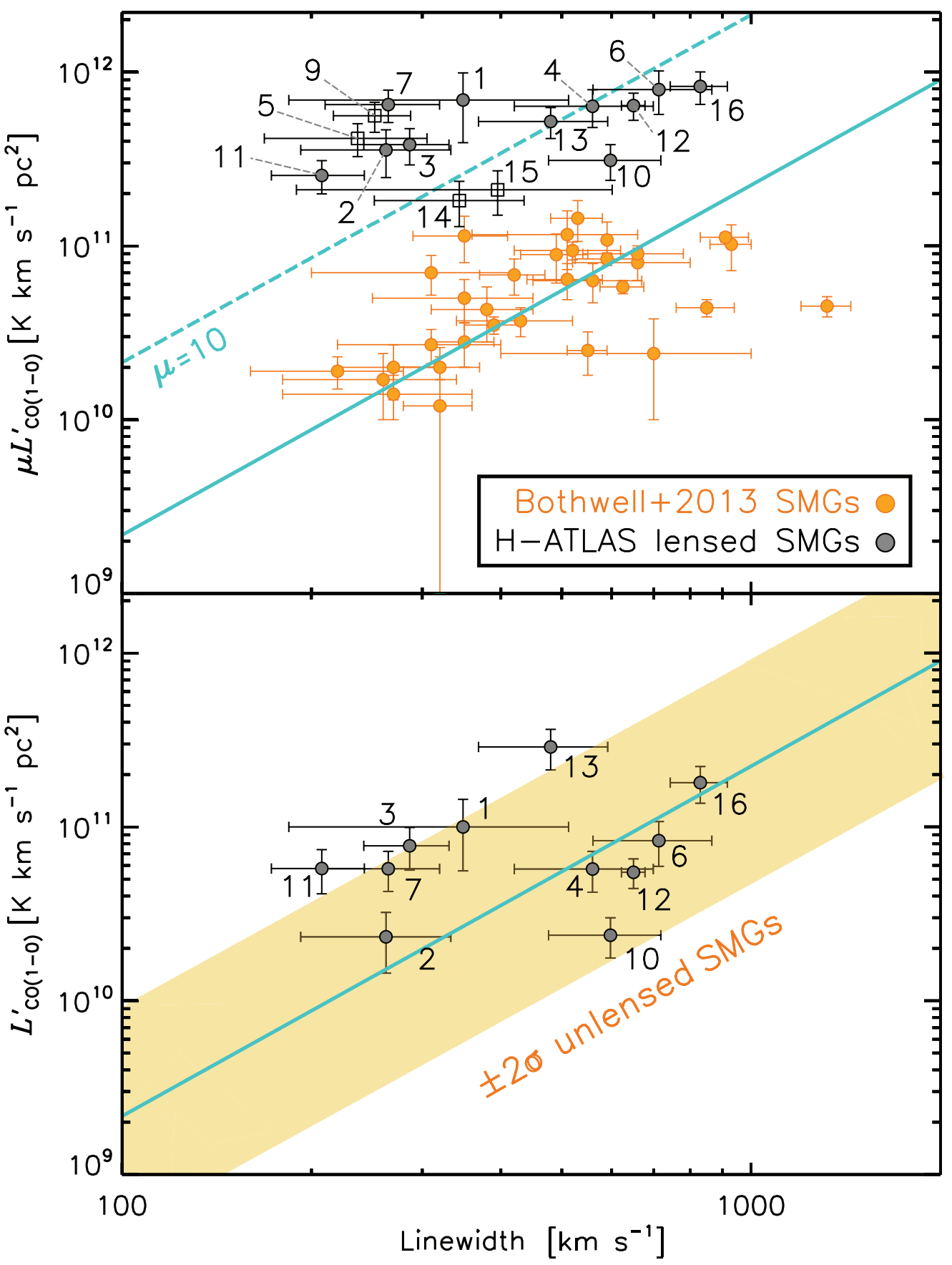

**Fig. 6:** *Upper panel*: $\mu L'_{\rm CO(1-0)}$ plotted against CO linewidth. Orange data points represent the unlensed SMGs from Bo13, while grey data points (indexed as in Table 2) are from this work, for which the filled circles show the sources with existing lensing model and the black open squares are the ones without lensing models. The solid turquoise line shows the fit proposed by Bo13 for the relation of $L'_{\rm CO}$ vs. linewidth as $L'_{\rm CO} \propto \Delta V^2_{\rm CO}$. The dashed turquoise line indicates the positions for $\mu = 10$ assuming this relation. As shown in the plot, the magnification factors of G12v2.43 (#5) and NAv1.177 (#9) are likely to be large. *Lower Panel*: Intrinsic $L'_{\rm CO(1-0)}$ plotted against the CO linewidth. The light orange region shows the $\pm 2\,\sigma$ range of the scatter derived from the SMGs of Bo13. Our sources generally agree with the correlation.

An effect of underestimating the CO linewidth would be to underestimate dynamical masses. In order to check this, Fig. 5 shows the plot of the relation of $M_{\rm gas}$ and $M_{\rm dyn}$ changes with CO linewidth and source size. For most broad-line sources, the values of the ratio $M_{gas}/M_{dyn}$ are from 0.2 to 0.5, which appear possible, although the high value for G15v2.235, 2.4 (source #13 as shown in Fig. 5 and 6) seems to point out a problem with its lensing model. On the other hand, for all narrow-line sources, values of $M_{gas}/M_{dyn}$ greater than 1, and even than 3 for most of them (which is equivalent to a 1.7–2 fold underestimation of the linewidth) point out a serious problem. The Spearman's correlation coefficient between CO linewidth and $M_{gas}/M_{dyn}$ is −0.83 with a $p$-value of 0.0016. The $r_{half}$ value of each source is also indicated by the symbol size. Four of the smallest sources (#1, #2, #7 and #11) have high $M_{gas}/M_{dyn}$ values, since the differential lensing could also potentially affect the estimation of the source size. However, we find a much weaker correlation between $r_{half}$ and $M_{gas}/M_{dyn}$, suggesting that the impact of differential lensing on the source size is much weaker compared to that of the linewidth. It is possible that the sources with narrow linewidth having higher values of $M_{gas}/M_{dyn}$ is partly due to differential lensing since the dynamical mass is proportional to $\Delta V_{CO}{}^2$, as identified in SDP 81 (Rybak et al. 2015b; Dye et al. 2015; Swinbank et al. 2015) and in GO9v1.97 (Yang et al. in prep.) and perhaps in SDP 17b whose CO(4–3) and $H_2O$ profiles are clearly asymmetric (O13). Also, it seems however that at least part of this problem reflects the fact that Eqs.(4) and (5) might underestimate the dynamical mass by a large factor, likely to be up to 5, as quoted, for example, by Tacconi et al. (2006) for unlensed SMGs. It is however obvious that further observation and modelling of high-spatial-resolution CO images is needed to progress in completely explaining such problems.

Acknowledging the possible bias from the narrow-linewidth sources, after excluding the sources with $\Delta V_{CO} < 400$ km s$^{-1}$, and also G15v2.235 as mentioned before, we derived an average value of $M_{gas}/M_{dyn,vir} = 0.34 \pm 0.10$, in line with the SPT sources (Aravena et al. 2016b), other unlensed SMGs (Bo13) and empirical model predictions (Béthermin et al. 2015). By assuming that the ISM is dominated by molecular content, and a small dark matter contribution within $r_{half}$, the ratio can serve as a proxy of molecular gas mass fraction. Then the molecular gas mass fraction of the $H$-ATLAS SMGs is thus ∼ 34% with a significant uncertainty, yet it is consistent with Bo13's average value computed from $M_{gas}/(M_{gas}+M_*)$, in which $M_*$ is the stellar mass.

It has been proposed that there exists a simple linear correlation between $L'_{CO}$ and $\Delta V$ of the CO(1–0) line (e.g. Harris et al. 2012; Bo13; Goto & Toft 2015; Dannerbauer et al. 2017), $L'_{CO} \propto (\sigma^2 R)/(\alpha_{CO} G)$, where $\sigma$ is the velocity dispersion of the CO line, $R$ is the CO emitting radius, $\alpha_{CO}$ is the CO luminosity to gas mass conversion factor and $G$ is the gravitational constant. We recall from Eqs. (4) and (5) that the dynamical mass is proportional to $\sigma^2 R$ and $M_{gas} = L'_{CO}\,\alpha_{CO}$; thus this correlation simply reflects the variation of the ratio between gas mass and dynamical mass. In Fig. 6, we overlay both the apparent CO line luminosity $\mu L'_{CO}$ and the intrinsic $L'_{CO}$ plotted against the CO linewidth on those of the Bo13's unlensed sources. The flat distribution of the lensed sources in the upper panel of Fig. 6 shows clearly the lensed feature. After correcting for the magnification, our sources are generally within the 2 $\sigma$ regions from Bo13's fit. However, it is clear that all the sources with $\Delta V_{CO} < 400$ km s$^{-1}$ are above the correlation and very close to the +2 $\sigma$ limit. Again, this supports our previous argument that these linewidths are likely being underestimated.

## 5. Physical properties of molecular gas

### 5.1. Multi-J CO line excitation and LVG modelling

As indicated by the histograms of $L'_{CO}/L_{IR}$ in Fig. 4, the shape of the average CO SLED of the $H$-ATLAS SMGs follows the trend of other high-redshift SMGs and both of them depart from the average CO SLED of local galaxies with $L_{IR} < 10^{12}\,L_\odot$ for the low-$J$ ($J_{up}$ = 3, 4, 5) part at the ∼ 1 $\sigma$ levels. Fig. 7 shows the $L_{IR}$-normalised CO SLED of the high-redshift SMGs (previous detections in the literature together with $H$-ATLAS ones) comparing with those of the local galaxies from Liu et al. (2015).

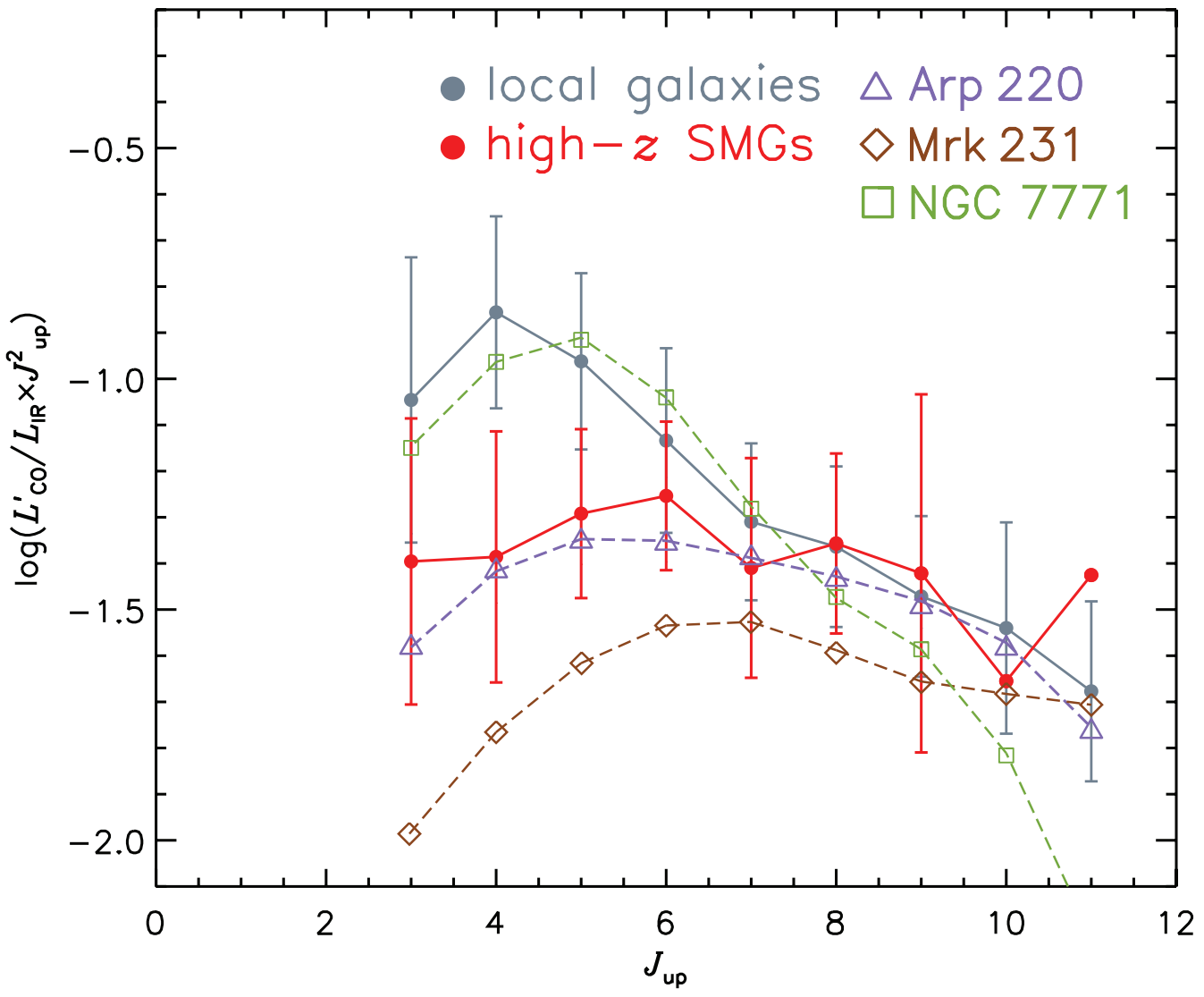

**Fig. 7:** The CO SLEDs of local galaxies and high-redshift SMGs normalised by $L_{IR}$.[4] Grey symbols indicate the Gaussian mean and deviation of the ratio $L'_{CO}/L_{IR} \times J^2_{up}$ for local galaxies mostly with $10^9\,L_\odot \leq L_{IR} \leq 10^{12}\,L_\odot$ (Liu et al. 2015). We also include a typical local ULIRG, Arp 220 (purple dashed line and open triangles, Rangwala et al. 2011), an AGN-dominated source, Mrk 231 (red dashed line and open diamond, van der Werf et al. 2010), and a LIRG, NGC 7771 (green dashed line and open square, Liu et al. 2015). Red symbols show high-redshift SMGs from both Carilli & Walter (2013) and from this work. The dots without error bars ($J_{up}$ = 10, 11) indicate that the volume of the $L'_{CO}/L_{IR} \times J^2_{up}$ values is insufficient for a normal-distribution fitting. We thus only indicate their mean value here. It is clear that the ratio of $L'_{CO}/L_{IR} \times J^2_{up}$ decreases with increasing $J_{up}$ for local galaxies while it remains flat for high-redshift SMGs and a typical starburst-dominated ULIRG, Arp 220.

Previous studies of global CO excitation in both local and high-redshift galaxies (e.g. Weiß et al. 2007; Rangwala et al. 2011; Deane et al. 2013; Papadopoulos et al. 2014; Zhang et al. 2014; Spilker et al. 2014; Liu et al. 2015; Daddi et al. 2015) show that there are most likely two excitation components dominating the CO emission from ground level up to $J_{up} = 11$, a low-excitation component peaking around $J_{up}$ =3 to $J_{up}$ =4 and a high-excitation component peaking at $J_{up} \gtrsim 6$. Rosenberg et al. (2015) further quantitively classify the local galaxies into three groups based on the shape of their CO SLEDs, which provides clues towards the dominant excitation conditions within. Com-paring the average $L_{\rm IR}$-normalised CO SLED of local galaxies (dominated by normal star-forming galaxies and LIRGs with $L_{\rm IR} = 10^9$–$10^{12}\,L_\odot$) with that of the SMGs in Fig. 7, it is clear that the low-excitation component is more prominent in local galaxies, resulting in the average CO SLED peaking at $J_{\rm up} = 3$ or $J_{\rm up} = 4$, and decreasing with increasing energy levels (as in the case of a local LIRG, NGC 7771, shown in Fig. 7). For the SMGs, the low-excitation component is rather weak while the high-excitation component is comparable to the local normal star-forming galaxies and LIRGs, resulting in a rather flat SLED. To compare the $L_{\rm IR}$-normalised CO SLED of high-redshift SMGs with that of the local ULIRGs, we also overplot a typical local non-AGN-dominated ULIRG (classified as a class II galaxy which is dominated by starburst as in Rosenberg et al. 2015), that is, Arp 220. It is found that the average $L_{\rm IR}$-normalised CO SLED of high-redshift SMGs agrees well with that of Arp 220. Since the average values of $J_{\rm up} = 10$ and $J_{\rm up} = 11$ of high-redshift SMGs are calculated based on only a few sources, the deviations between Arp 220 and high-redshift SMGs for these two lines are not significant from a statistical point of view. Nevertheless, for the AGN-dominated ULIRG, for example, Mrk 231 as shown in Fig.7 (classified as a class-III galaxy which is dominated by AGN powering, Rosenberg et al. 2015), the $L_{\rm IR}$-normalised CO SLED is below that of the high-redshift SMGs. This shows that AGN are contributing much less of the $L_{\rm IR}$ luminosity of our high-redshift SMGs compared to the AGN-dominated ULIRGs. Thus, in the high-redshift SMGs, the average CO gas excitation conditions are likely to be similar to those of local non-AGN-dominated ULIRGs.

[4] The actual value used in y-axis is $L'_{CO}/L_{IR} \times J^2_{up}$. $J^2_{up}$ is included so that the unit of $L'_{CO} \times J^2_{up}$ is comparable to velocity integrated flux density, which is Jy km s$^{-1}$.

To further investigate the CO line excitation and extract the information of physical conditions of the molecular gas, we apply a large velocity gradient (LVG) statistical equilibrium method (e.g. Sobolev 1960; Goldreich & Kwan 1974; Scoville & Solomon 1974) for modelling the fluxes of multiple CO lines. We adopt a one-dimensional (1D) non-LTE radiative transfer code developed by van der Tak et al. (2007), that is, `RADEX`, with an escape probability of $\beta = (1 - {\rm e}^{\tau})/\tau$ derived from an expanding sphere geometry. The CO collisional data are taken from the LAMDA database (Schöier et al. 2005).

As a first step, we use one excitation component in the LVG modelling. Similar to Weiß et al. (2007), the inputs of the code are the molecular gas kinetic temperature ($T_{\rm k}$), the volume density of the molecular hydrogen ($n_{\rm H_2}$), the column density of the CO molecule ($N_{\rm CO}$), and the solid angle ($\Omega_{\rm app}$, note that this solid angle includes the lensing magnification factor) of the source which scales with the resulting fluxes from each CO transition equally, so that the shape of the CO SLED only depends on $T_{\rm k}$, $n_{\rm H_2}$ and $N_{\rm CO}$. We fix the velocity gradient to 1 km s$^{-1}$, so that the actual input of $N_{\rm CO}$ is column density per unit velocity gradient $N_{\rm CO}/{\rm d}v$ instead.

A Bayesian approach is used to fit our observed flux to the fluxes generated from `RADEX` models given the parameters $p$ (model parameter $p$ includes $T_{\rm k}$, $n_{\rm H_2}$, $N_{\rm CO}/{\rm d}v$ and $\Omega_{\rm app}$). We use the code `emcee` (Foreman-Mackey et al. 2013, see `https://github.com/dfm/emcee`) to perform the Markov chain Monte Carlo (MCMC) calculation with the affine-invariant ensemble sampler (Goodman & Weare 2010). The Bayesian posterior probability of the model parameters given our data $I^{\rm data}$ can thus be written as following (e.g. the notation in Wall & Jenkins 2012),

$$Pr(p|I^{\rm data}) = \frac{Pr(p)Pr(I^{\rm data}|p)}{Pr(I^{\rm data})}, \tag{6}$$

in which $Pr(p|I^{\rm data})$ is the posterior probability of the parameter $p$ given: the prior probability of $p$ as $Pr(p)$; the likelihood of the resulting CO flux $I^{\rm data}$ given the parameter inputs $p$ as $Pr(I^{\rm data}|p)$; and the probability of the data $Pr(I^{\rm data})$, also called evidence, which is commonly treated as a normalising factor. By assuming the noise is independent Gaussian centred, we can write the likelihood as the product of Gaussian probability distributions,

$$Pr(I^{\rm data}|p) = \prod_i \frac{1}{\sqrt{2\pi\sigma_i^2}} \exp\left(-\frac{\left(I_i^{\rm data} - I_i^{\rm model}(p)\right)^2}{2\sigma_i^2}\right), \tag{7}$$

where $\sigma_i$ is the error associated with each set of measured fluxes $I_i^{\rm data}$, and the `RADEX`-generated results given a set of input parameters $p$ is given by $I_i^{\rm model}(p)$. We note that we use the logarithmic form of Eq.(7) in our practical calculations for convenience, so that the resulting parameters are all in logarithmic form.

**Table 6:** Single-component MCMC-resulting molecular gas properties of the *H*-ATLAS SMGs.

| Source | $\log(n_{\rm H_2})$ $\log({\rm cm^{-3}})$ | | $\log(T_{\rm k})$ $\log({\rm K})$ | | $\log(N_{\rm CO}/{\rm d}v)$ $\log({\rm cm^{-2}\,km^{-1}\,s})$ | |
|---|---|---|---|---|---|---|
| | ${\rm med}_{\pm1\sigma}$ | ${\rm max_{post.}}$ | ${\rm med}_{\pm1\sigma}$ | ${\rm max_{post.}}$ | ${\rm med}_{\pm1\sigma}$ | ${\rm max_{post.}}$ |
| G09v1.97 | $3.3^{+0.8}_{-0.9}$ | 3.2 | $2.30^{+0.47}_{-0.47}$ | 2.24 | $17.13^{+0.85}_{-0.90}$ | 17.38 |
| G09v1.40 | $2.9^{+0.6}_{-0.5}$ | 2.4 | $2.61^{+0.29}_{-0.49}$ | 2.89 | $17.16^{+0.48}_{-0.57}$ | 17.26 |
| SDP17b | $3.2^{+2.1}_{-0.8}$ | 3.1 | $2.01^{+0.64}_{-0.64}$ | 2.57 | $17.95^{+0.19}_{-0.67}$ | 17.34 |
| SDP81 | $2.8^{+0.4}_{-0.5}$ | 3.2 | $2.53^{+0.09}_{-0.10}$ | 2.52 | $17.53^{+0.37}_{-0.35}$ | 17.22 |
| G12v2.43 | $2.9^{+0.5}_{-0.5}$ | 2.7 | $2.59^{+0.30}_{-0.43}$ | 2.88 | $17.47^{+0.61}_{-0.34}$ | 17.43 |
| G12v2.30 | $3.0^{+0.3}_{-0.3}$ | 2.8 | $2.88^{+0.09}_{-0.17}$ | 2.97 | $17.43^{+0.29}_{-0.29}$ | 17.50 |
| NCv1.143 | $3.0^{+0.4}_{-0.5}$ | 2.7 | $2.75^{+0.18}_{-0.29}$ | 2.93 | $17.31^{+0.40}_{-0.42}$ | 17.43 |
| NAv1.195 | – | – | – | – | – | – |
| NAv1.177 | $3.0^{+0.6}_{-0.8}$ | 3.6 | $2.50^{+0.18}_{-0.17}$ | 2.43 | $16.73^{+0.85}_{-0.71}$ | 16.06 |
| NBv1.78 | $4.8^{+1.4}_{-1.5}$ | 4.3 | $1.71^{+0.49}_{-0.33}$ | 1.47 | $18.22^{+0.88}_{-1.10}$ | 17.63 |
| NAv1.144 | $3.3^{+1.3}_{-0.8}$ | 3.3 | $1.88^{+0.53}_{-0.36}$ | 1.98 | $18.05^{+0.81}_{-0.62}$ | 17.98 |
| NAv1.56 | $2.5^{+0.5}_{-0.4}$ | 2.4 | $2.35^{+0.43}_{-0.48}$ | 2.75 | $16.38^{+0.72}_{-0.61}$ | 15.82 |
| G15v2.235 | $2.8^{+0.8}_{-0.6}$ | 3.1 | $1.93^{+0.39}_{-0.34}$ | 1.87 | $17.30^{+0.56}_{-0.63}$ | 17.29 |
| G12v2.890 | $3.2^{+1.3}_{-0.8}$ | 3.4 | $2.15^{+0.58}_{-0.57}$ | 2.43 | $17.43^{+1.06}_{-1.38}$ | 17.49 |
| G12v2.257 | $4.0^{+1.8}_{-1.3}$ | 4.4 | $1.35^{+0.37}_{-0.20}$ | 1.39 | $17.84^{+1.02}_{-0.84}$ | 17.38 |
| G15v2.779 | $4.1^{+1.5}_{-1.2}$ | 5.5 | $1.40^{+0.22}_{-0.14}$ | 1.33 | $17.80^{+1.05}_{-0.91}$ | 17.17 |

**Notes.** The value of ${\rm med}_{\pm1\sigma}$ shows the median value and $\pm1\sigma$ range of the values from the marginal probability distribution of each parameter. The set of parameters with the maximum probability of the posterior distribution within the $\pm1\sigma$ range are shown by the values of ${\rm max_{post.}}$.

Rather than generating a grid of line fluxes for a range of input parameters from the LVG models (e.g. Kamenetzky et al. 2011; Krips et al. 2011; Spinoglio et al. 2012), we directly use the `Python` package `emcee` to call `pyradex` (a `Python` wrapper of `RADEX` written by A. Ginsburg, see `https://github.com/keflavich/pyradex`) in each iteration for computing the `RADEX` results and passing them to `Python`, and sample the posterior probability distribution function. This can avoid calculations in the unfavourable part of the parameter space, thus saving the total running time of the codes. Following previous works (e.g. Spilker et al. 2014), we adopted flat log-prior within physically reasonable ranges, which are boundaries of the parameter space that we explored. The prior possibilities outside the boundary are set to 0. The parameter-space boundaries are as follows: $n_{\rm H_2} = 10^2$–$10^7$ cm$^{-3}$, $T_{\rm k} = T_{\rm CMB}$–$10^3$ K, $N_{\rm CO}/{\rm d}v = 10^{15.5}$–$10^{19.5}$ cm$^{-2}$ km$^{-1}$ s, in which $T_{\rm CMB}$ is the CMB temperature at the redshift of the source, which can be derived from $T_{\rm CMB} = 2.7315(1+z)$. We also adopt the range of $dv/dr$ to be 0.1–1000 km s$^{-1}$ pc$^{-1}$ (e.g. Tunnard & Greve 2016), which limits the range of the ratio between $N_{\rm CO}$ and $n_{\rm H_2}$. This prior also puts limits on the ratio between the LVG-solved $dv/dr$ and the $dv/dr$ derived from the virialised state, that is, $K_{\rm vir} = (dv/dr)_{\rm LVG}/(dv/dr)_{\rm vir}$, in which $(dv/dr)_{\rm vir} = 0.65\alpha^{0.5}(n_{\rm H_2}/(10^3\ {\rm cm}^{-3}))^{-0.5}$ km s$^{-1}$ pc$^{-1}$, where $\alpha = 0.5$–3 depending on the density profile (Papadopoulos & Seaquist 1999). Additionally, we set the priors to limit the column length to be smaller than the diameter of the entire SMG, which is about 7 kpc. This yields a constraint of the ratio $N_{\rm CO}/n_{\rm H_2}$ that is well outside the range given by the prior from $dv/dr$. Lastly, the molecular gas mass traced by the CO lines should not exceed the dynamical mass, for example, $\sim 10^{12}\ M_\odot$ (see Sect. 4.2). This yields an upper limit of $N_{\rm CO}/dv$ to be smaller than $\sim 10^{20}$ cm$^{-2}$ km$^{-1}$ s (e.g. Rangwala et al. 2011), which is well outside the parameter space as well.

A total of 400 walkers have been deployed to explore the parameter space initiated from the point of solution acquired by the quasi-Newton solver. We ensured proper convergence of the MCMC chains by a burn-in period of 100 iterations and 1000 subsequent iterations. The resulting posterior probability distributions and the marginal distribution of the parameters (generated by `corner.py`, Foreman-Mackey 2016) are shown by the blue density-contour plots and the blue histograms in Fig. C.1. We also indicate the 39% and 68% quantiles of the marginalised probability distribution of the parameter with dashed lines. The solutions with the maximum posterior probability within the 39% and 68% quantiles (the $\pm 1\,\sigma$ range) are marked with orange lines and points. The corresponding fit to the CO SLED is also shown in the Figure with an orange line overlaid on the black data points. All the results, the median value, the $\pm 1\,\sigma$ range and the maximum posterior probability, are summarised in Table 6, except for NAv1.195 because only one CO line of it has been observed, leading to an unreliable fitting.

From the single excitation component fitting, the range of $n_{\rm H_2}$ is found to be $\approx 10^{2.5}$–$10^{4.1}$ cm$^{-3}$, $T_{\rm k}$ is from 22 K to 750 K and $N_{\rm CO}/dv \approx 10^{17.13}$–$10^{18.22}$ cm$^{-2}$ km$^{-1}$ s for the *H*-ATLAS SMGs. In most cases, the values are close to those found by single-component LVG modelling of local ULIRGs (e.g. Ao et al. 2008) and high-redshift SMGs (e.g Lestrade et al. 2010; Combes et al. 2012; Riechers et al. 2013). Nevertheless, one should note that the observed CO SLEDs are dominated by the excitation from dense and warm molecular gas as suggested by the CO SLEDs peaking around $J_{\rm up} = 4$–7. Our observed CO SLEDs are biased towards the mid- and high-$J$ CO lines, and thus a single component fit is biased towards the high-excitation component seen in local ULIRGs. Indeed, most of our values of $T_{\rm k}$ from the single component analysis are higher than the low-excitation component seen in local ULIRGs but close to the values of the high-excitation component, for example, the warm molecular gas as found in Arp 220 (Rangwala et al. 2011). The lack of $J_{\rm up} \leq 2$ data will likely lead to overestimations of the values of $T_{\rm k}$ for our SMGs. The under-presentation of the low-excitation component is also shown in the fitted CO SLED to the observed flux: the modelled fluxes of CO(3–2) and CO(1–0) are often underestimated, especially in the cases of G09v1.40, SDP 17b, G12v2.43, NAv1.144 and G12v2.890 shown in Fig. C.1.

Therefore, to fully consider both the low-excitation component with a cooler temperature and the warmer, dense high-excitation component, we perform a two-excitation-component LVG modelling with the CO SLEDs: a low-excitation component with a lower value of $T_{\rm k}$ and a high-excitation component with a higher $T_{\rm k}$. An MCMC method similar to the aforementioned single-component LVG modelling is adopted. We assign two sets of $n_{\rm H_2}$, $T_{\rm k}$, $N_{\rm CO}/dv$ and $\Omega_{\rm app}$ to the two excitation components, so that the two components can have different physical conditions. Similar priors are also applied to help constrain the posterior distribution. For the two-excitation-components fit, the number of free parameters can be reduced significantly in some sources, therefore we carefully add some more informative priors. Besides the similar priors such as those used in the single-component analysis, we put additional prior constraints on the two-component LVG modelling as follows: (i) The size of the cooler low-excitation component should be larger than that of the high-excitation component (this has been suggested by the observations of the sizes of the emitting regions of different transitions of the CO lines, e.g. Ivison et al. 2011); and (ii) most importantly, we assume that the temperature of the low-excitation component should be close to the cold dust temperature. At high densities ($n_{\rm H_2} \geq 10^{4.5}$ cm$^{-3}$), the temperatures of the dust and the gas can be well coupled (Goldsmith 2001). However, the different heating mechanisms do not necessarily produce thermally balanced dust and gas temperature, especially at lower densities ($n_{\rm H_2} \leq 10^{3.5}$ cm$^{-3}$). So we use rather loose priors for the $T_{\rm k}$ of the cool low-excitation component, that is, a normal distribution with the value of mean and standard deviation equal to the cold dust temperature $Pr(T_{\rm k})_{\rm cool} \sim \mathcal{N}(T_{\rm d}, {T_{\rm d}}^2)$. This prior offers a reasonable guess of the $T_{\rm k}$ for the cooler component in the range of $\sim 0$ K–90 K (which will be further reduced into the range of $T_{\rm CMB}$–90 K by the aforementioned priors).

We have applied similar MCMC walkers in the parameter space and generated the posterior probability distributions (Fig. C.1). The dark green contours and histograms are the posterior probability distribution and marginal probability distribution for $n_{\rm H_2}$, $T_{\rm k}$ and $N_{\rm CO}/dv$ of the low-excitation component, while those shown in light green are for the high-excitation component. The solutions with the maximum posterior probability are shown by the dotted-dash pink line and red dashed line for the low- and high-excitation components, respectively. The corresponding CO SLEDs are over-plotted in Fig. C.1 with dotted-dash pink line and the dashed red lines. The values are summarised in Table 7. We show the histograms of the derived parameters for low- and high-excitation components in Fig.8, together with those of the single component LVG modelling. For the low-excitation component, the density ranges from $\sim 10^{2.8}$ cm$^{-3}$ to $10^{4.6}$ cm$^{-3}$ with large uncertainties, the gas temperature ranges from $\sim 20$ K to 30 K, and the CO column density per unit velocity gradient ranges from $10^{15.7}$ cm$^{-2}$ km$^{-1}$ s to $10^{17.9}$ km$^{-1}$ s with significant uncertainties. For the high-excitation component, the density ranges from $\sim 10^{2.8}$ cm$^{-3}$ to $10^{4.2}$ cm$^{-3}$, the gas temperature ranges from $\sim 60$ K to 400 K, and the CO column density per unit velocity gradient ranges from $10^{17.1}$ cm$^{-2}$ km$^{-1}$ s to $10^{18.1}$ cm$^{-2}$ km$^{-1}$ s. As a comparison, the gas densities of the two components are close, and the differences are within uncertainties. The gas temperatures of the high-excitation component are higher (peaking around $\sim 200$ K) than those of the low-excitation ones (peaking around $\sim 25$ K). The $T_{\rm k}$ of the cooler component is $\sim 10$–15 K lower than the dust temperature $T_{\rm d}$ as shown in Table 3.

As shown in the two-component-model-produced CO SLEDs in Fig. C.1, after including the low-excitation component, the flux of the low-$J$ CO lines can be better reproduced, especially in G09v1.40, SDP 17b, G12v2.43, and NAv1.144. This indicates that to fully explain the CO SLEDs, at least two excitation components are needed. The gas temperatures also decreased comparing the $T_{\rm k}$ from single component LVG modelling with the $T_{\rm k}$ of the high-excitation component from two-

**Table 7:** Two-component MCMC-resulted molecular gas properties of the *H*-ATLAS SMGs.

| Source | low-excitation | | | | | | high-excitation | | | | | |
|---|---|---|---|---|---|---|---|---|---|---|---|---|
| | $\log(n_{H_2})$ $\log(\mathrm{cm^{-3}})$ | | $\log(T_k)$ $\log(\mathrm{K})$ | | $\log(N_{CO}/dv)$ $\log(\mathrm{cm^{-2}\,km^{-1}\,s})$ | | $\log(n_{H_2})$ $\log(\mathrm{cm^{-3}})$ | | $\log(T_k)$ $\log(\mathrm{K})$ | | $\log(N_{CO}/dv)$ $\log(\mathrm{cm^{-2}\,km^{-1}\,s})$ | |
| | $\mathrm{med}_{\pm1\sigma}$ | $\mathrm{max_{post.}}$ | $\mathrm{med}_{\pm1\sigma}$ | $\mathrm{max_{post.}}$ | $\mathrm{med}_{\pm1\sigma}$ | $\mathrm{max_{post.}}$ | $\mathrm{med}_{\pm1\sigma}$ | $\mathrm{max_{post.}}$ | $\mathrm{med}_{\pm1\sigma}$ | $\mathrm{max_{post.}}$ | $\mathrm{med}_{\pm1\sigma}$ | $\mathrm{max_{post.}}$ |
| G09v1.97 | $2.8^{+1.5}_{-0.6}$ | 2.1 | $1.30^{+0.32}_{-0.16}$ | 1.55 | $15.70^{+1.51}_{-0.89}$ | 14.61 | $2.8^{+1.1}_{-0.6}$ | 3.1 | $2.17^{+0.51}_{-0.40}$ | 2.21 | $17.86^{+0.84}_{-1.10}$ | 17.58 |
| G09v1.40 | $3.3^{+1.3}_{-0.9}$ | 3.6 | $1.41^{+0.31}_{-0.30}$ | 1.55 | $16.08^{+1.17}_{-1.07}$ | 15.22 | $3.2^{+1.0}_{-0.8}$ | 3.5 | $2.41^{+0.39}_{-0.38}$ | 2.46 | $17.44^{+0.89}_{-0.69}$ | 17.20 |
| SDP17b | $3.2^{+1.6}_{-0.9}$ | 2.5 | $1.43^{+0.32}_{-0.30}$ | 1.65 | $16.49^{+1.71}_{-1.41}$ | 15.69 | $3.4^{+1.5}_{-0.9}$ | 3.8 | $2.31^{+0.45}_{-0.45}$ | 2.38 | $17.84^{+0.92}_{-0.90}$ | 17.30 |
| SDP81 | $2.8^{+1.0}_{-0.6}$ | 2.3 | $1.38^{+0.32}_{-0.24}$ | 1.52 | $15.96^{+1.26}_{-1.00}$ | 15.83 | $2.9^{+0.7}_{-0.6}$ | 3.2 | $2.43^{+0.16}_{-0.23}$ | 2.40 | $17.76^{+0.46}_{-0.58}$ | 17.45 |
| G12v2.43 | $4.1^{+1.3}_{-1.3}$ | 4.0 | $1.30^{+0.28}_{-0.19}$ | 1.44 | $15.52^{+1.29}_{-0.72}$ | 14.74 | $3.4^{+1.1}_{-0.9}$ | 4.2 | $1.91^{+0.38}_{-0.28}$ | 1.60 | $18.02^{+0.81}_{-0.83}$ | 17.90 |
| G12v2.30 | $3.6^{+1.3}_{-1.1}$ | 4.3 | $1.42^{+0.31}_{-0.20}$ | 1.35 | $17.22^{+1.04}_{-1.02}$ | 17.06 | $3.3^{+0.8}_{-0.7}$ | 3.3 | $2.57^{+0.33}_{-0.30}$ | 2.69 | $17.70^{+0.83}_{-0.62}$ | 17.77 |
| NCv1.143 | $4.1^{+1.5}_{-1.4}$ | 4.3 | $1.30^{+0.21}_{-0.13}$ | 1.28 | $17.00^{+1.17}_{-0.94}$ | 16.94 | $4.2^{+1.3}_{-1.2}$ | 5.1 | $1.80^{+0.26}_{-0.20}$ | 1.65 | $18.14^{+0.90}_{-0.96}$ | 17.70 |
| NAv1.195 | – | – | – | – | – | – | – | – | – | – | – | – |
| NAv1.177 | $3.9^{+1.2}_{-1.3}$ | 4.4 | $1.40^{+0.27}_{-0.23}$ | 1.20 | $16.96^{+1.18}_{-1.10}$ | 17.19 | $3.0^{+0.9}_{-0.7}$ | 3.0 | $2.60^{+0.28}_{-0.32}$ | 2.78 | $17.25^{+0.58}_{-0.93}$ | 17.33 |
| NBv1.78 | $4.6^{+1.5}_{-1.6}$ | 4.1 | $1.48^{+0.29}_{-0.28}$ | 1.55 | $17.66^{+1.24}_{-1.52}$ | 17.07 | $3.8^{+1.6}_{-1.2}$ | 4.6 | $2.27^{+0.43}_{-0.42}$ | 2.49 | $17.63^{+1.06}_{-1.37}$ | 16.01 |
| NAv1.144 | $3.3^{+1.9}_{-0.1}$ | 2.6 | $1.34^{+0.36}_{-0.29}$ | 1.34 | $16.19^{+1.91}_{-1.19}$ | 16.16 | $3.1^{+0.9}_{-0.7}$ | 3.4 | $2.30^{+0.38}_{-0.34}$ | 2.09 | $17.87^{+0.74}_{-0.62}$ | 17.90 |
| NAv1.56 | $3.0^{+1.0}_{-0.6}$ | 2.3 | $1.42^{+0.34}_{-0.31}$ | 1.63 | $16.24^{+1.08}_{-1.06}$ | 16.63 | $2.7^{+0.9}_{-0.5}$ | 2.9 | $2.31^{+0.45}_{-0.48}$ | 2.14 | $17.44^{+0.83}_{-0.95}$ | 17.51 |
| G15v2.235 | $3.9^{+1.8}_{-1.2}$ | 3.7 | $1.24^{+0.29}_{-0.18}$ | 1.23 | $17.63^{+1.07}_{-1.26}$ | 17.78 | $2.8^{+1.2}_{-0.6}$ | 3.0 | $2.19^{+0.53}_{-0.48}$ | 2.23 | $17.14^{+0.82}_{-1.09}$ | 17.56 |
| G12v2.890 | $4.1^{+1.9}_{-1.5}$ | 2.8 | $1.35^{+0.25}_{-0.25}$ | 1.48 | $17.10^{+1.57}_{-1.59}$ | 15.80 | $3.8^{+1.8}_{-1.2}$ | 5.1 | $2.25^{+0.49}_{-0.49}$ | 2.55 | $17.80^{+1.04}_{-1.36}$ | 17.47 |
| G12v2.257 | $4.5^{+1.5}_{-1.4}$ | 3.7 | $1.28^{+0.26}_{-0.20}$ | 1.36 | $17.54^{+1.10}_{-1.03}$ | 18.01 | $3.5^{+1.8}_{-1.0}$ | 4.4 | $2.17^{+0.55}_{-0.54}$ | 2.59 | $17.34^{+1.19}_{-1.48}$ | 15.86 |
| G15v2.779 | $4.3^{+1.2}_{-1.1}$ | 4.0 | $1.33^{+0.16}_{-0.11}$ | 1.38 | $17.91^{+0.95}_{-0.91}$ | 17.94 | $3.4^{+1.9}_{-1.0}$ | 5.0 | $2.15^{+0.58}_{-0.53}$ | 2.48 | $17.05^{+1.58}_{-1.66}$ | 16.25 |

**Notes.** See caption of Table 6.

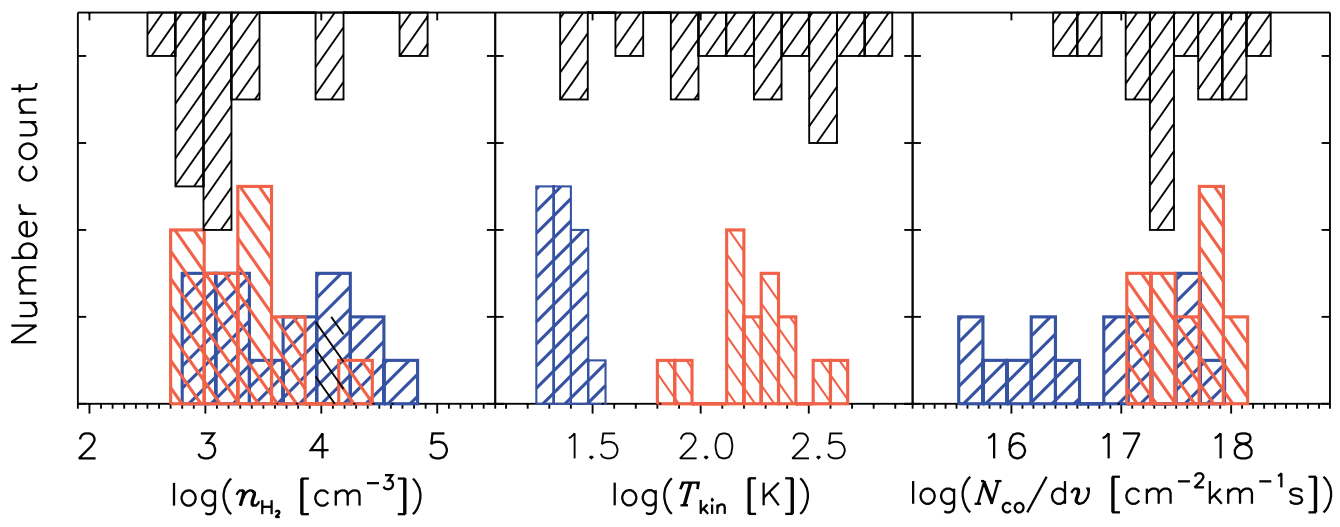

**Fig. 8:** Histograms of $n_{H_2}$, $T_k$ and $N_{CO}/dv$ derived from single-component LVG modelling (black), and those of the low- and high-excitation components from two-component LVG modelling as shown in blue and red, respectively.

component analysis, as in the case of G09v1.97, G09v1.40, G12v2.43, G12v2.30, NCv1.143 and NAv1.56. This again suggests that bias could be introduced using only a single excitation component to explain the full CO SLED. The physical properties of both components agree well with other studies of high-redshift SMGs (e.g. Scott et al. 2011; Danielson et al. 2013; Spilker et al. 2014). One has a low gas temperature, while another is smaller in size but with a warmer gas temperature, supporting the idea that there is likely more than one molecular gas excitation component. The latter is thought to be more closely related to the on-going star-forming activity compared with the cooler component. We also note that in the cases of NBv1.78, NAv1.56, G12v2.890 and G15v2.779, due to the very limited number of data points, the two-component fitting is highly reliant upon the priors and does not produce better fitting results than the single-component fittings. Thus, those individual two-component fitting results should be used with caution. Nevertheless, here, for the purpose of a statistical analysis of the physical properties of the gas excitation, we include these results in the discussion below.

We have investigated whether or not the CO linewidth correlates with the derived gas excitation condition. By comparing the linewidth with the parameters derived from LVG modelling, no significant correlation is found, with the absolute values of the correlation coefficient $\lesssim 0.3$.

To further investigate the physical properties from a statistical point of view, we plot the gas thermal pressure $P_{th}$ (defined by $P_{th} \equiv n_{H_2} \times T_k$) sampled from the MCMC posterior probability distributions versus the star formation efficiency (SFE) proxy defined as the ratio between $L_{IR}$ and molecular gas mass in Fig. 9. Because $L_{IR}$ is proportional to the star formation rate (e.g. Kennicutt & Evans 2012), the ratio between $L_{IR}$ and $M_{gas}$ is thus a good representative of SFE as defined by

$$\mathrm{SFE} \equiv \mathrm{SFR}/M_{gas}, \tag{8}$$

representing the SFR per unit molecular gas mass. As displayed in the left panel of Fig. 9, for comparison, we also include the values of the Milky Way (Draine 2011), the Tuffy galaxies (Zhu et al. 2007), the Antennae galaxies (Gao et al. 2001; Zhu et al. 2003), Arp 220 (Rangwala et al. 2011) and the $z \sim 6$ SMG HFLS3 (Riechers et al. 2013) for comparison. For the gas thermal pressure $P_{th}$ versus SFE of the *H*-ATLAS SMGs only (as indicated by the grey dashed square in the left panel of Fig. 9), we find the Pearson's correlation coefficient $R_P = 0.68$ (with the p-value $p = 0.003$), indicating the existence of a strong correlation. The values of the local ULIRG Arp 220 and the high-redshift SMG HFLS3 also follow the similar relation we find in the plot, with similar values of $P_{th}$ and SFE as for the *H*-ATLAS ULIRGs. However, the dynamical range of the $P_{th}$ is small for the SMGs and local ULIRGs; thus the correlation may be biased by a few sources with either very large or very small values of $P_{th}$. Therefore, we include some nearby galaxies with lower SFE: The Milky Way, the Tuffy galaxies and the Antennae galaxies. After including these sources, we find the Pearson's correlation coefficient increases to $R_P = 0.89$ ($p = 0.001$). With a fit to all the data points from the single component LVG modelling, we get a slope of $1.1 \pm 0.5$. We have also tested whether this correlation could arise from any relation between $n_{H_2}$ and SFE or between $T_k$ and SFE. For these two pairs of quantities, the correlation coefficient is much weaker, $R_P < 0.33$ ($p > 0.14$), compared with the one between the gas thermal pressure $P_{th}$ and SFE. This rules out the possibility that either $n_{H_2}$ or $T_k$ is dominating the correlation. All these pieces of evidence point out that there is likely a strong close-to-linear correlation between $P_{th}$ of the molecular gas and SFE, suggesting that the thermal pressure of the bulk of molecular gas is playing an important role in regulating the star formation at galactic scale across a range of redshifts. This is in-line with the theoretical works discussing the relation between gas pressure and SFE (e.g. Elmegreen & Efremov 1997; Wong & Blitz 2002).

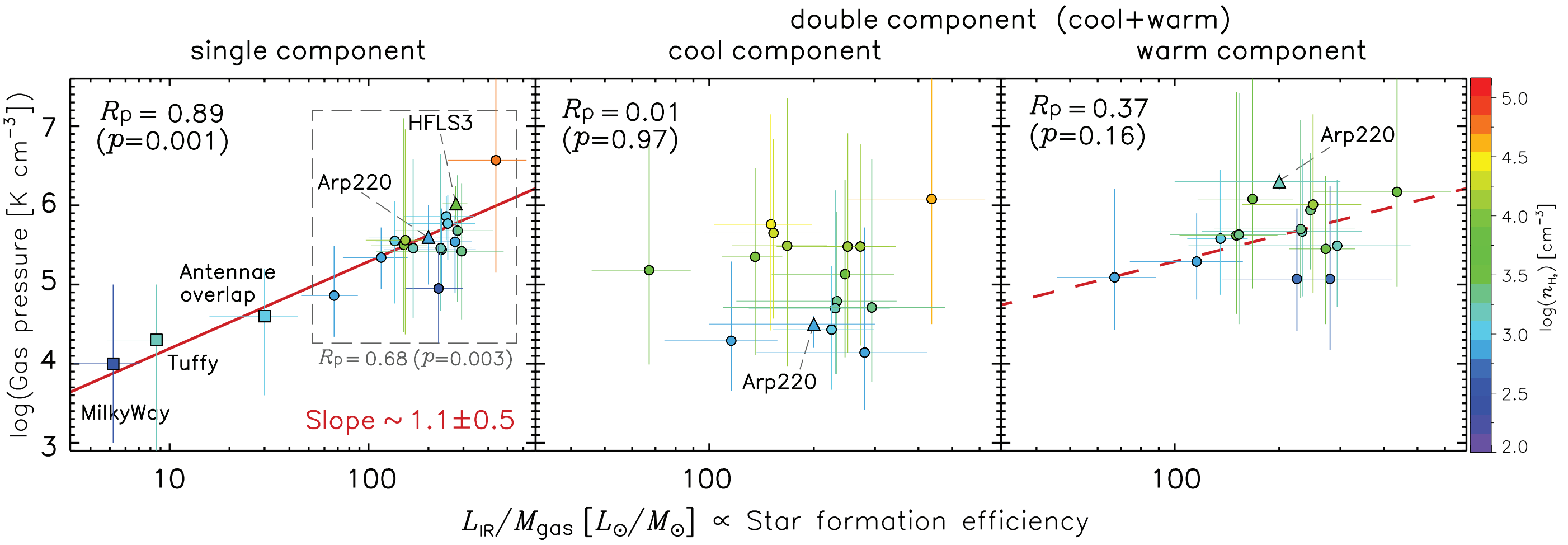

**Fig. 9:** *Left*: Thermal gas pressure plotted against star formation efficiency as indicated by $L_{IR}/M_{gas}$. Filled circles are the *H*-ATLAS SMGs. We also plot the values of the Milky Way, the Tuffy galaxies, the Antennae galaxies, Arp 220 and HFLS3 in filled triangles for comparison. The colour is coded based on their values of $n_{H_2}$. The red line shows the fit to the correlation, which yields a slope of $1.1 \pm 0.5$ for all the sources. The grey dashed square shows the region where *H*-ATLAS SMGs reside. *Middle*: Similar correlation plot as in the left panel but only for the cool component from the two-component LVG modelling. There is no correlation found ($R_P = 0.01$ and $p = 0.97$). *Right*: Similar correlation plot as in the left panel but only for the warm component in the two-component LVG modelling. The dashed red line is an overlay of exactly the same red line plotted in the left panel. The data points follow the same correlation found in the single component fit with an $R_P = 0.37$ and $p = 0.16$. In all the three panels, the legends show the Pearson's correlation coefficient $R_P$ with corresponding $p$-value. The values of $n_{H_2}$ for each point are indicated by the colour bar.

After decomposing the CO excitation into the low-excitation component and the high-excitation one, we also plot their $P_{th}$ versus SFE separately. The results of the (cool) low-excitation component and the (warm) high-excitation component are shown in the middle and right panel of Fig. 9, respectively. There is likely no correlation between $P_{th}$ and SFE for the low-excitation component, as suggested by the low coefficient $R_P = 0.01$ ($p = 0.97$). Nevertheless, for the high-excitation component there is still evidence of a correlation between $P_{th}$ and SFE with $R_P = 0.37$ ($p = 0.16$). Although, a large dynamical range and smaller uncertainties of $P_{th}$ and SFE are needed to further confirm this correlation. We also overlay the red line from the correlation fit to $P_{th}$ versus SFE of the single component modelling. It is clearly shown in the plot that, the data points of the high-excitation component follow this correlation well. This again shows that the high-excitation component is more closely related to the on-going star formation activity, while the low-excitation gas is much less tied to star formation.

### 5.2. The [C I](2–1) emission line in the high-redshift SMGs

The $^3$P fine structure [C I] lines of atomic carbon, [C I](1–0) at 492.2 GHz and [C I](2–1) at 809.3 GHz, are in a simple three-level energy system. The critical densities $n_{crit}$ for the [C I](1–0) and [C I](2–1) lines are both $\sim$0.5–1 $\times 10^3$ cm$^{-3}$, which is comparable to those of the CO(1–0) and CO(2–1) lines (Table 1). The energy levels for [C I]$^3$P$_2$ and [C I]$^3$P$_1$ are 23.6 K and 62.6 K, respectively. Atomic carbon is found to be well-mixed with the bulk of the $H_2$ gas, making it a promising molecular gas tracer together with low-$J$ CO lines (e.g. Papadopoulos & Greve 2004). From these two optically thin [C I] lines, one can derive the excitation temperature and gas density without relying on other complementary information (e.g. Weiß et al. 2003). In the high-redshift universe, [C I] has only been detected in a small number of systems, mainly gravitationally lensed SMGs and quasars (see the references in Walter et al. 2011; see Bothwell et al. 2017 for recent detections of [C I](1–0) lines in a sample of SPT-selected lensed SMGs; see also Wilson et al. 2017 for detections of [C I](2–1) by stacking *Herschel* SPIRE/FTS spectral data of galaxies at moderate redshifts). In this work, we have detected seven [C I](2–1) lines out of eight observations in our lensed high-redshift SMGs.

Papadopoulos & Greve (2004) find a good agreement between the total molecular gas mass derived from [C I] and CO lines and dust continuum in local ULIRGs. [C I] likely traces $H_2$ even more robustly than the low-$J$ CO lines (plus standard conversion factor) in extreme conditions on galactic scales (e.g. Zhang et al. 2014). High-redshift observations (e.g. Weiß et al. 2003; Alaghband-Zadeh et al. 2013) also support such an agreement, although Bothwell et al. (2017) recently found that either a larger $\alpha_{CO}$ or a high [C I] abundance is needed to balance the gas mass derived from CO and from [C I] lines. Since the observations of CO(1–0) lines at high-redshift becomes observationally difficult, acquiring the intensities of the [C I] lines, which are usually brighter than the CO(1–0) lines and residing in favourable bands for observation, could help us to better determine the total molecular mass in high-redshift SMGs. Zhang et al. (2016) show that at high-redshift the CO(1–0) line will also suffer observing against the CMB, making the line more difficult to observed. Moreover, they also found that CO can be destroyed by the cosmic rays coming from the intense star-forming activities, suggesting the [C I] lines to be a better tracer of the total molecular gas mass in such environments (Bisbas et al. 2015).

Using *Herschel* spectral data of a sample of local (U)LIRGs, Jiao et al. (2017) find a tight correlation between the CO(1–0)

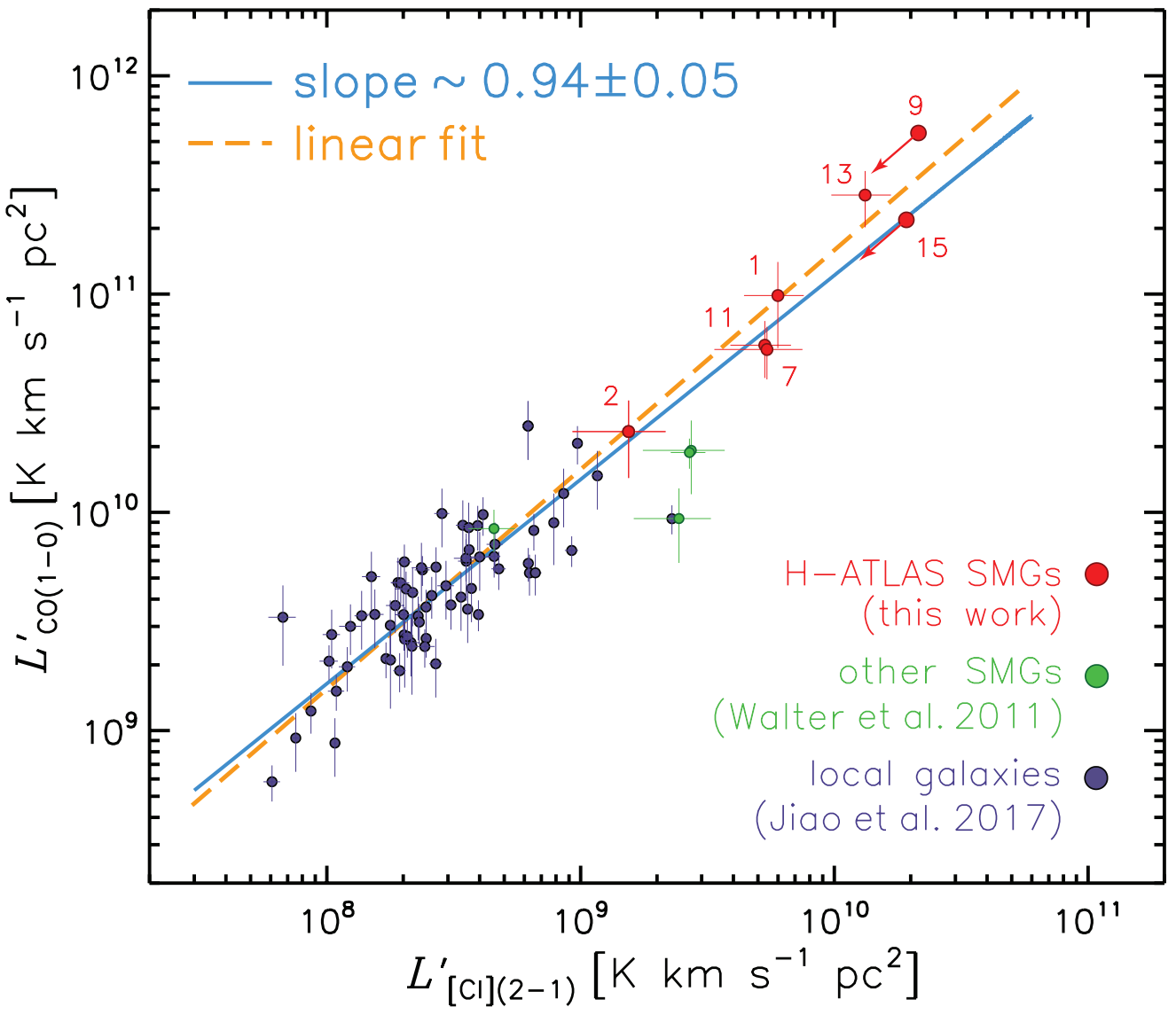

**Fig. 10:** $L'_{\rm [C\,I](2-1)}$ versus $L'_{\rm CO(1-0)}$ including (U)LIRGs from Jiao et al. (2017) (dark purple) *H*-ATLAS SMGs (red) and other SMGs from Walter et al. (2011) (green). Two of our sources, NAv1.177 (#9) and G12v2.257 (#15) are lacking lens modelling, thus we plot the limit, and the arrows show the direction for lensing correction. A fit to the correlation from the local to high-redshift galaxies is indicated by the blue line, with a slope of 0.94 ± 0.05. We also show a fit with a fixed slope of 1, that is, a linear fit, by the orange dashed line. The source index ID is indicated in Fig. 1 and Table 3.

line luminosity and the [C I] line luminosities. The tight correlations suggest that the [C I] lines trace the total molecular gas mass as the CO(1–0) line does. Using the empirical correlations, Jiao et al. (2017) derive a conversion factor $\alpha_{\rm [C\,I](2-1)}$ that converts $L'_{\rm [C\,I](2-1)}$ into $M_{\rm gas}$ to be $\alpha_{\rm [C\,I](2-1)} = 27.5 \pm 1.3\,(\rm K\,km\,s^{-1}\,pc^2)^{-1}$. To test whether this correlation can be extended to high-redshift SMGs, we plot the $L'_{\rm [C\,I](2-1)}$ versus $L'_{\rm CO(1-0)}$ including our lensed SMGs in Fig. 10. The fit to all the galaxies across different redshifts indicates a tight correlation close to linear, that is, the slope equals 0.94 ± 0.05. When fixing the slope to one, we find the fitted linear correlation to be valid both for local sources and high-redshift SMGs. Nevertheless, one would need to know the abundance of [C I] and the excitation temperature for the [C I] lines to properly derive the molecular gas mass from the [C I] lines as in Weiß et al. (2003). This would require the detection of both the [C I](1–0) and [C I](2–1) lines for the *H*-ATLAS SMGs, while we have only detection of the 2–1 line.

### *5.3. Star formation and the molecular gas content in high-redshift GMGs*

One of the key interests of studying star formation at galactic scale across all redshifts from the observation point of view is to quantify the 'star formation law', which is the correlation between $M_{\rm gas}$ and SFR (or the surface SFR density as originally defined by Kennicutt 1998b). This law is not only an essential input for the theoretical models of galaxy formation and evolution but also an important tool to study the relation between star formation and the molecular gas content. As we mentioned in Section 5.1, the SFR per unit molecular gas mass (i.e. SFE) shows how efficiently each unit of molecular gas mass is converted into stars (Eq.(8)). The inverse of SFE can be defined as the molecular gas depletion time,

$$t_{\rm dep} \equiv M_{\rm gas}/{\rm SFR}, \tag{9}$$

which indicates the exhausting time-scale of the molecular gas mass with the current SFR. The gas depletion time is a good way of representing the variations of the star formation properties. The value of $t_{\rm dep}$ varies from $\sim$1.3–1.5 Gyr for local star-forming galaxies (e.g. Kennicutt 1998b; Daddi et al. 2010; Genzel et al. 2010; Saintonge et al. 2011) to smaller values for high-redshift ones ('main sequence'): $\sim$0.7 Gyr (e.g. Tacconi et al. 2013; Sargent et al. 2014) or even smaller values (Saintonge et al. 2013). Indeed, there seems to be a cosmic evolution trend for star-forming galaxies as found by Saintonge et al. (2013). In Fig. 11 (extended from Fig. 7 of Aravena et al. 2016b), we plot the gas depletion time of our sources (see values in Table 5) and compare it with different kinds of galaxies across a range of redshifts, including; the aforementioned evolution track with $z$ of star-forming galaxies (Saintonge et al. 2013); nearby ULIRGs (Solomon et al. 1997) and $z = 0.6$–1 ULIRGs (Combes et al. 2013); other lensed/unlensed SMGs/ULIRGs studied (see Aravena et al. 2016b and the references within). As shown in Fig. 11, unlike star-forming galaxies, for all types of ULIRGs, the depletion time is found to be much smaller, from $\sim$10 to 100 Myr (roughly 10 times smaller than that of the star-forming main sequence galaxies), and there seems to be no evidence of a cosmic evolution of $t_{\rm dep}$. The *H*-ATLAS SMGs show no difference in the $t_{\rm dep}$ compared with SPT lensed ones and other SMGs/ULIRGs studied. They are well below the values for star-forming main sequence galaxies and show no evidence of variation across redshifts.

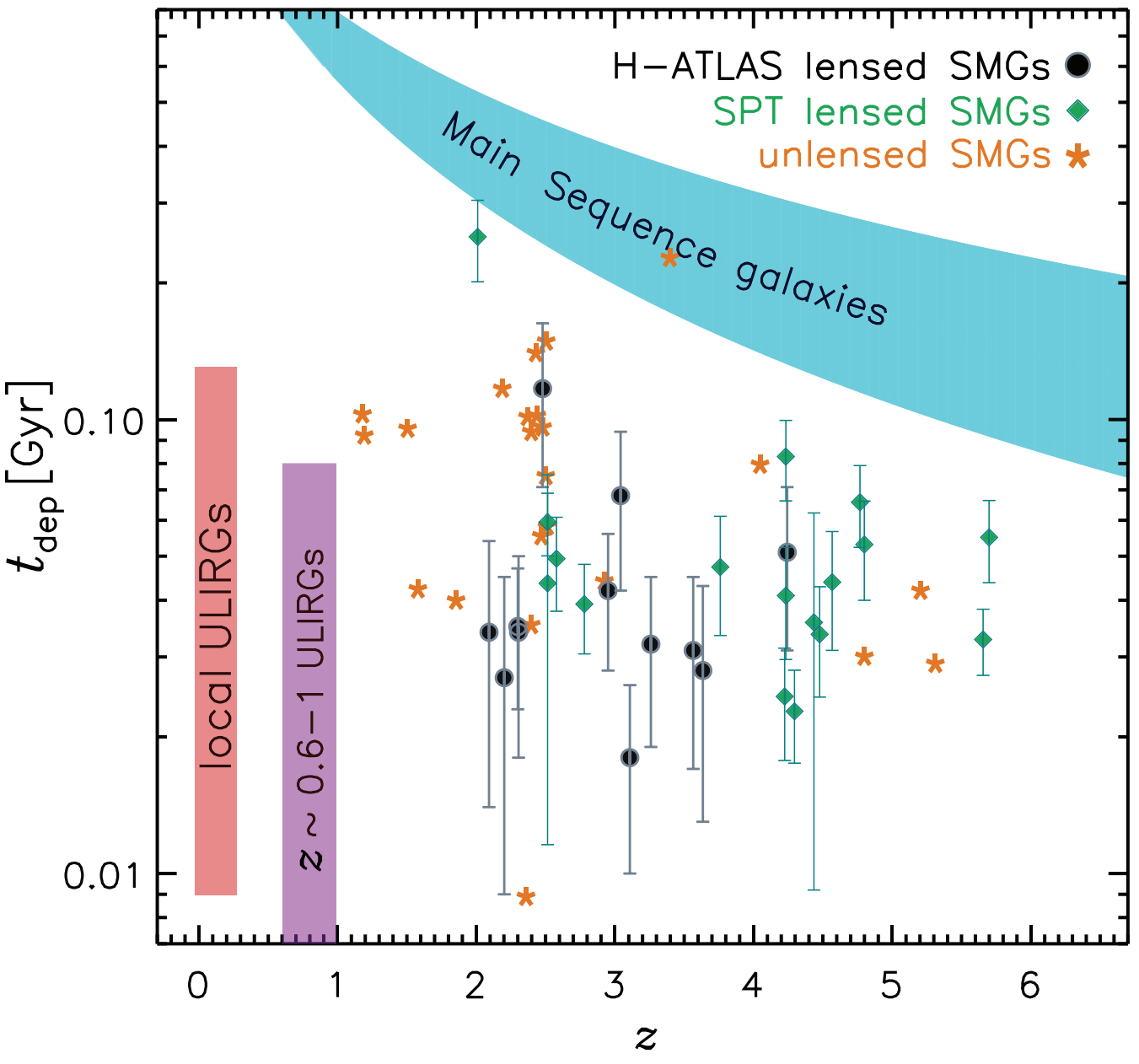

**Fig. 11:** Extended from Fig. 7 of Aravena et al. 2016b: Molecular gas depletion time of our lensed SMGs (black circles), SPT lensed SMGs (green diamonds), and other unlensed SMGs (orange stars). The ranges of local ULIRGs (Solomon et al. 1997) and $z = 0.6$–1 ULIRGs (Combes et al. 2013) are also included as the red and purple regions, respectively. The cyan region indicates the main sequence galaxies as described by Saintonge et al. (2013) through the formula $t_{\rm dep} = 1.5(1+z)^{\alpha}$, in which $\alpha$ is from $\alpha = -1.5$ (Davé et al. 2012) to $\alpha = -1.0$ (Magnelli et al. 2013).

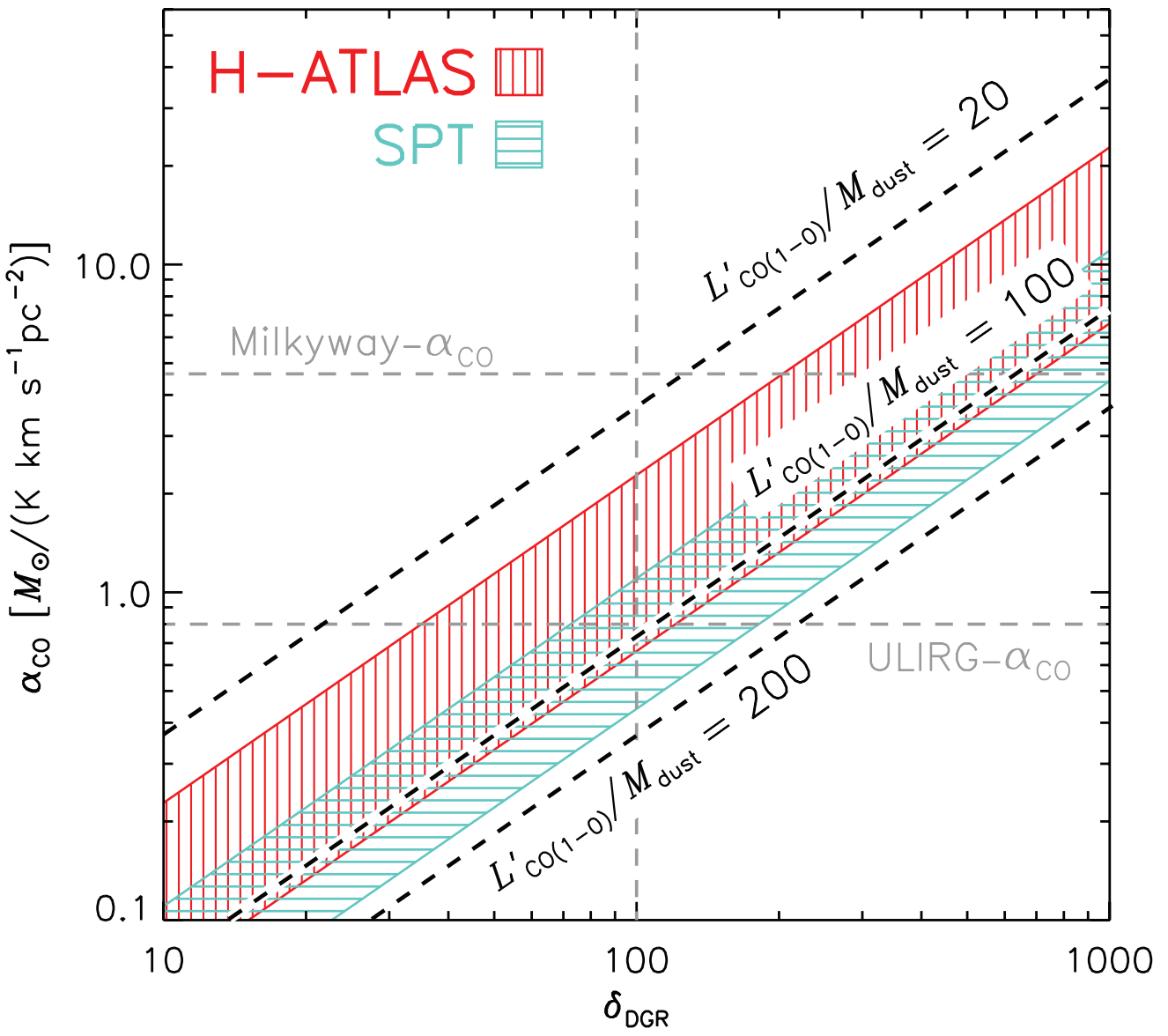

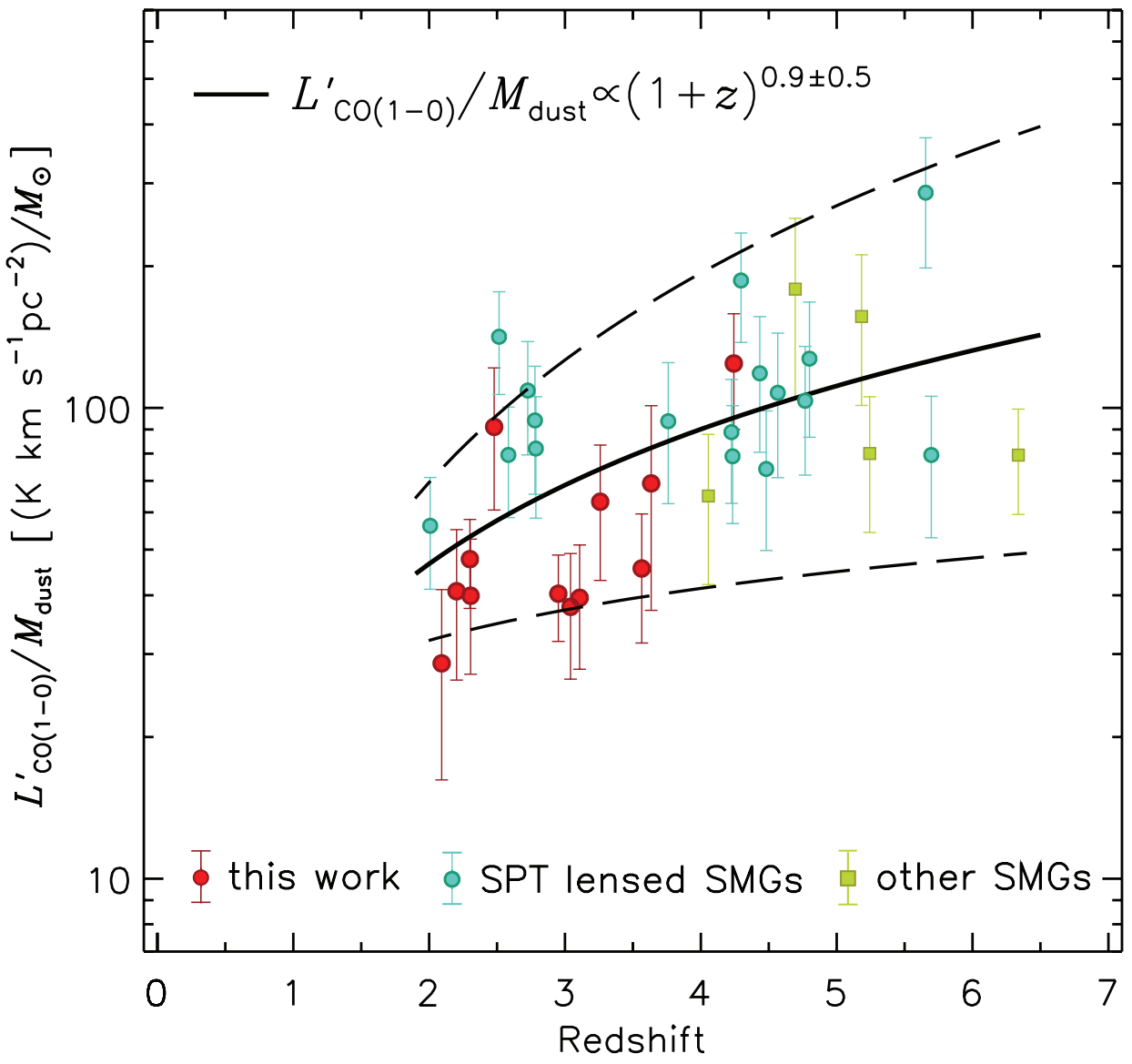

**Fig. 12:** *Left panel*: The $\pm 1\sigma$ range of $\delta_{GDR}$ and $\alpha_{CO}$ derived from observed $M_{dust}$ and $L'_{CO}$ using Eq. (10). Red region is from this work while the green region is from Aravena et al. (2016b). Our results generally agree with the SPT sources' albeit having slightly larger $\alpha_{CO}$ or smaller $\delta_{GDR}$. *Right panel*: The ratio between $L'_{CO}$ and $M_{dust}$ of SMGs across the different redshifts. The SPT lensed SMGs in green points are from Aravena et al. (2016b), while our sources are in red. We also plot other SMGs from the literature in yellow filled squares (see references in text). The black line shows a best-fit to all the data points and dashed black lines show the $\pm 1\sigma$ ranges (for the slope only) from the fit. The resulting relation is shown in the legend.

However, one should note that the values of $M_{gas}$ derived above are based on the assumption of $\alpha_{CO} = 0.8\,M_\odot\,(\mathrm{K\,km\,s^{-1}\,pc^2})^{-1}$. This conversion factor may be different in various types of galaxies. To derive reliable value of $\alpha_{CO}$, one needs to estimate the molecular gas mass from other methods rather than converting from CO fluxes. One of the most convenient ways is to compute the molecular gas mass from the dust mass by assuming a value for the gas-to-dust mass ratio $\delta_{GDR}$, since the dust continuum is much easier to measure in practical observations (e.g. Scoville et al. 2014, 2017). Nevertheless, $\delta_{GDR}$ has no constant value for different galaxies. In fact, both $\delta_{GDR}$ and $\alpha_{CO}$ are functions of the metallicity of different galaxies based on both observational results (e.g. Wilson 1995; Leroy et al. 2011; Magdis et al. 2011; Genzel et al. 2012) and theoretical works (e.g. Narayanan et al. 2012). To avoid the uncertainties caused by assuming single values for $\delta_{GDR}$ and $\alpha_{CO}$, we explore a combination of both, which is derived directly from observational quantities. From Eq.(2) and Eq.(3), we can derive

$$\delta_{GDR}/\alpha_{CO} = L'_{CO(1-0)}/M_{dust}, \qquad (10)$$

in which $L'_{CO(1-0)}$ is from CO line observations (converted mostly from CO(3–2) in our cases as mentioned in Section 4.1), and $M_{dust}$ is calculated from the observed submm/mm dust continuum flux densities based on a modified black body model (both for our work and the SPT lensed SMGs).

The left panel of Fig. 12 shows the possible $\pm 1\,\sigma$ range of $\delta_{GDR}$ versus $\alpha_{CO}$ for *H*-ATLAS SMGs computed using Eq.(10), together with the one from SPT-lensed SMGs for comparison. By taking a common $\delta_{GDR}$ of 100, the corresponding $\alpha_{CO}$ for our *H*-ATLAS SMGs is $\sim 0.7$–$2\,M_\odot\,(\mathrm{K\,km\,s^{-1}\,pc^2})^{-1}$, while for the SPT sources, $\alpha_{CO}$ is slightly smaller, $\sim 0.4$–$1\,M_\odot\,(\mathrm{K\,km\,s^{-1}\,pc^2})^{-1}$. Nevertheless, one need to accurately measure either $\delta_{GDR}$ or $\alpha_{CO}$ in order to break the degeneracies between these two quantities. Normally, one could use the calibration of the relation between the metallicity $Z$ and $\delta_{GDR}$ (e.g. Magdis et al. 2011), to derive the latter from the former by observing the optical emission lines. But for the high-redshift SMGs, due to the extreme dust obscuration, it is very challenging to measure the metallicity from optical observation, making it thus difficult to accurately pinpoint the value of $\delta_{GDR}$ and $\alpha_{CO}$.

By including other published SMGs at different redshifts (Magdis et al. 2011; Combes et al. 2012; Salomé et al. 2012; Walter et al. 2012; Riechers et al. 2013), we have plotted the $L'_{CO(1-0)}/M_{dust}$ versus redshift for all of them in the right panel of Fig. 12 similarly to Tan et al. (2014). There seems to be an increasing trend for the SMGs across the cosmic time, from nearby to the high-redshift Universe. The best fit to this trend is $L'_{CO(1-0)}/M_{dust} \propto (1+z)^{0.9\pm0.5}$, in the redshift range from $z \approx 2$ up to $z \approx 6$. This agrees with the fact that the average SPT sources have higher values of $\delta_{GDR}/\alpha_{CO}$ (i.e. $L'_{CO(1-0)}/M_{dust}$), since the SPT sources have a higher average redshift compared with the *H*-ATLAS one. It has been suggested that $\delta_{GDR}$ is linearly anti-correlated with $Z$, that is, $\delta_{GDR} \propto Z^{-1}$ (e.g. Santini et al. 2010; Leroy et al. 2011; Magdis et al. 2011), while the dependence of $\alpha_{CO}$ seems to be steeper (e.g. $\alpha_{CO} \propto Z^{-1.4}$, calibration from Leroy et al. 2011). Combining these two calibrations, one would derive a correlation between the ratio $\delta_{GDR}/\alpha_{CO}$ ($L'_{CO(1-0)}/M_{dust}$) and $Z$ (e.g. Magdis et al. 2012). This offers one possible explanation for the observed correlation, if the higher-redshifted SMGs in this plot have larger values of metallicity $Z$. Considering the cosmic enrichment of metallicity at high-redshift (e.g. Troncoso et al. 2014), the higher-redshifted SMGs in the plot are unlikely to be evolutionally linked with the lower redshifted ones. At given redshifts, $Z$ increases with stellar masses (see a summary in Tan et al. 2014). Therefore, this increasing value of $L'_{CO(1-0)}/M_{dust}$ could rather be explained by a selection bias towards the high-mass systems which correspond to those higher-redshifted SMGs in the Figure.

### 5.4. CO gas and $H_2O$ gas comparison

The systematic study of local galaxies (from normal star-forming galaxies to nearby ULIRGs) shows the close relation between the submm $H_2O$ emission and the star-forming activity (Yang et al. 2013). A similar conclusion has been extended to the high-redshift by the study of a group of lensed SMGs (O13 and Y16). The submm $H_2O$ lines are dominated by FIR pumping, which is closely related to the warm dust ($T_d \sim 40$–90 K, see e.g. González-Alfonso et al. 2014). One may compare the gas content traced by these submm $H_2O$ lines with the CO lines we observed. The $H_2O$ column density derived is around $\sim 0.3 \times 10^{15}$ cm$^{-2}$ km$^{-1}$ s to $\sim 2 \times 10^{16}$ cm$^{-2}$ km$^{-1}$ s (Y16), which is about several tens up to several thousand times smaller than the CO column density (from single component LVG modelling).

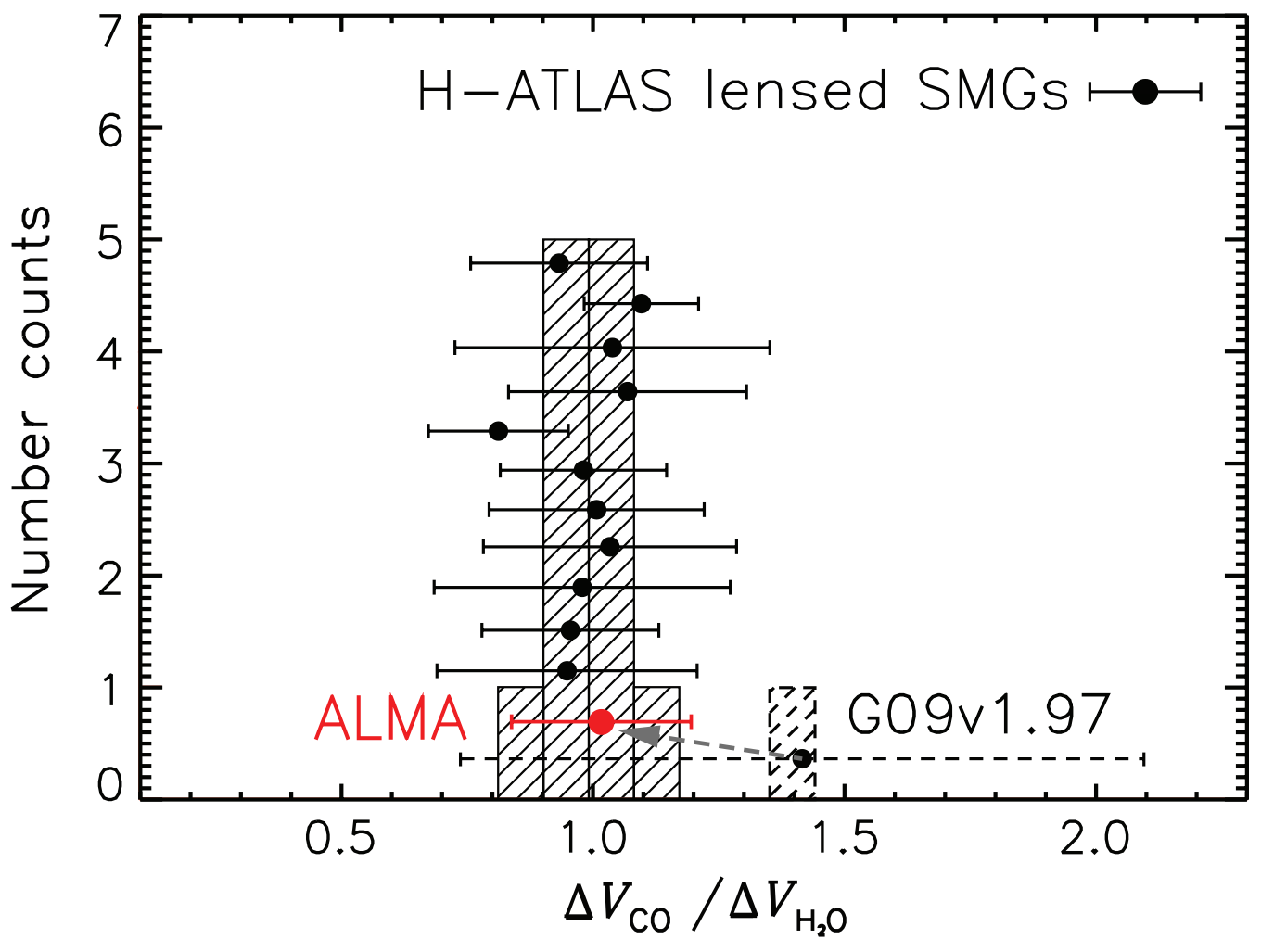

**Fig. 13:** Histogram showing the distribution of the ratio (indicated by black data points) between the linewidths of CO and $H_2O$. The $H_2O$ linewidth are from Y16 and O13. It is clear that most of our sources have similar CO and $H_2O$ linewidth, with an average ratio of 1.0 ± 0.1. For G09v1.97, the IRAM-30m spectrum shows that the asymmetric CO line is $\sim 1.5$ times larger than that of the $H_2O$ line (Y16) as shown by the dashed error bar and histogram. However, using the ALMA high S/N CO and $H_2O$ spectral data of G09v1.97 (Yang et al. in prep.), we find a ratio of the linewidth to be very close to 1 as indicated by the red point. The source of SDP 81 is not shown here because it is a special case as discussed in the text.

Comparing the linewidths between CO and $H_2O$ is instructive. As shown in Fig. 13, most of the *H*-ATLAS sources have similar CO and $H_2O$ linewidths. This suggests that they are possibly coming from similar regions as found in O13 and Y16. However, for SDP 81 and G09v1.97, the CO linewidth obtained using the IRAM-30m data is approximately two times larger than that of the $H_2O$ lines. As shown in O13, the blue component of the CO line of SDP 81 is much weaker compared with the red component, and this component is not even detected by ALMA long baseline observations (ALMA Partnership, Vlahakis et al. 2015). Thus, here the linewidth ratio only accounts for the $H_2O$ linewidth of the red component, which results in the CO linewidth being about twice larger than that of the $H_2O$ line in SDP 81. However, if the S/N of the spectral data is improved, one will detect the weakly magnified velocity component as described in Section 4.3. From the ALMA data of CO(6–5) and $J = 2$ $H_2O$ lines of G09v1.97 (Yang et al., in prep.), we find asymmetrical CO and $H_2O$ line profiles that are very similar, showing a dominant red velocity component with an approximately six times weaker blue velocity component (see Fig. A.1 for the ALMA CO(6–5) spectrum). The linewidth ratio, acquired by the high S/N ALMA data, is found very close to 1 as indicated by the red data point in Fig. 13. The case of G09v1.97 clearly shows that the low/mid-S/N spectral data may suffer from the bias caused by differential lensing effect as mentioned in Section 4.3. Besides the line profiles between $H_2O$ and CO, Oteo et al. (2017) find similar line profiles for $H_2O$ and mid-*J* CO and HCN lines in a $z \sim 1.6$ lensed SMG, SDP 9.

The similarity of the line profile strongly suggests that the emitting regions are co-spatially located. Both the gas tracers are closely linked to the on-going star formation activities. However, further detailed studies combining the gas excitation, dust emission, and the far-infrared pumping of the $H_2O$ lines is needed to fully incorporate the complex physical processes in these SMGs.

## 6. Summary

In this work, we report a survey of multiple $J_{up}$ (mostly from $J_{up} = 3$ up to $J_{up} = 8$) CO lines in a sample of *Herschel*-selected lensed SMGs at $z \sim 2$–4. We have detected 47 CO lines and 7 [C I] lines in these SMGs using the IRAM-30m telescope.

Comparing the CO linewidth with those of the SPT lensed SMGs and an unlensed sample, we find evidence of a significant bias introduced by differential lensing that distorts the line profile, resulting in underestimation of the linewidth (usually by a factor of 2). This induces underestimation of the dynamical mass if one uses the observed linewidth measured for the lensed SMGs blindly. Differential lensing also slightly affects flux estimates of the lowest-*J* CO lines. This is mostly due to the difference in spatial distribution, since the lowest-*J* lines, especially CO(1–0), are usually several times more extended than the mid-*J* to high-*J* CO lines. By comparing the CO(1–0) to CO(3–2) flux ratio of the SMGs for our lensed SMGs and the ones of the unlensed SMGs, we find that the differential lensing could cause a $\sim$ 1.3 ± 0.4 times overestimation of the magnification of the CO(1–0).

We also calculate the dynamical mass of the *H*-ATLAS SMGs; if one did not correct for differential lensing effects, this would lie in the range of $0.1$–$7 \times 10^{11}$ $M_\odot$, with a large uncertainty due to the unknown extension and sky inclination (if assuming a rotating-disk scenario) of the CO emission. But for the sources with a narrow linewidth, due to the aforementioned differential lensing effect, their dynamical masses might be underestimated. Nevertheless, for the sources with broad linewidth, the ratio between gas mass and dynamical mass equals about 0.34 ± 0.10, that is, a gas fraction of $\sim 34\%$, which is close to the values of other SMGs and empirical model predictions.

The multiple-transition CO line data allow us to study the molecular gas excitation of the SMGs. The CO SLEDs are mostly peaking around $J_{up} = 5$–7, thus dominated by the warm dense gas. Using LVG modelling by fitting the SLEDs with a single excitation component via a Bayesian approach, we derive a gas density $n_{H_2} \approx 10^{2.5}$–$10^{4.1}$ cm$^{-3}$, gas temperature $T_k \approx 20$–750 K and CO column density per unit velocity gradient $N_{CO}/dv \approx 10^{16.4}$–$10^{17.8}$ cm$^{-2}$ km$^{-1}$ s. But we notice that the $J_{up} \leq 3$ CO lines are likely under-predicted by the single-component LVG model. This indicates the existence of a low-excitation component. We have thus performed a two-component LVG modelling and derived the gas density, temperature, and the CO column density per unit velocity gradient for each component. The low-excitation component has a cooler gas temperature of about 20–30 K with a gas density of about $10^{2.8}$–$10^{4.6}$ cm$^{-3}$, while for the high-excitation component, the gas temperature is higher, ~ 60–400 K, with a gas density of about $10^{2.7}$–$10^{4.2}$ cm$^{-3}$. We find a correlation between the gas thermal pressure $P_{\rm th}$ derived from single component LVG modelling and the star formation efficiency. The warm high-excitation component also follows a similar trend while the cool low-excitation component is much less correlated with the on-going star formation. This shows that the high-excitation warm dense gas is more closely related to the star formation than the cool gas in the SMGs. This agrees with previous studies of a large sample of local galaxies.

We also study the global properties of the molecular gas and their relation with star formation. By assuming a conversion from CO luminosity to gas mass, $\alpha_{\rm CO} = 0.8\,M_{\odot}\,({\rm K\,km\,s^{-1}\,pc^2})^{-1}$, we derive gas to dust mass ratios in the approximate range from 30 to 100. The gas depletion time $t_{\rm dep}$ of the *H*-ATLAS SMGs shows no difference compared with other lensed/unlensed SMGs; the value is also close to that of the local ULIRG. Furthermore, no cosmic evolution trend is found for $t_{\rm dep}$. However, in order to avoid the uncertainties from the assumptions of the values of $\alpha_{\rm CO}$ (and $\delta_{\rm GDR}$), we study the quantity $L'_{\rm CO(1-0)}/M_{\rm dust}$ which is proportional to the ratio $\delta_{\rm GDR}/\alpha_{\rm CO}$. We find for the SMGs, this ratio is increasing with increasing redshift, which could be caused by a selection bias.

With the detections of seven [C I](2–1) lines in our *H*-ATLAS lensed SMGs, we extend the correlation between the luminosity of the CO(1–0) line and the [C I](2–1) line found in a sample of local (U)LIRGs. All of them can be well fitted with a single linear correlation, indicating that the [C I] lines can be good indicators of the total molecular gas mass, both for nearby (U)LIRGs and the high-redshift SMGs. However, [C I](1–0) data are needed to properly derive the gas mass from the [C I] lines.

Finally, we compare the linewidths of the CO and $H_2O$ emission lines for a sample of 13 SMGs, using multi-*J* CO emission lines. We find that the linewidths of the CO lines and $H_2O$ lines agree very well in most cases. This supports our previous argument that the emitting regions of the CO and $H_2O$ lines are likely to be co-spatially located.

This work reports for the first time, a systematic study of the CO gas excitation in a sample of high-redshift lensed SMGs. However, deriving accurate values for the physical properties of the molecular gas is challenging. To model the CO excitation with two excitation components, eight free parameters are needed. A rather complete sample of the CO SLED is thus required to better constrain the models. Also, for strongly lensed systems, due to differential lensing, the line profile can be easily distorted (e.g. the case of SDP 81 Dye et al. 2015), namely the magnification factor at different velocity components of the emission line can be rather different. One could overcome this disadvantage by increasing the S/N of the spectrum using a telescope with better sensitivities such as ALMA and future NOEMA.

*Acknowledgements.* We would like to thank the anonymous referee for constructive comments. The authors are grateful to the IRAM staff for their support. CY thanks Claudia Marka and Nicolas Billot for their help of the IRAM-30m/EMIR observation. The authors also thank Yinhe Zhao for kindly sharing the local [C I] line data. CY and YG acknowledge support by National Key R&D Program of China (2017YFA0402700) and the CAS Key Research Program of Frontier Sciences. CY, AO and YG acknowledge support by NSFC grants 11311130491 and 11420101002. CY, AO, AB and YG acknowledge support from the Sino-French LIA-Origins joint exchange program. CY is supported by the China Scholarship Council grant (CSC No.201404910443). E.G-A is a Research Associate at the Harvard-Smithsonian Center for Astrophysics, and thanks the Spanish Ministerio de Economía y Competitividad for support under projects FIS2012-39162-C06-01 and ESP2015-65597-C4-1-R, and NASA grant ADAP NNX15AE56G. ZYZ, RJI and IO acknowledges support from ERC in the form of the Advanced Investigator Programme, 321302, COSMICISM. H.D. acknowledges financial support from the Spanish Ministry of Economy and Competitiveness (MINECO) under the 2014 Ramón y Cajal program MINECO RYC-2014-15686. MJM acknowledges the support of the National Science Centre, Poland through the POLONEZ grant 2015/19/P/ST9/04010. This project has received funding from the European Union's Horizon 2020 research and innovation programme under the Marie Skłodowska-Curie grant agreement No. 665778. IRS acknowledges support from STFC (ST/P000541/1), the ERC Advanced Grant DustyGal (321334) and a Royal Society/Wolfson Merit Award. US participants in *H*-ATLAS acknowledge support from NASA through a contract from JPL. Italian participants in *H*-ATLAS acknowledge a financial contribution from the agreement ASI-INAF I/009/10/0. SPIRE has been developed by a consortium of institutes led by Cardiff Univ. (UK) and including: Univ. Lethbridge (Canada); NAOC (China); CEA, LAM (France); IFSI, Univ. Padua (Italy); IAC (Spain); Stockholm Observatory (Sweden); Imperial College London, RAL, UCL-MSSL, UKATC, Univ. Sussex (UK); and Caltech, JPL, NHSC, Univ. Colorado (USA). This development has been supported by national funding agencies: CSA (Canada); NAOC (China); CEA, CNES, CNRS (France); ASI (Italy); MCINN (Spain); SNSB (Sweden); STFC, UKSA (UK); and NASA (USA). Extensive use was made of the NASA IDL (Interactive Data Language) Astronomy User's Library (https://idlastro.gsfc.nasa.gov) and the Coyote Graphics Library.

[1] Purple Mountain Observatory/Key Lab of Radio Astronomy, Chinese Academy of Sciences, Nanjing 210008, PR China
e-mail: yangcht@pmo.ac.cn
[2] Institut d'Astrophysique Spatiale, CNRS, Univ. Paris-Sud, Université Paris-Saclay, Bât. 121, 91405 Orsay cedex, France
[3] Graduate University of the Chinese Academy of Sciences, 19A Yuquan Road, Shijingshan District, 10049, Beijing, PR China
[4] CNRS, UMR 7095, Institut d'Astrophysique de Paris, F-75014, Paris, France
[5] UPMC Univ. Paris 06, UMR 7095, Institut d'Astrophysique de Paris, F-75014, Paris, France
[6] Leiden Observatory, Leiden University, P.O. Box 9513, NL-2300 RA Leiden, The Netherlands
[7] Institute for Astronomy, University of Edinburgh, Royal Observatory, Blackford Hill, Edinburgh, EH9 3HJ, UK
[8] European Southern Observatory, Karl Schwarzschild Straße 2, 85748, Garching, Germany
[9] Max Planck Institute for Astronomy, Konigstuhl 17, D-69117 Heidel- berg, Germany
[10] Universidad de Alcalá, Departamento de Física y Matemáticas, Campus Universitario, 28871 Alcalá de Henares, Madrid, Spain
[11] Instituto de Astrofísica de Canarias (IAC), E-38205 La Laguna, Tenerife, Spain
[12] Universidad de La Laguna, Dpto. Astrofísica, E-38206 La Laguna, Tenerife, Spain
[13] Joint ALMA Observatory, Alonso de Córdova 3107, Vitacura, Santiago, Chile
[14] Institut de Radioastronomie Millimétrique (IRAM), 300 rue de la Piscine, 38406 Saint-Martin-d'Hères, France
[15] Astronomy Department, Cornell University, 220 Space Sciences Building, Ithaca, NY 14853, USA
[16] Department of Physics and Astronomy, Rutgers, The State University of New Jersey, 136 Frelinghuysen Road, Piscataway, NJ 08854-8019, USA
[17] Astronomical Observatory Institute, Faculty of Physics, Adam Mickiewicz University, ul. Słoneczna 36, 60-286 Poznań, Poland
[18] Department of Physics and Astronomy, University of California, Irvine, Irvine, CA 92697, USA
[19] Centre for Extragalactic Astronomy, Durham University, Department of Physics, South Road, Durham DH1 3LE, UK

## Appendix A: Spectral data.

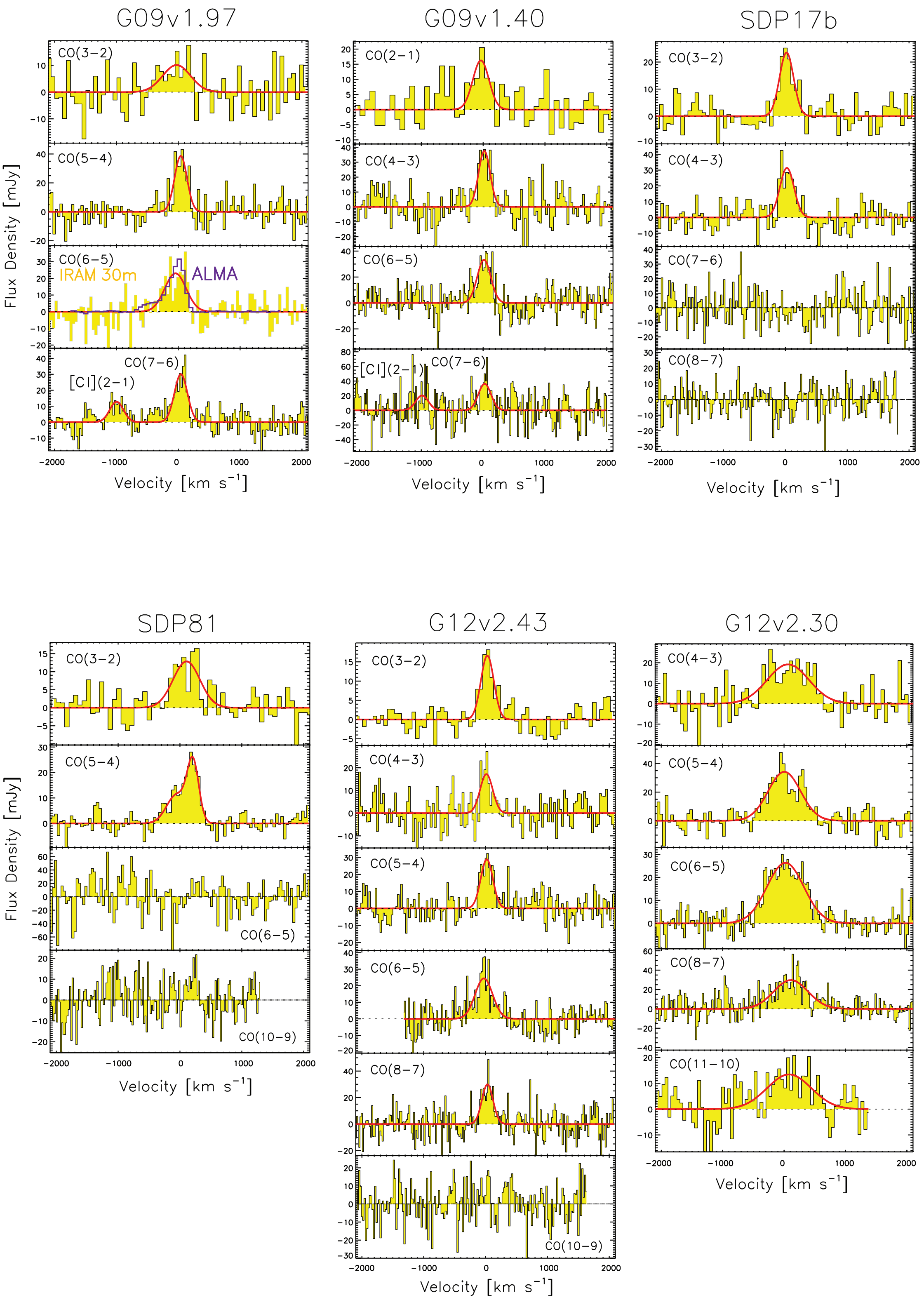

**Fig. A.1:** Spatially integrated spectra of CO lines in the *H*-ATLAS sources. The red lines represent the Gaussian fitting to the emission lines. The detections are $\gtrsim 3\,\sigma$. Zero velocity is set to the CO line sky frequency according to the previously measured spectroscopy redshifts $z_{\rm spec}$ given in Table 3. For CO(6–5) in G09v1.97, we also overlaid the ALMA spectral data (Yang et al., in prep.) in purple for comparison.

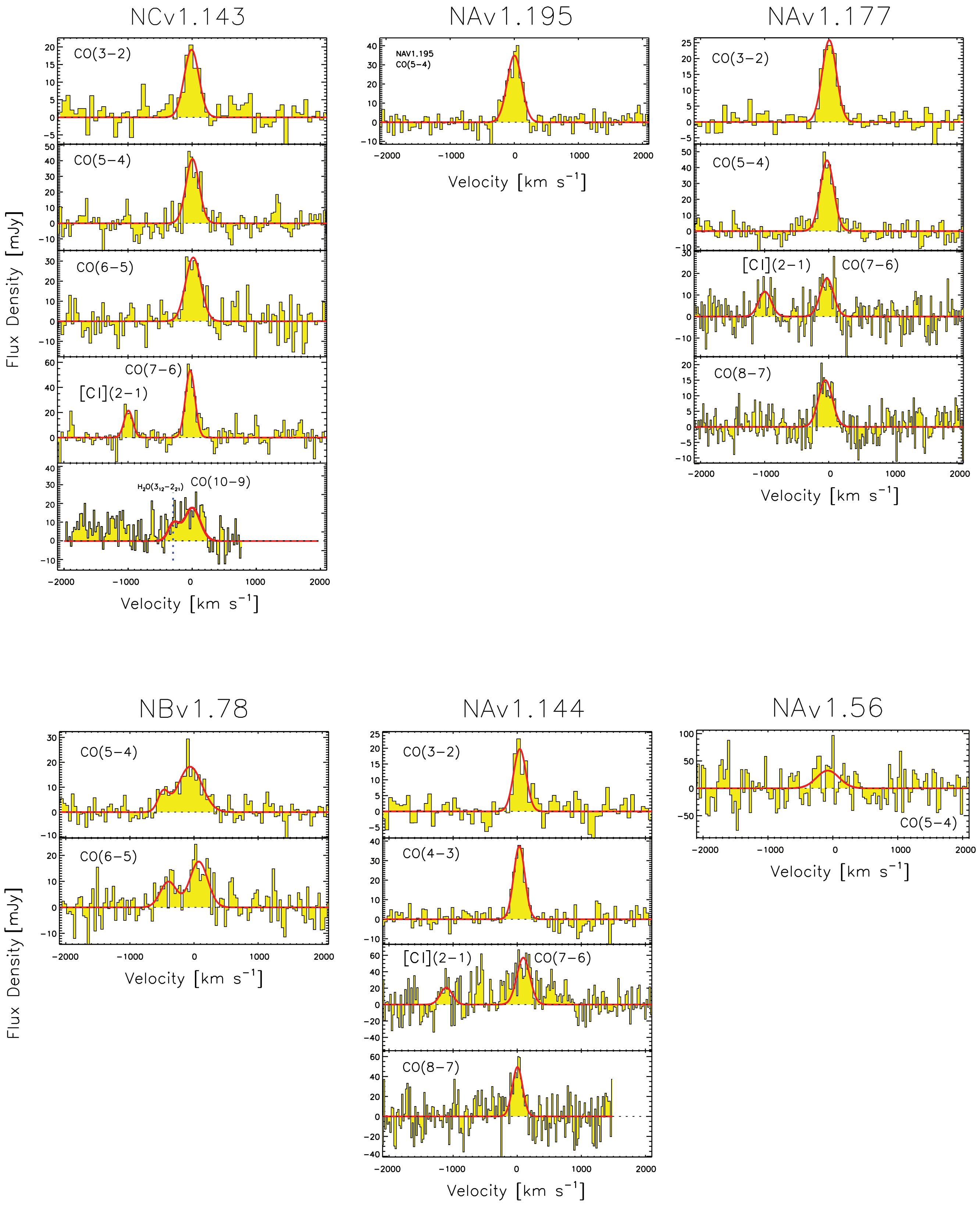

**Fig.** A.1. Continued.

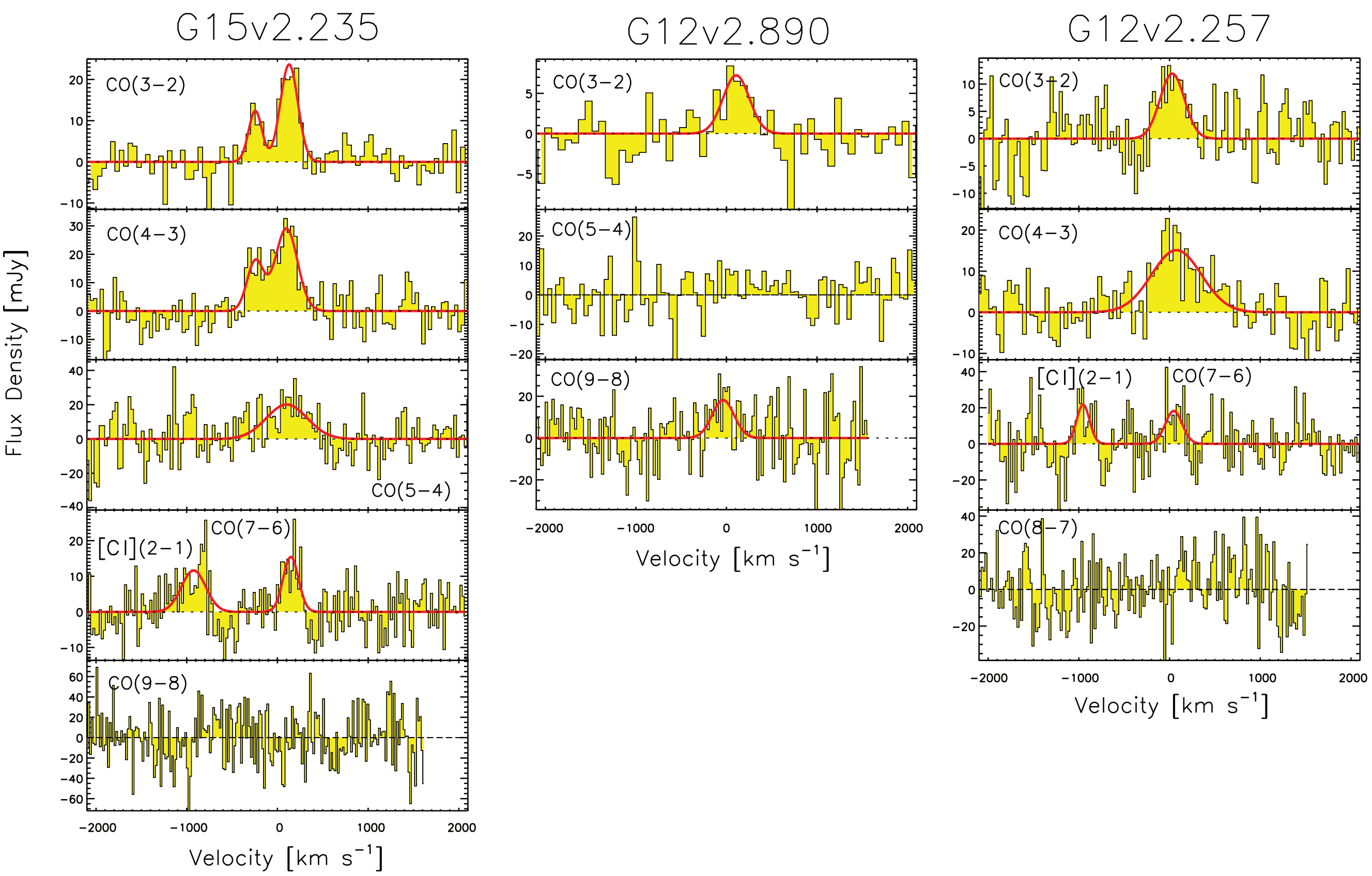

**Fig. A.1.** Continued.

## Appendix B: Tables

**Table B.1:** The observed parameters of the CO and [C I] emission lines.

| Source | Line | $\nu_{sky}$ (GHz) | $t_{on}$ (hour) | Band | $z_{Line}$ | $z_{CO}$ | $D_L$ (Mpc) | $\Delta V$ (km s$^{-1}$) | $I_{Line}$ (Jy km s$^{-1}$) | $\mu L_{IR}$ ($10^{13}$ $L_\odot$) | $L_{IR}$ ($10^{12}$ $L_\odot$) | $\mu L_{Line}$ ($10^8$ $L_\odot$) | $\mu L'_{Line}$ ($10^{10}$ K km s$^{-1}$ pc$^2$) | $L_{Line}$ ($10^7$ $L_\odot$) | $L'_{Line}$ ($10^9$ K km s$^{-1}$ pc$^2$) |
|---|---|---|---|---|---|---|---|---|---|---|---|---|---|---|---|
| G09v1.97 | CO(3–2) | 74.621 | 0.9 | E090 | 3.634 ± 0.002 | 3.6345 ± 0.0001 | 32751 ± 588 | 529 ± 241 | 5.7 ± 2.2 | 15.5 ± 4.3 | 22.5 ± 6.5 | 4.7 ± 1.9 | 35.9 ± 14.2 | 6.9 ± 2.8 | 52.0 ± 21.0 |
| | CO(5–4) | 124.356 | 1.4 | E150 | 3.6346 ± 0.0002 | | | 237 ± 33 | 9.7 ± 1.2 | | | 13.5 ± 1.6 | 22.1 ± 2.7 | 19.6 ± 2.9 | 32.0 ± 4.8 |
| | CO(6–5) | 149.217 | 2.0 | E150 | 3.6333 ± 0.0005 | | | 408 ± 79 | 10.0 ± 1.7 | | | 16.6 ± 2.8 | 15.7 ± 2.6 | 24.1 ± 4.5 | 22.8 ± 4.3 |
| | CO(7–6) | 174.072 | 1.6 | E150 | 3.6340 ± 0.0005 | | | 249 ± 33 | 8.0 ± 0.9 | | | 15.5 ± 1.8 | 9.2 ± 1.1 | 22.4 ± 3.2 | 13.3 ± 1.9 |
| | [C I](2–1) | 174.653 | | E150 | 3.6347 ± 0.0002 | | | 261 ± 79 | 3.6 ± 0.9 | | | 6.9 ± 1.8 | 4.1 ± 1.1 | 10.0 ± 2.8 | 6.0 ± 1.7 |
| G09v1.40 | CO(2–1) | 74.552 | 1.6 | E090 | 2.0919 ± 0.0004 | 2.0924 ± 0.0001 | 16835 ± 283 | 309 ± 93 | 5.3 ± 1.4 | 6.6 ± 2.5 | 4.3 ± 1.9 | 1.2 ± 0.3 | 29.9 ± 7.8 | 0.8 ± 0.3 | 19.6 ± 6.8 |
| | CO(4–3) | 149.108 | 3.6 | E150 | 2.0925 ± 0.0002 | | | 222 ± 35 | 8.8 ± 1.2 | | | 3.9 ± 0.5 | 12.4 ± 1.7 | 2.5 ± 0.7 | 8.1 ± 2.2 |
| | CO(6–5) | 223.611 | 1.2 | E230 | 2.0924 ± 0.0001 | | | 259 ± 33 | 9.2 ± 1.0 | | | 6.0 ± 0.7 | 5.7 ± 0.6 | 3.9 ± 1.0 | 3.7 ± 0.9 |
| | CO(7–6) | 260.858 | 1.6 | E230 | 2.0924 ± 0.0005 | | | 240 | 9.3 ± 1.7 | | | 7.2 ± 1.3 | 4.3 ± 0.8 | 4.7 ± 1.4 | 2.8 ± 0.8 |
| | [C I](2–1) | 261.729 | | E230 | 2.0926 ± 0.0003 | | | 240 | 5.2 ± 1.7 | | | 4.0 ± 1.3 | 2.4 ± 0.8 | 2.6 ± 1.1 | 1.6 ± 0.6 |
| SDP17b | CO(3–2) | 104.628 | 2.8 | E090 | 2.3052 ± 0.0002 | 2.3053 ± 0.0001 | 18942 ± 322 | 269 ± 49 | 6.7 ± 1.1 | 7.1 ± 2.6 | 14.5 ± 5.7 | 2.6 ± 0.4 | 19.8 ± 3.1 | 5.4 ± 1.1 | 40.5 ± 8.6 |
| | CO(4–3) | 139.498 | 2.8 | E150 | 2.3053 ± 0.0002 | | | 271 ± 42 | 9.0 ± 1.2 | | | 4.7 ± 0.6 | 15.0 ± 2.0 | 9.6 ± 1.9 | 30.6 ± 6.0 |
| | CO(7–6) | 244.070 | 1.8 | E230 | – | | | (300) | < 19.0 | | | < 17.6 | < 10.6 | < 35.9 | < 21.8 |
| | [C I](2–1) | 244.885 | 1.8 | E230 | – | | | (300) | < 19.0 | | | < 17.6 | < 10.6 | < 35.9 | < 21.8 |
| | CO(8–7) | 278.911 | 1.8 | E330 | – | | | (300) | < 15.7 | | | < 16.3 | < 6.5 | < 33.3 | < 13.1 |
| SDP81 | CO(3–2) | 85.593 | 2.3 | E090 | 3.0414 ± 0.0006 | 3.0413 ± 0.0005 | 26469 ± 466 | 513 ± 101 | 7.0 ± 1.2 | 5.9 ± 1.5 | 5.3 ± 1.5 | 4.4 ± 0.7 | 33.0 ± 5.6 | 3.9 ± 0.8 | 29.8 ± 5.9 |
| | CO(5–4) | 142.641 | 1.8 | E150 | 3.0393 ± 0.0016 | | | 222 ± 49 | 5.4 ± 2.9 | | | 5.6 ± 3.0 | 9.2 ± 4.9 | 5.1 ± 2.7 | 8.3 ± 4.5 |
| | | 142.641 | | | 3.0428 ± 0.0003 | | | 384 ± 199 | 4.6 ± 3.0 | | | 4.8 ± 3.1 | 7.8 ± 5.1 | 4.3 ± 2.9 | 7.1 ± 4.7 |
| | CO(6–5) | 171.157 | 2.3 | E150 | – | | | (500) | < 26.2 | | | < 32.2 | < 30.4 | < 28.9 | < 27.4 |
| | CO(10-9) | 285.145 | 1.8 | E330 | – | | | (500) | < 8.8 | | | < 18.6 | < 3.8 | < 16.8 | < 3.4 |
| G12v2.43 | CO(3–2) | 83.789 | 1.8 | E090 | 3.1274 ± 0.0002 | 3.1271 ± 0.0001 | 27367 ± 484 | 247 ± 35 | 4.4 ± 0.5 | 9.0 ± 0.2 | – | 2.9 ± 0.3 | 21.6 ± 2.6 | – | – |
| | CO(4–3) | 111.713 | 3.8 | E090 | 3.1270 ± 0.0001 | | | 240 | 5.6 ± 1.2 | | | 4.8 ± 1.1 | 15.5 ± 3.2 | – | – |
| | CO(5–4) | 139.634 | 3.8 | E150 | 3.1273 ± 0.0002 | | | 232 ± 34 | 7.1 ± 0.9 | | | 7.7 ± 1.0 | 12.6 ± 1.6 | – | – |
| | CO(6–5) | 167.549 | 1.6 | E150 | 3.1266 ± 0.0003 | | | 334 ± 47 | 8.7 ± 1.1 | | | 11.4 ± 1.4 | 10.7 ± 1.3 | – | – |
| | CO(8–7) | 223.358 | 1.8 | E230 | 3.1274 ± 0.0002 | | | 212 ± 33 | 6.7 ± 0.9 | | | 11.7 ± 1.6 | 4.7 ± 0.6 | – | – |
| | CO(10-9) | 279.134 | 1.6 | E330 | – | | | (300) | < 10.0 | | | < 21.5 | < 4.2 | – | – |
| G12v2.30 | CO(4–3) | 108.251 | 1.4 | E090 | 3.2599 ± 0.0009 | 3.2596 ± 0.0002 | 28761 ± 511 | 801 ± 151 | 16.4 ± 2.7 | 15.6 ± 4.1 | 16.4 ± 4.4 | 15.2 ± 2.5 | 48.6 ± 7.9 | 16.0 ± 2.8 | 51.1 ± 8.9 |
| | CO(5–4) | 135.306 | 1.4 | E150 | 3.2591 ± 0.0004 | | | 621 ± 64 | 22.6 ± 2.0 | | | 26.3 ± 2.3 | 42.9 ± 3.8 | 27.4 ± 3.0 | 45.2 ± 4.9 |
| | CO(6–5) | 162.356 | 1.8 | E150 | 3.2593 ± 0.0003 | | | 704 ± 43 | 19.8 ± 1.1 | | | 27.6 ± 1.5 | 26.1 ± 1.4 | 29.1 ± 2.4 | 27.5 ± 2.3 |
| | CO(8–7) | 216.436 | 1.8 | E230 | 3.2606 ± 0.0004 | | | 655 ± 67 | 20.7 ± 1.8 | | | 38.6 ± 3.3 | 15.4 ± 1.4 | 40.6 ± 4.4 | 16.2 ± 1.8 |
| | CO(11-10) | 297.491 | 1.8 | E330 | 3.260 ± 0.001 | | | 848 ± 199 | 12.1 ± 2.5 | | | 31.0 ± 6.0 | 4.8 ± 1.0 | 32.7 ± 6.9 | 5.0 ± 1.1 |
| NCv1.143 | CO(3–2) | 75.749 | 2.1 | E090 | 3.5649 ± 0.0002 | 3.5650 ± 0.0004 | 32007 ± 574 | 270 ± 36 | 5.5 ± 0.6 | 12.9 ± 4.0 | 11.4 ± 3.9 | 4.5 ± 0.5 | 33.7 ± 3.9 | 3.9 ± 0.7 | 29.9 ± 5.7 |
| | CO(5–4) | 126.236 | 1.2 | E150 | 3.5651 ± 0.0002 | | | 243 ± 24 | 10.7 ± 0.9 | | | 14.3 ± 1.2 | 23.4 ± 2.0 | 12.7 ± 2.2 | 20.7 ± 3.6 |
| | CO(6–5) | 151.473 | 1.4 | E150 | 3.5653 ± 0.0003 | | | 283 ± 41 | 9.6 ± 1.2 | | | 15.5 ± 1.9 | 14.6 ± 1.8 | 13.7 ± 2.7 | 13.0 ± 2.5 |
| | CO(7–6) | 176.704 | 5.7 | E150 | 3.5654 ± 0.0005 | | | 250 | 11.5 ± 0.9 | | | 21.7 ± 1.7 | 12.9 ± 1.0 | 19.2 ± 3.3 | 11.4 ± 1.9 |
| | [C I](2–1) | 177.293 | | E150 | 3.5646 ± 0.0002 | | | 250 | 4.4 ± 0.9 | | | 8.2 ± 1.7 | 4.9 ± 1.0 | 7.3 ± 1.9 | 4.2 ± 1.1 |
| | CO(10-9) | 252.352 | 4.3 | *NOEMA* | – | | | 288 ± 65 | 5.4 ± 1.2 | | | 14.6 ± 2.9 | 3.0 ± 0.6 | 12.8 ± 5.1 | 2.6 ± 1.0 |
| NAv1.195 | CO(5–4) | 145.854 | 4.0 | E150 | 2.9510 ± 0.0001 | 2.9510 ± 0.0001 | 25528 ± 448 | 266 ± 19 | 9.9 ± 0.6 | 7.5 ± 2.0 | 18.3 ± 5.1 | 9.8 ± 0.6 | 16.0 ± 1.0 | 23.8 ± 2.3 | 38.9 ± 3.7 |
| NAv1.177 | CO(3–2) | 91.529 | 1.4 | E090 | 2.7781 ± 0.0001 | 2.7778 ± 0.0001 | 23736 ± 414 | 262 ± 26 | 7.2 ± 0.6 | 6.2 ± 0.2 | – | 3.9 ± 0.3 | 29.1 ± 2.5 | – | – |
| | CO(5–4) | 152.533 | 1.4 | E150 | 2.7777 ± 0.0001 | | | 248 ± 19 | 11.8 ± 0.8 | | | 10.5 ± 0.7 | 17.2 ± 1.2 | – | – |
| | CO(7–6) | 213.513 | 2.9 | E230 | 2.7780 ± 0.0004 | | | 244 ± 51 | 4.6 ± 0.8 | | | 5.8 ± 1.1 | 3.4 ± 0.6 | – | – |
| | [C I](2–1) | 214.225 | 2.9 | E230 | 2.7776 ± 0.0003 | | | 233 ± 77 | 2.9 ± 0.8 | | | 3.6 ± 1.0 | 2.1 ± 0.6 | – | – |
| | CO(8–7) | 243.991 | 3.7 | E230 | 2.7773 ± 0.0002 | | | 255 ± 34 | 4.1 ± 0.5 | | | 5.8 ± 0.7 | 2.3 ± 0.3 | – | – |
| NBv1.78 | CO(5–4) | 140.170 | 2.5 | E150 | 3.1102 ± 0.0003 | 3.1080 ± 0.0003 | 27167 ± 480 | 196 ± 89 | 1.6 ± 0.8 | 10.8 ± 3.9 | 8.4 ± 3.1 | 1.7 ± 0.8 | 2.8 ± 1.3 | 1.3 ± 0.6 | 2.1 ± 1.1 |
| | | 140.170 | | | 3.1044 ± 0.0005 | | | 434 ± 63 | 8.4 ± 1.0 | | | 9.1 ± 1.1 | 14.8 ± 1.7 | 7.0 ± 1.1 | 11.4 ± 1.9 |
| | CO(6–5) | 168.193 | 5.4 | E150 | 3.1055 ± 0.0007 | | | 328 ± 78 | 3.1 ± 1.2 | | | 7.9 ± 1.6 | 7.5 ± 1.4 | 6.1 ± 1.4 | 5.8 ± 1.3 |
| | | 168.193 | | | 3.1120 ± 0.0004 | | | 297 ± 131 | 6.1 ± 1.2 | | | 4.0 ± 1.5 | 3.8 ± 1.4 | 3.1 ± 1.2 | 2.9 ± 1.2 |
| NAv1.144 | CO(3–2) | 107.994 | 2.2 | E090 | 2.2024 ± 0.0002 | 2.2023 ± 0.0001 | 17918 ± 303 | 230 ± 34 | 4.8 ± 0.6 | 6.0 ± 3.5 | 13.0 ± 8.3 | 1.7 ± 0.2 | 13.2 ± 1.7 | 4.0 ± 0.9 | 30.0 ± 6.7 |
| | CO(4–3) | 143.985 | 2.2 | E150 | 2.2023 ± 0.0001 | | | 205 ± 19 | 8.1 ± 0.6 | | | 3.9 ± 0.3 | 12.5 ± 1.0 | 8.9 ± 1.8 | 28.4 ± 5.6 |
| | CO(7–6) | 251.921 | 1.3 | E230 | 2.2008 ± 0.0005 | | | 205 | 14.6 ± 1.6 | | | 12.3 ± 1.4 | 7.3 ± 0.8 | 28.0 ± 6.0 | 16.7 ± 3.5 |

**Table B.1:** continued.

| Source | Line | $\nu_{\rm sky}$ (GHz) | $t_{\rm on}$ (hour) | Band | $z_{\rm Line}$ | $z_{\rm CO}$ | $D_{\rm L}$ (Mpc) | $\Delta V$ (km s$^{-1}$) | $I_{\rm Line}$ (Jy km s$^{-1}$) | $\mu L_{\rm IR}$ ($10^{13}$ $L_\odot$) | $L_{\rm IR}$ ($10^{12}$ $L_\odot$) | $\mu L_{\rm Line}$ ($10^8$ $L_\odot$) | $\mu L'_{\rm Line}$ ($10^{10}$ K km s$^{-1}$ pc$^2$) | $L_{\rm Line}$ ($10^7$ $L_\odot$) | $L'_{\rm Line}$ ($10^9$ K km s$^{-1}$ pc$^2$) |
|---|---|---|---|---|---|---|---|---|---|---|---|---|---|---|---|
| | [C I](2–1) | 252.762 | 1.3 | E230 | 2.2030 ± 0.0002 | | | 205 | 4.8 ± 1.6 | | | 4.1 ± 1.4 | 2.4 ± 0.8 | 9.3 ± 3.5 | 5.5 ± 2.1 |
| | CO(8–7) | 287.882 | 3.8 | E330 | 2.2020 ± 0.0002 | | | 205 | 11.8 ± 1.9 | | | 10.4 ± 1.7 | 4.1 ± 0.7 | 23.6 ± 5.8 | 9.4 ± 2.3 |
| NAv1.56 | CO(5–4) | 174.574 | 3.7 | E150 | 2.3001 ± 0.0009 | 2.3001 ± 0.0009 | 18890 ± 321 | 409 ± 201 | 14.1 ± 6.0 | 11.5 ± 3.1 | 9.8 ± 2.8 | 9.1 ± 3.9 | 14.9 ± 6.4 | 7.8 ± 3.4 | 12.7 ± 5.5 |
| G15v2.235 | CO(3–2) | 99.418 | 1.9 | E090 | 2.4797 ± 0.0001 | 2.4789 ± 0.0001 | 20686 ± 355 | 176 ± 47 | 2.3 ± 0.6 | 2.8 ± 0.7 | 15.6 ± 4.7 | 1.0 ± 0.2 | 7.7 ± 1.8 | 5.7 ± 1.6 | 42.9 ± 12.2 |
| | | 99.418 | | | 2.4754 ± 0.0002 | | | 229 ± 28 | 5.8 ± 0.6 | | | 2.5 ± 0.3 | 19.3 ± 2.0 | 14.2 ± 2.8 | 107.0 ± 21.1 |
| | CO(4–3) | 132.552 | 1.3 | E150 | 2.4793 ± 0.0003 | | | 216 ± 73 | 4.0 ± 1.3 | | | 2.4 ± 0.8 | 7.6 ± 2.4 | 13.2 ± 4.8 | 42.2 ± 15.3 |
| | | 132.552 | | | 2.4754 ± 0.0004 | | | 289 ± 55 | 8.9 ± 1.4 | | | 5.3 ± 0.8 | 16.8 ± 2.7 | 29.3 ± 6.8 | 93.3 ± 21.6 |
| | CO(5–4) | 165.680 | 0.7 | E150 | 2.4794 ± 0.0006 | | | 531 ± 131 | 11.3 ± 2.4 | | | 8.4 ± 1.8 | 13.7 ± 2.9 | 46.5 ± 12.6 | 75.9 ± 20.6 |
| | CO(7–6) | 231.916 | 1.8 | E230 | 2.4791 ± 0.0004 | | | 201 ± 45 | 3.3 ± 0.6 | | | 3.4 ± 0.7 | 2.0 ± 0.4 | 18.9 ± 4.9 | 11.2 ± 2.9 |
| | [C I](2–1) | 232.690 | 1.8 | E230 | 2.4799 ± 0.0002 | | | 316 ± 76 | 3.9 ± 0.8 | | | 4.1 ± 0.8 | 2.4 ± 0.5 | 22.5 ± 6.0 | 13.4 ± 3.6 |
| | CO(9–8) | 298.118 | 1.6 | E330 | – | | | (300) | < 24.1 | | | < 32.0 | < 9.0 | < 177.9 | < 50.0 |
| G12v2.890 | CO(3–2) | 96.653 | 1.7 | E090 | 2.5790 ± 0.0005 | 2.5783 ± 0.0003 | 21694 ± 375 | 344 ± 92 | 2.7 ± 0.6 | 2.5 ± 0.3 | – | 1.3 ± 0.3 | 9.5 ± 2.2 | – | – |
| | CO(5–4) | 161.072 | 1.7 | E150 | – | | | (300) | < 7.4 | | | < 5.7 | < 9.6 | – | – |
| | CO(9–8) | 289.827 | 1.8 | E330 | 2.5773 ± 0.0006 | | | 286 ± 109 | 5.5 ± 1.8 | | | 7.8 ± 2.6 | 2.2 ± 0.7 | – | – |
| G12v2.257 | CO(3–2) | 108.369 | 1.4 | E090 | 2.1912 ± 0.0004 | 2.1914 ± 0.0001 | 17810 ± 301 | 318 ± 82 | 4.0 ± 0.9 | 2.6 ± 0.3 | – | 1.4 ± 0.3 | 10.9 ± 2.4 | – | – |
| | CO(4–3) | 144.486 | 2.5 | E150 | 2.1917 ± 0.0005 | | | 648 ± 106 | 10.4 ± 1.5 | | | 5.0 ± 0.7 | 15.8 ± 2.2 | – | – |
| | CO(7–6) | 252.798 | 1.2 | E230 | 2.1914 ± 0.0003 | | | 220 ± 95 | 4.3 ± 1.6 | | | 3.6 ± 1.3 | 2.1 ± 0.8 | – | – |
| | [C I](2–1) | 253.641 | | E230 | 2.1914 ± 0.0004 | | | 168 ± 70 | 3.9 ± 1.4 | | | 3.2 ± 1.2 | 1.9 ± 0.7 | – | – |
| | CO(8–7) | 288.884 | 1.5 | E330 | – | | | (300) | < 15.4 | | | < 14.6 | < 5.8 | – | – |

**Notes.** $\nu_{\rm sky}$ is the observed frequency at zero velocity in each spectrum; $t_{\rm on}$ is the on source time spent during the observation; E090, E150, E230 and E330 EMIR bands were used, and the total time used during the observations is less than the sum of all the $t_{\rm on}$ because some of the lines were observed simultaneously utilising different EMIR receiver band combinations; $z_{\rm line}$ is the redshift derived according to the detected lines; $\Delta V$ is the linewidth; $z_{\rm CO}$ is the error-weighted average CO line redshift for each source; $\Delta V$ with no errors are fixed values used for more accurately inferring the integrated flux density of each line and the $\Delta V$ in brackets are the one used for inferring the upper limits for non-detections; $I_{\rm line}$ is the velocity-integrated flux density.

## Appendix C: Bayesian approach of the radiative transfer modelling.

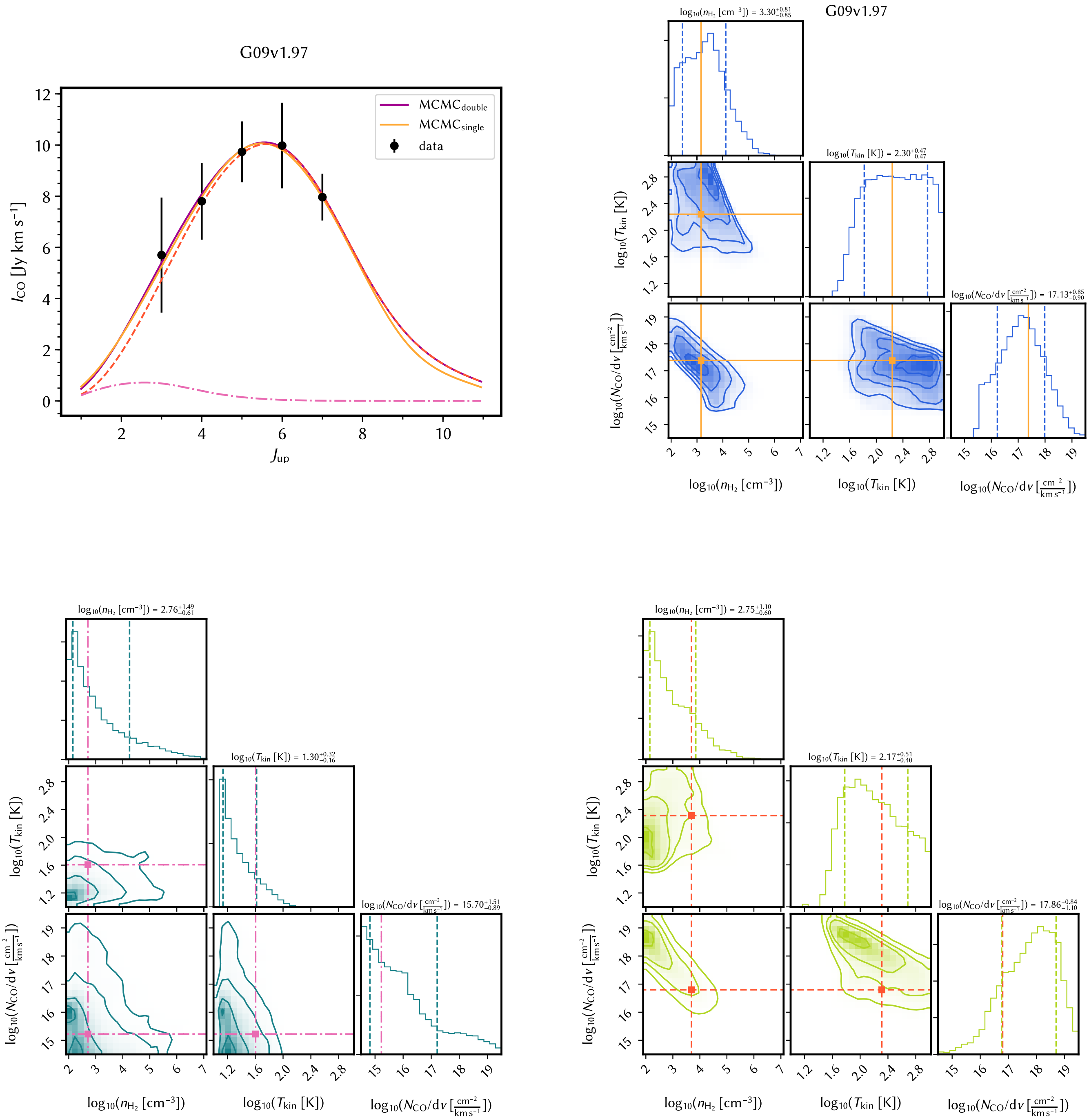

**Fig. C.1:** *Upper left panel*: The CO SLED (without correcting for lensing magnification) of each source is plotted in black. The flux of CO(1–0) has been corrected for differential lensing effect as discussed in the text. The solid orange curve shows the best fit from the single component model corresponding to the maximum posterior possibility, while the solid purple line shows the best fit of the two-component model. Dashed red line shows the warmer component and the dashed-dotted line shows the cooler component fit. The upper limits are shown in grey open circles with downward arrows. *Upper right panel*: The posterior probability distributions of molecular gas density $n_{\mathrm{H_2}}$, gas temperature $T_{\mathrm{kin}}$ and CO column density per velocity $N_{\mathrm{CO}}/\mathrm{d}v$ of the source, with the maximum posterior possibility point in the parameter space shown in orange lines and points. The contours are in steps of $0.5\,\sigma$ starting from the centre. *Lower panels*: The posterior probability distributions for $n_{\mathrm{H_2}}$, $T_{\mathrm{kin}}$ and $N_{\mathrm{CO}}/\mathrm{d}v$ of the cooler (dark green) and warmer (light green) component of the two-component model.

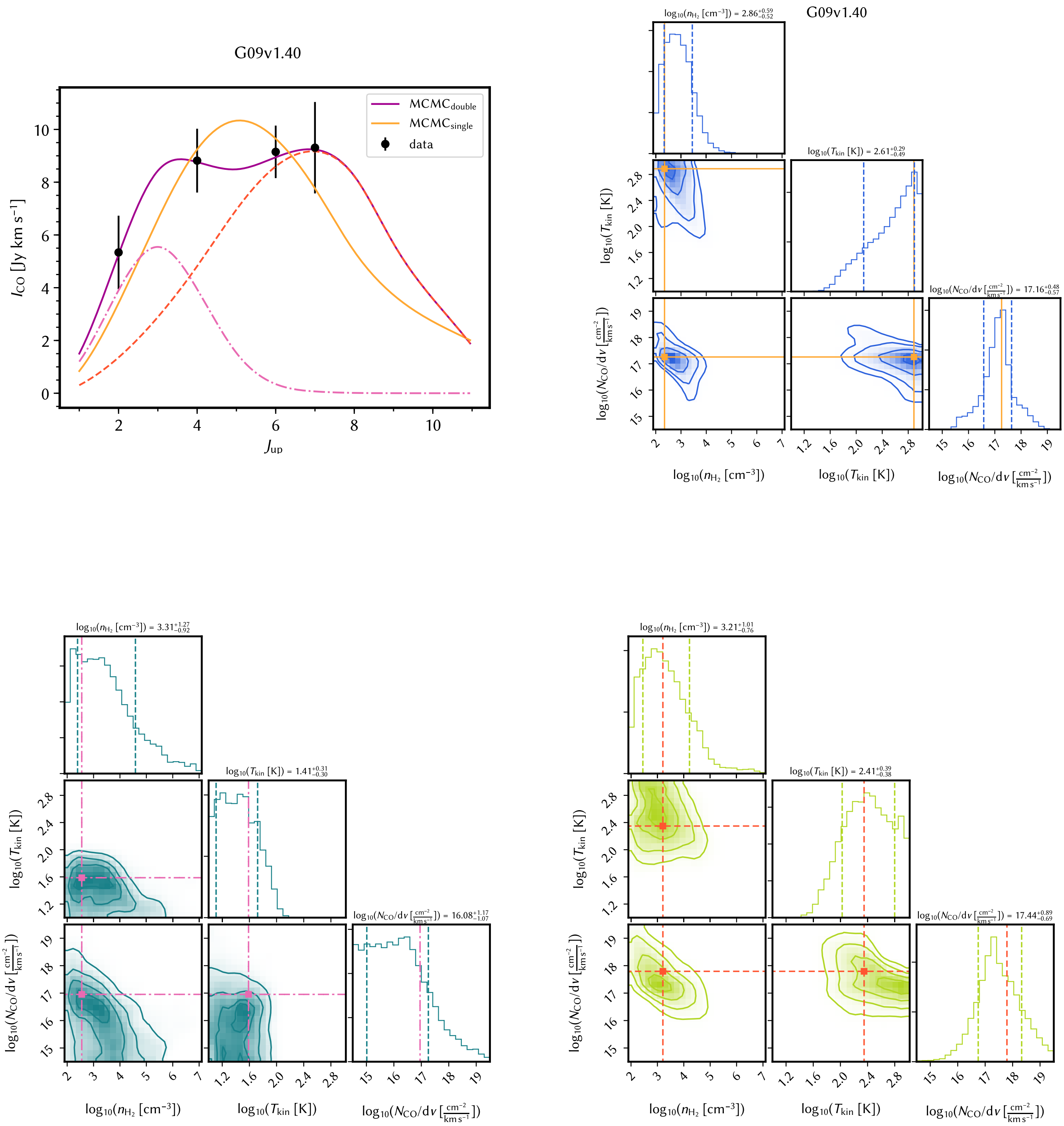

**Fig. C.1.** Continued.

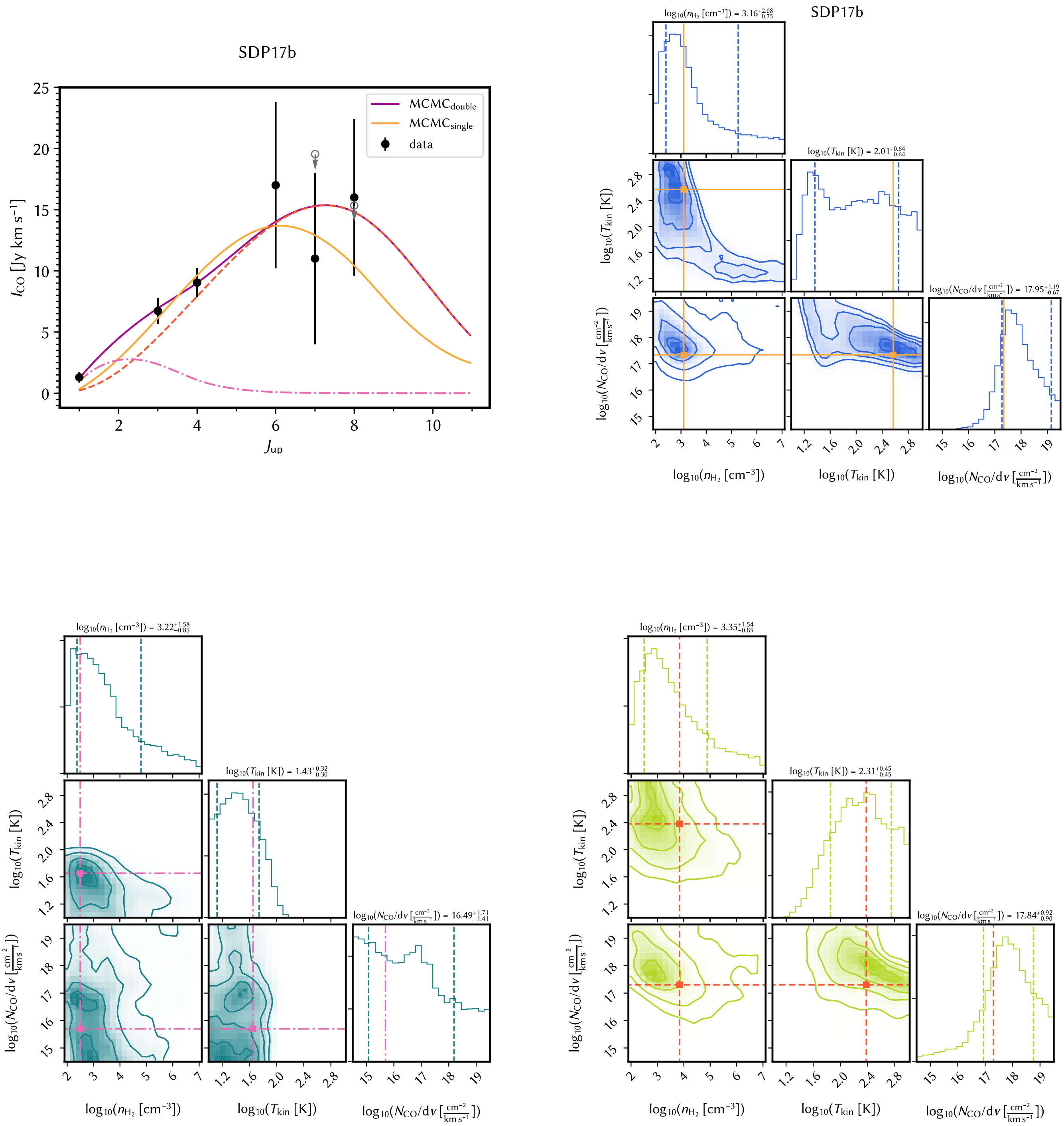

**Fig. C.1.** Continued.

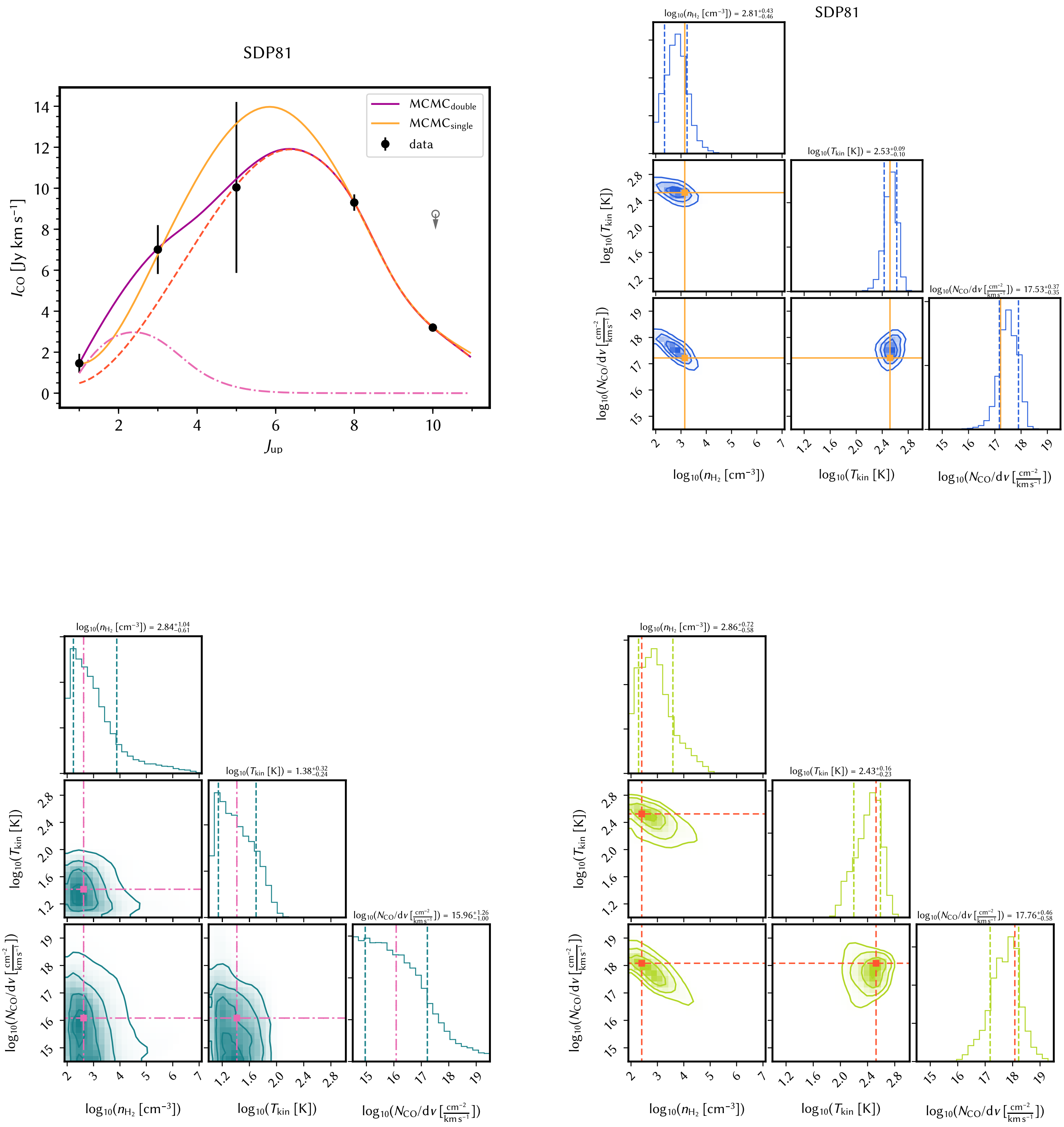

**Fig. C.1.** Continued.

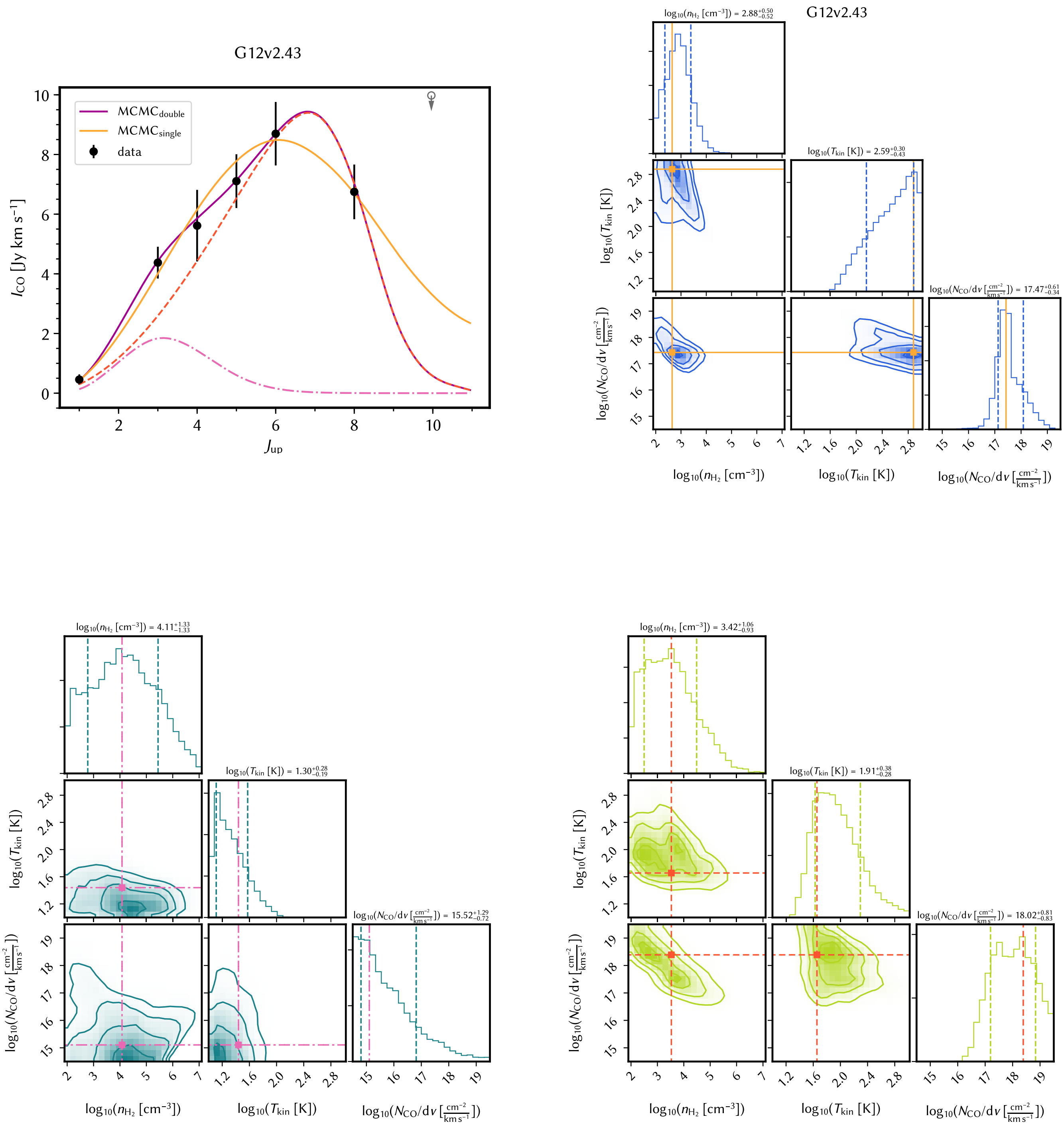

**Fig.** C.1. Continued.

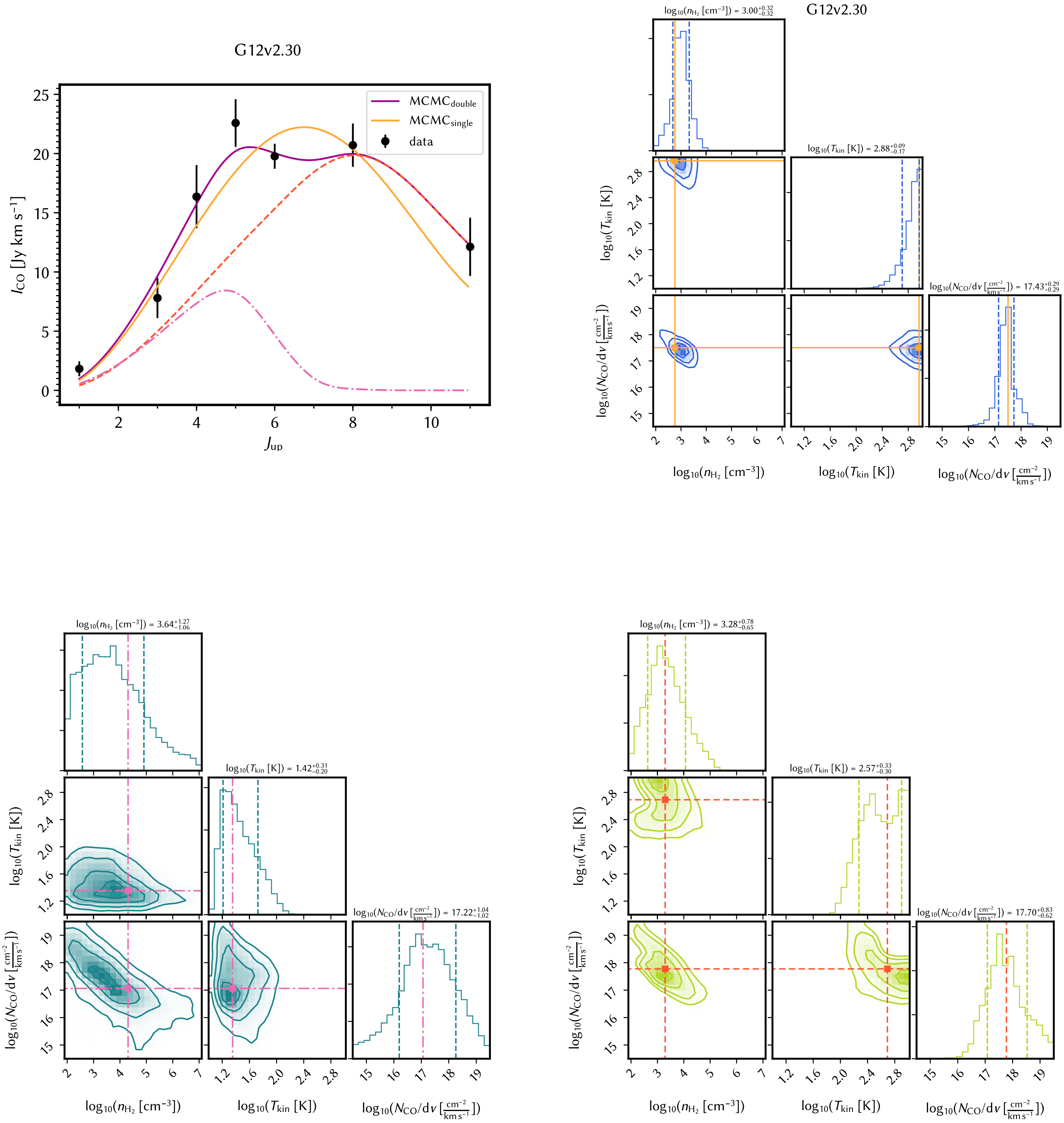

**Fig. C.1.** Continued.

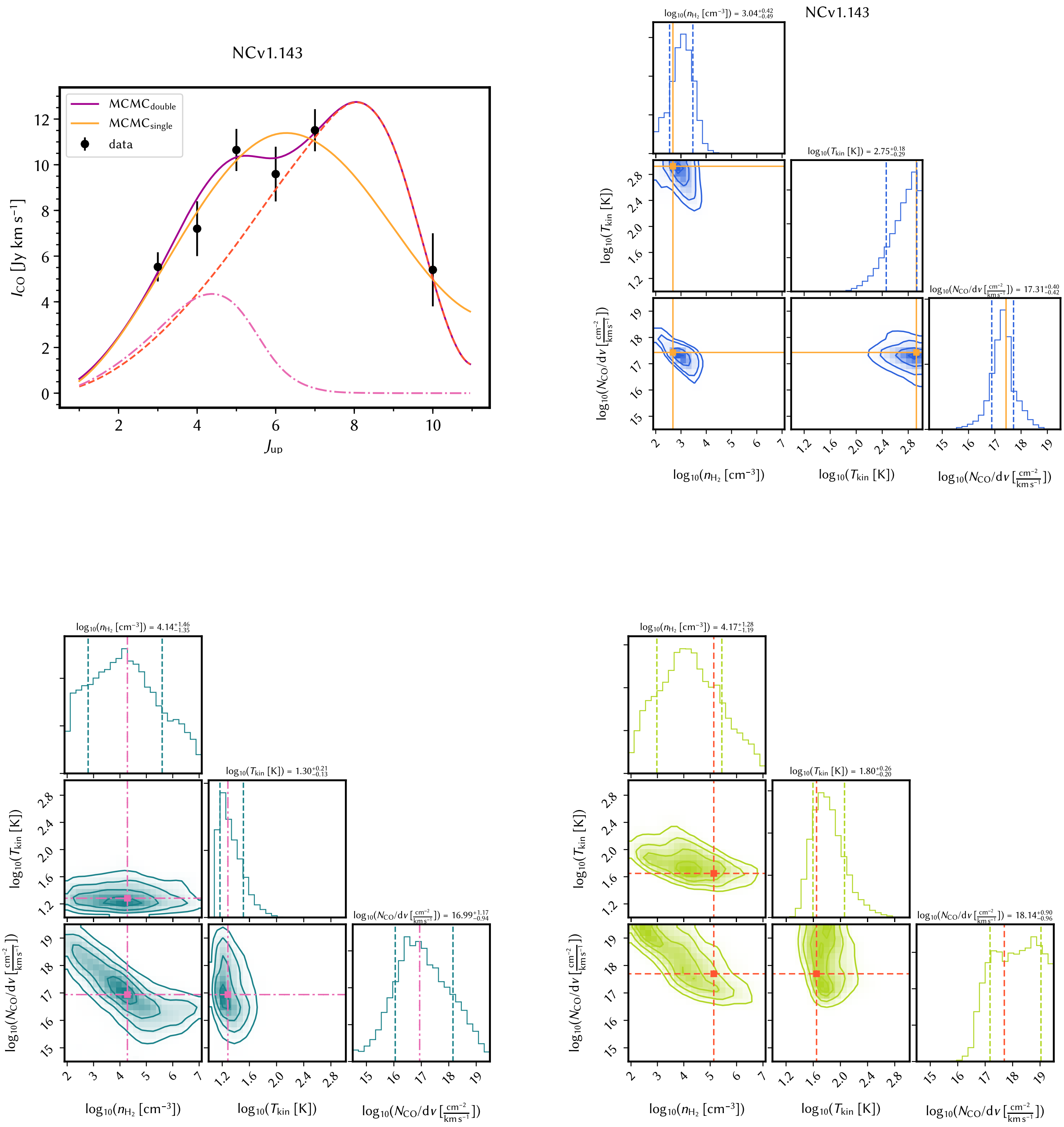

**Fig.** C.1. Continued.

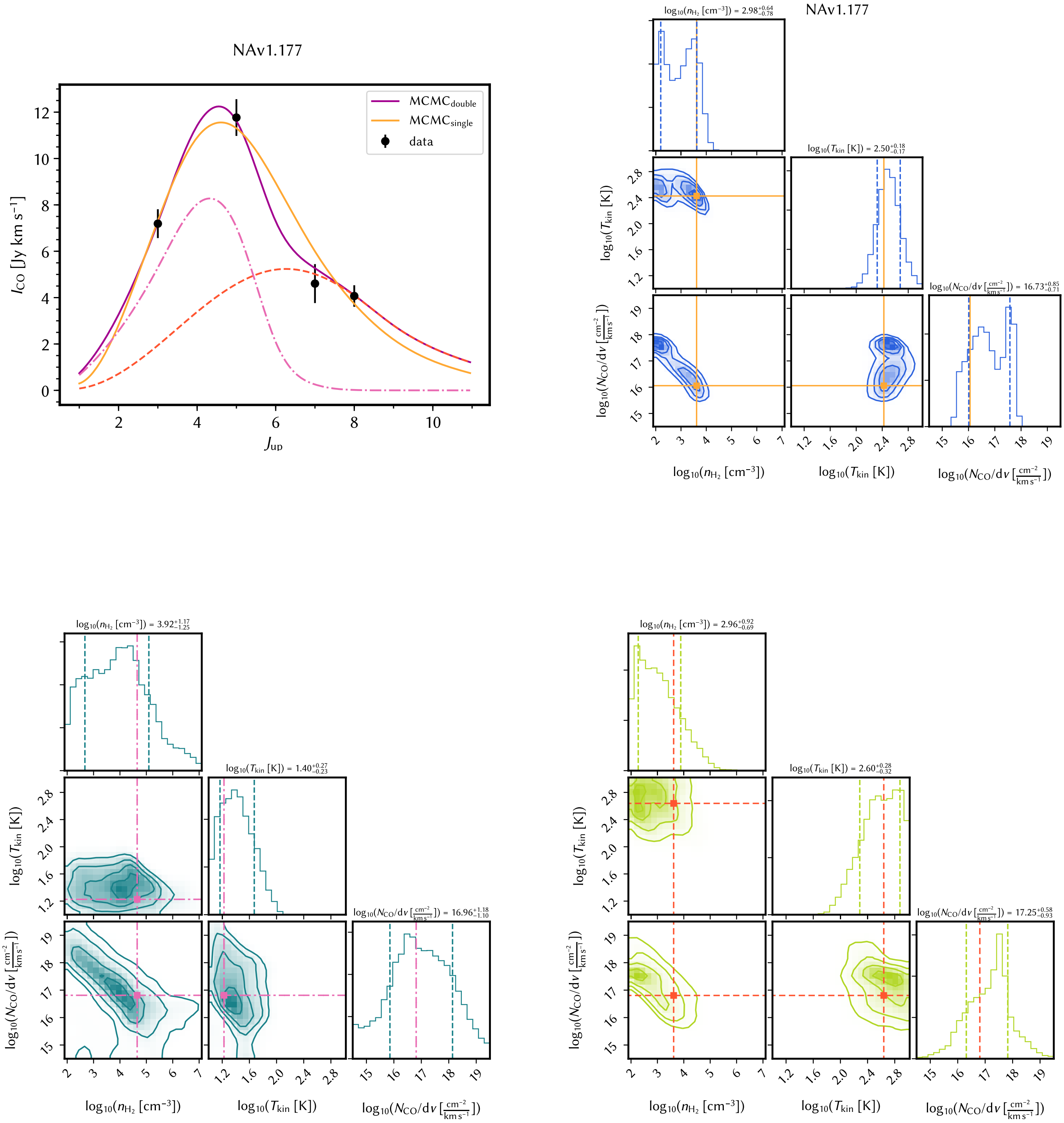

**Fig. C.1.** Continued.

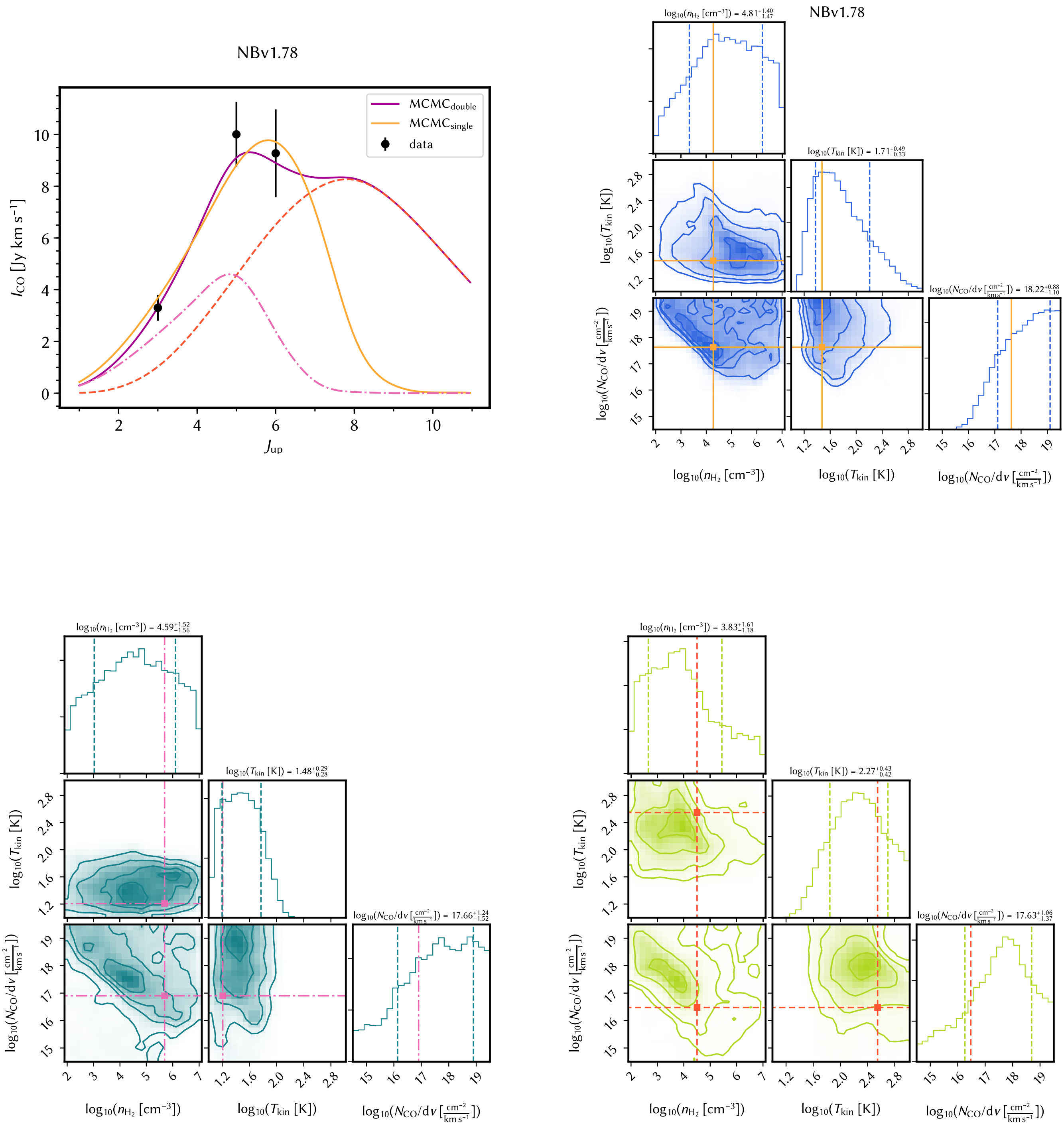

**Fig. C.1.** Continued.

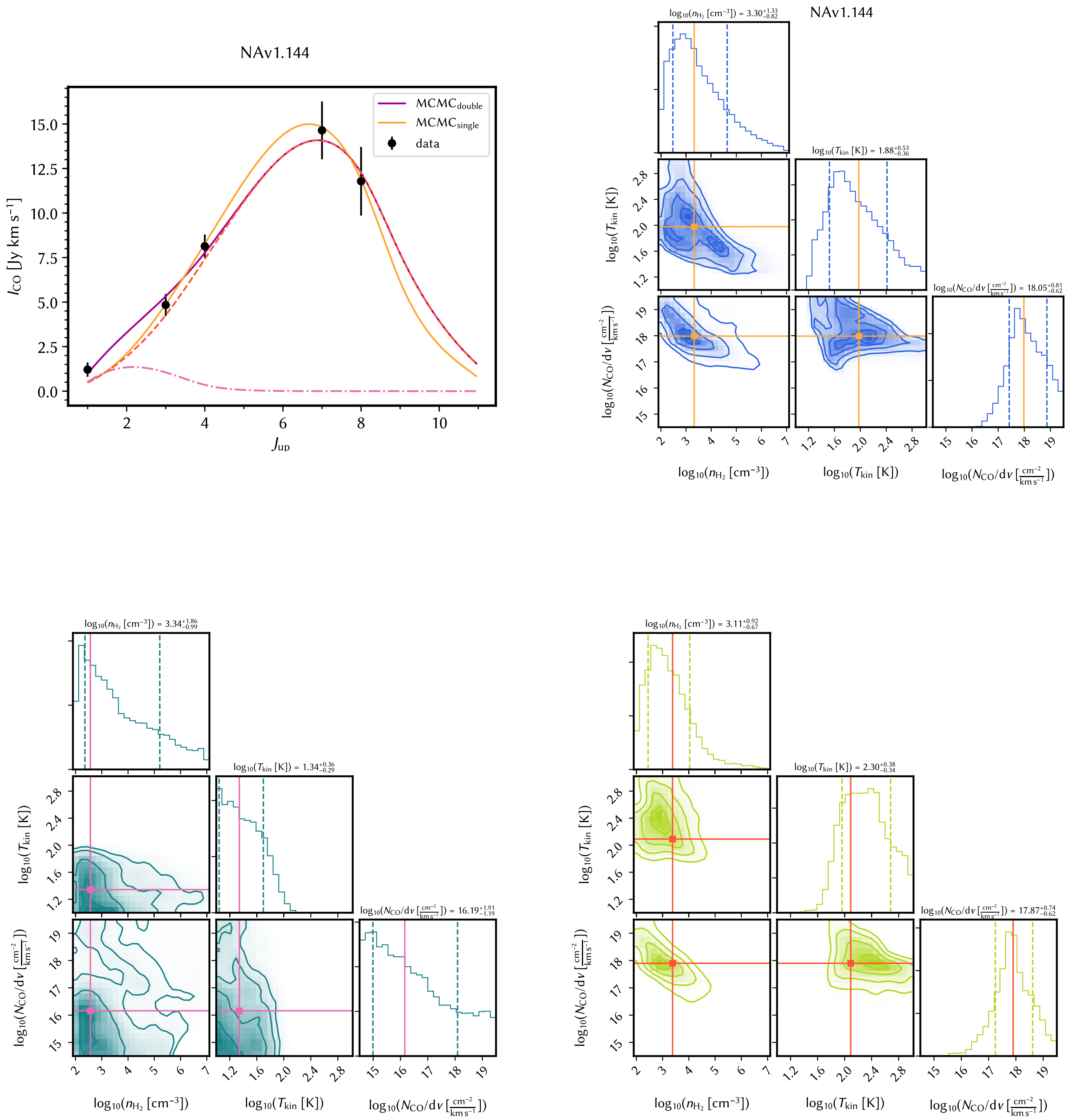

**Fig. C.1.** Continued.

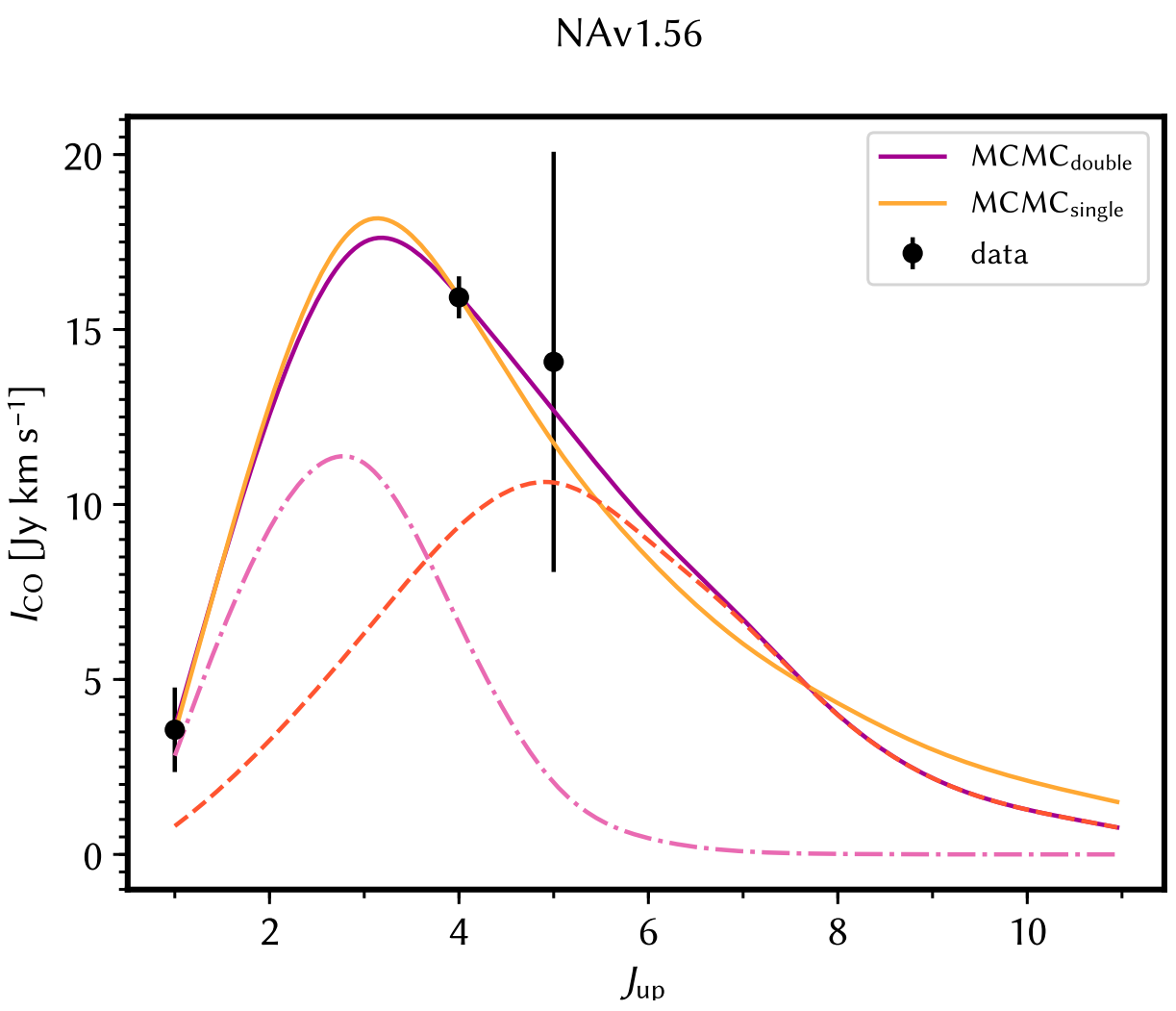

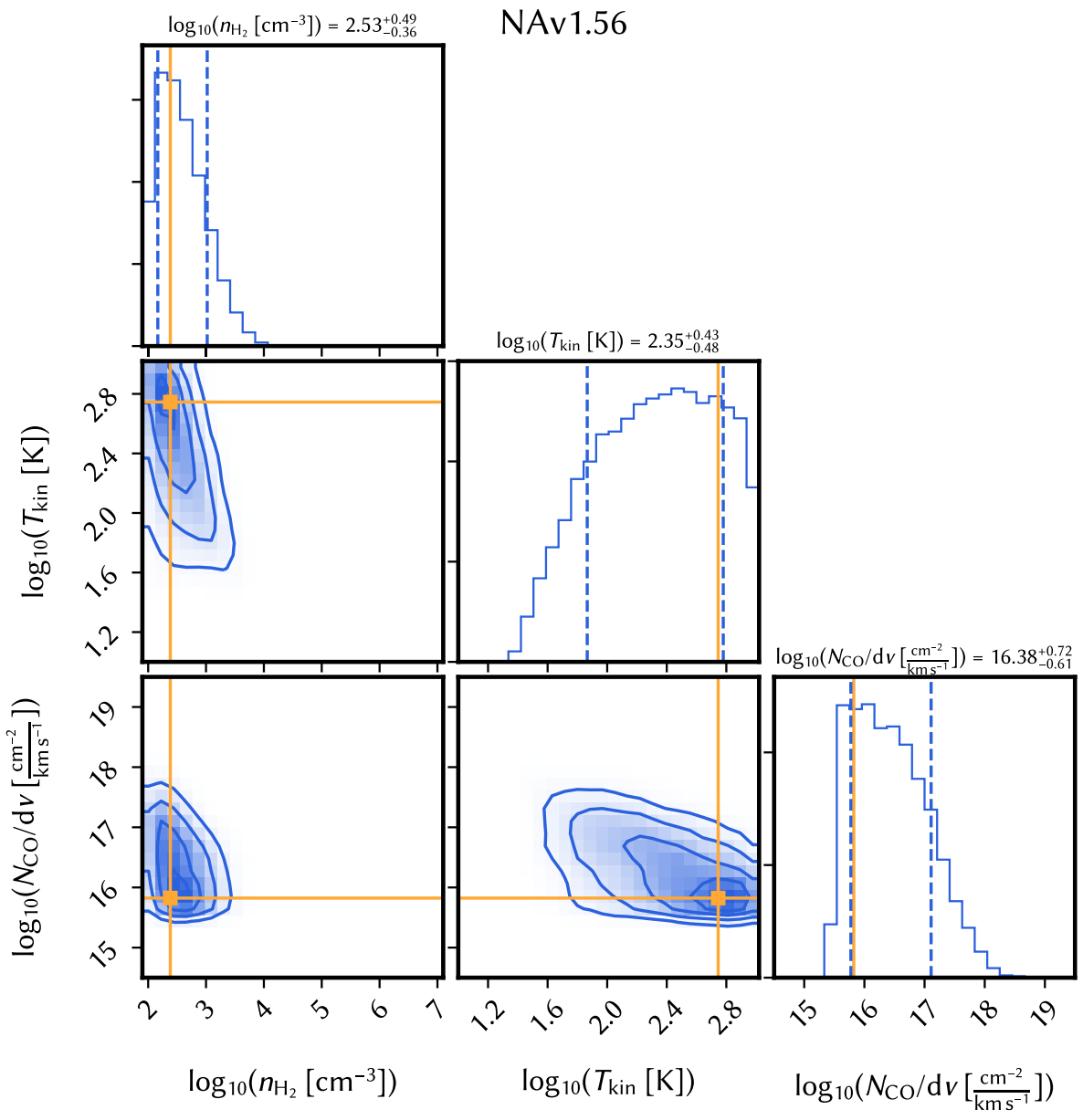

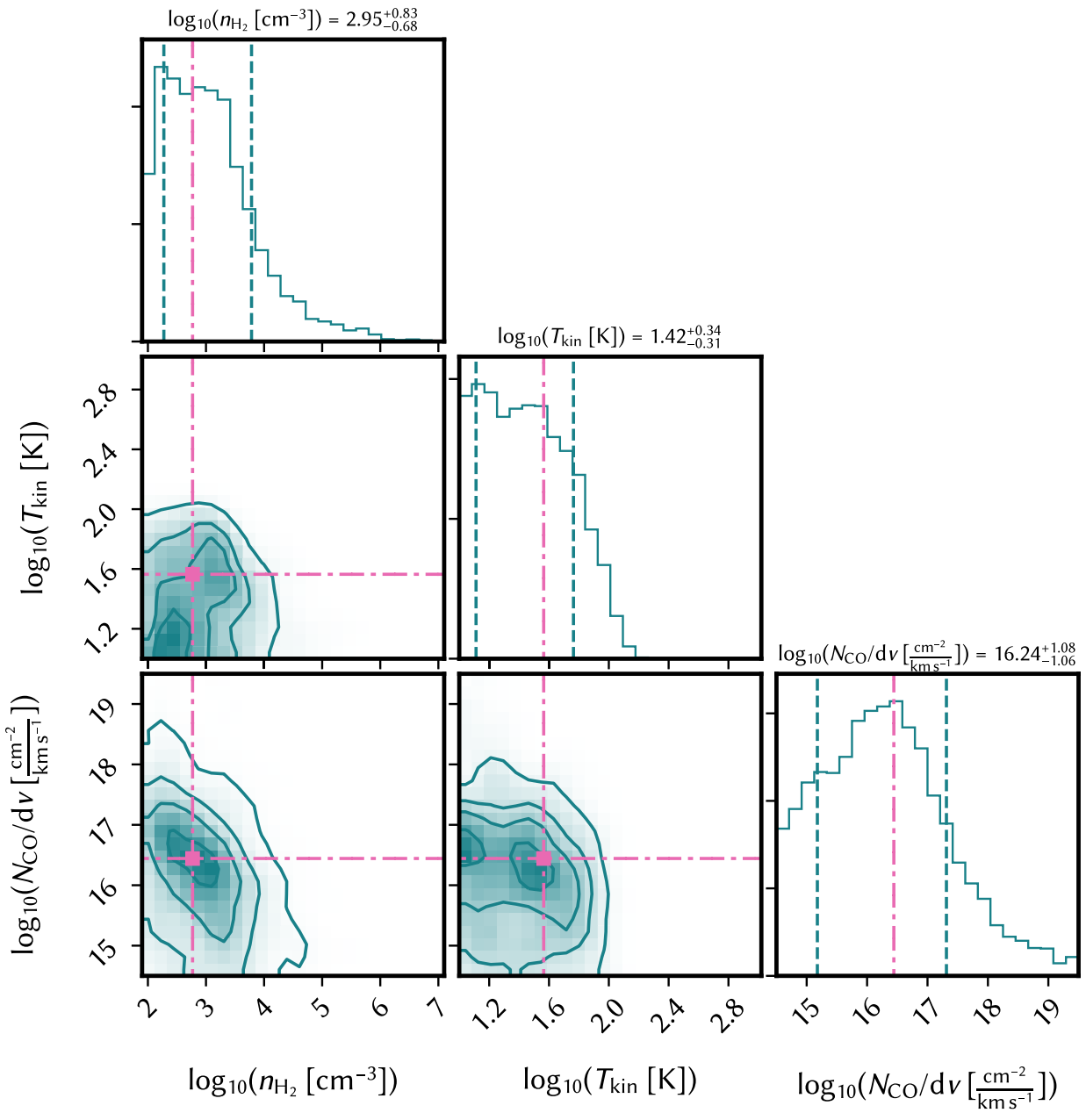

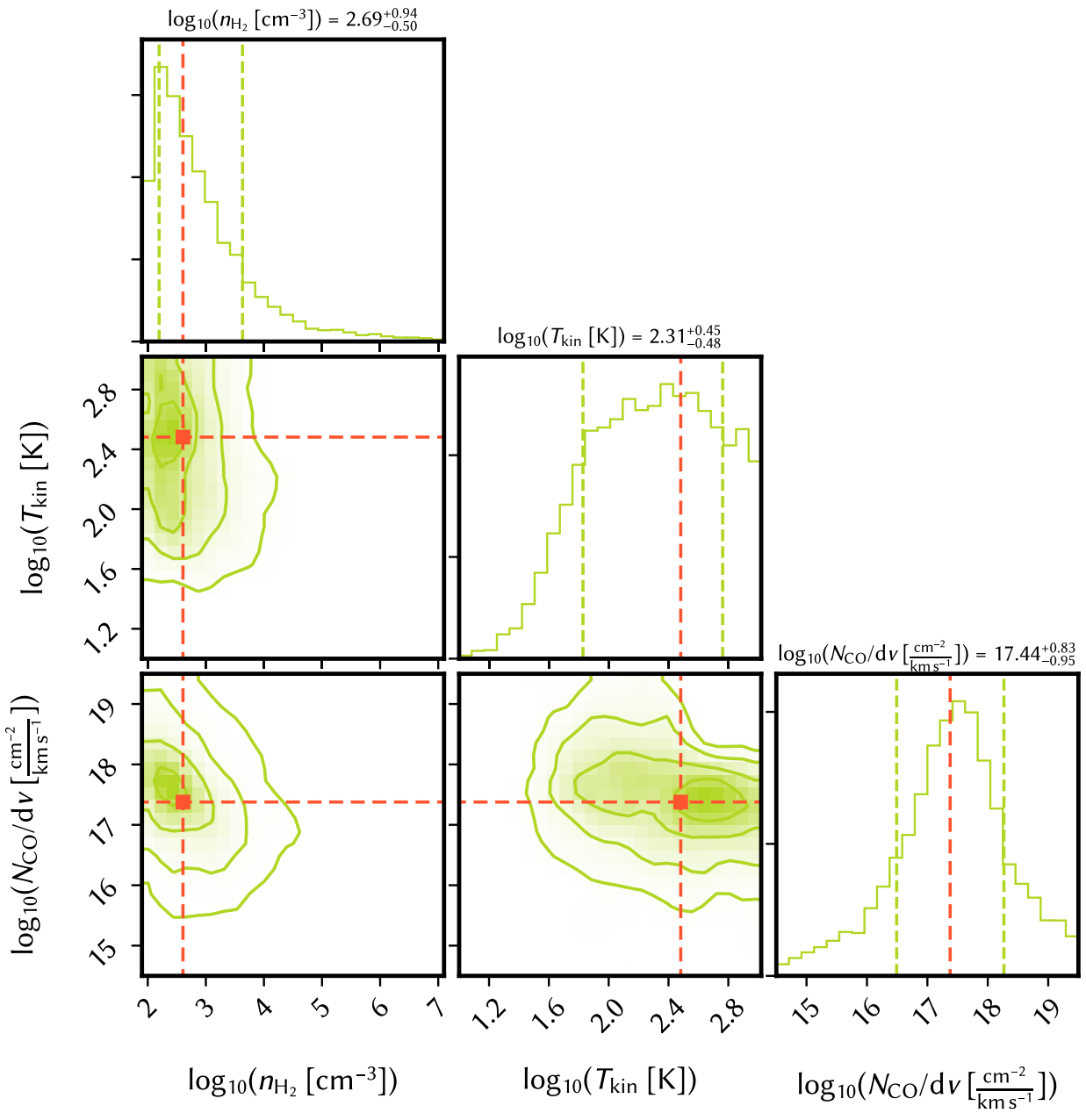

**Fig. C.1.** Continued.

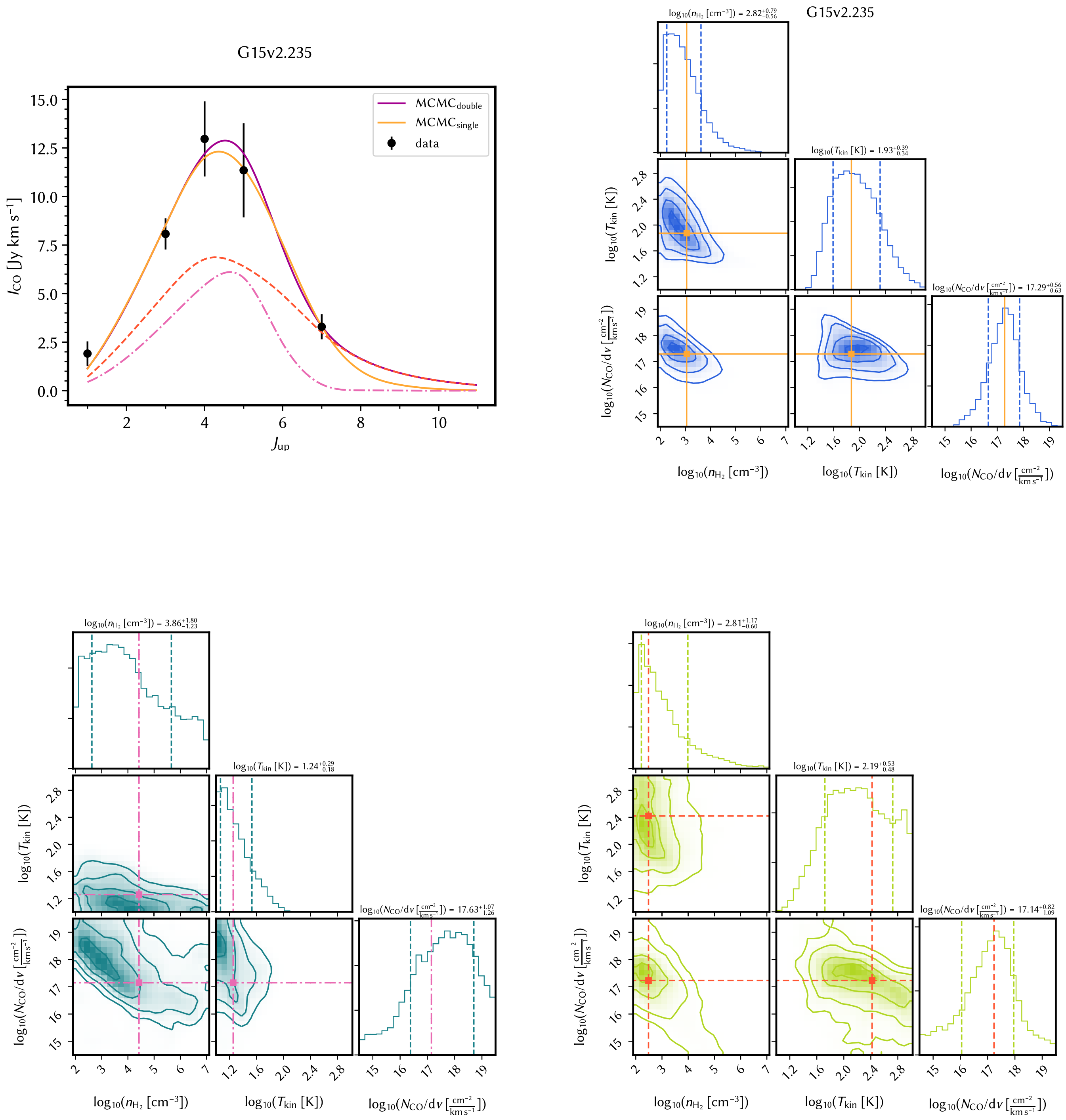

**Fig. C.1.** Continued.

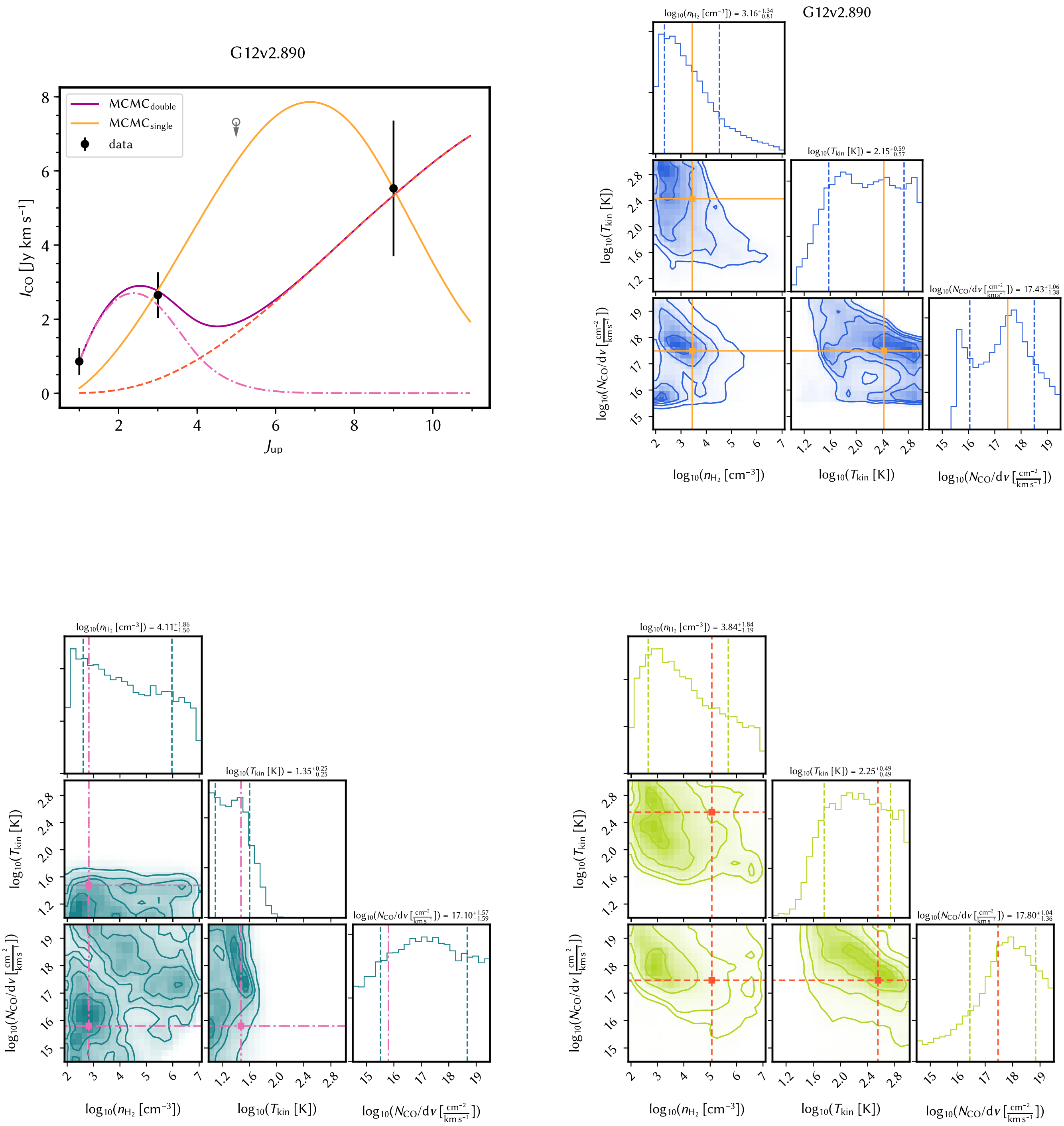

**Fig.** C.1. Continued.

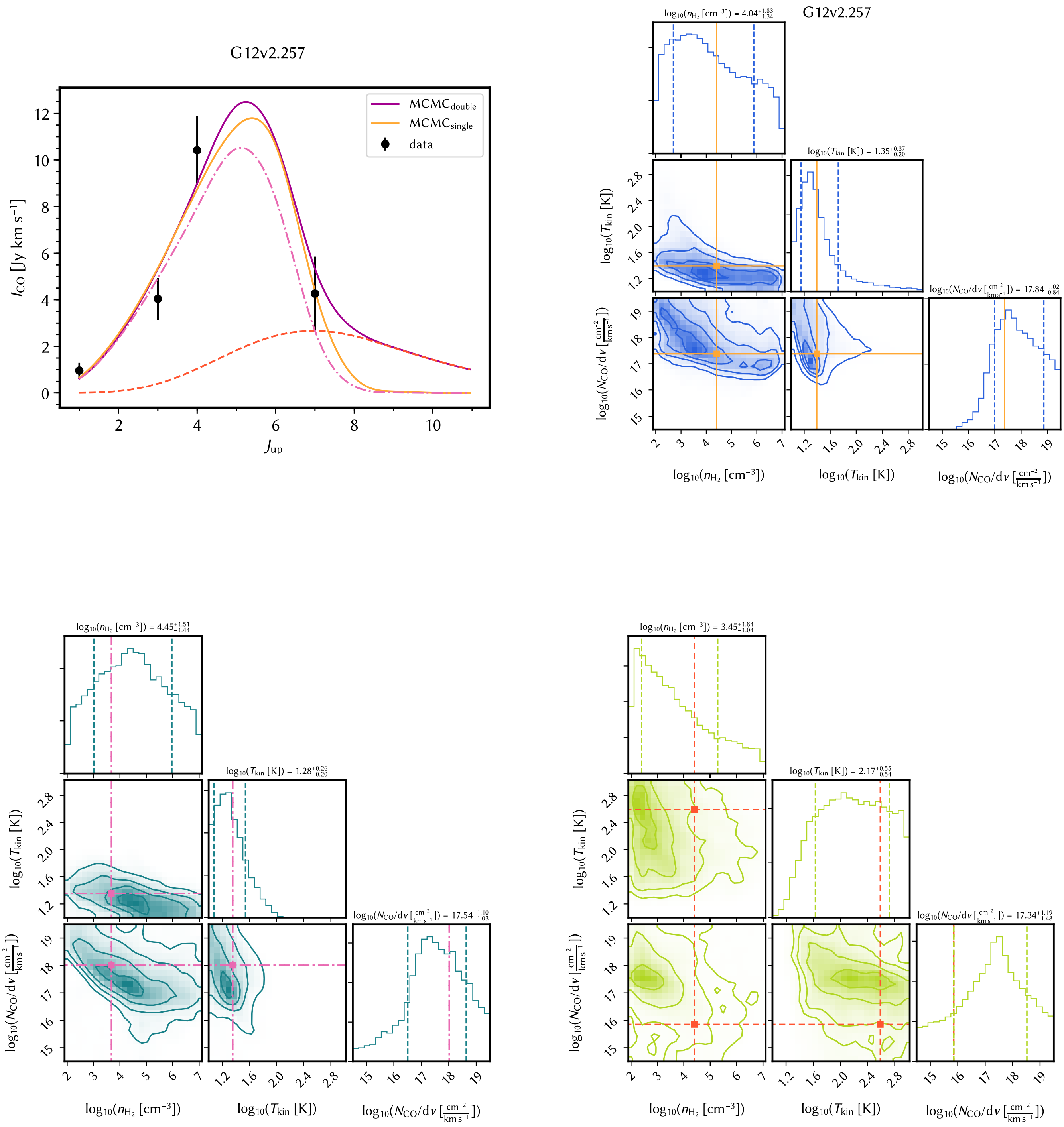

**Fig. C.1.** Continued.

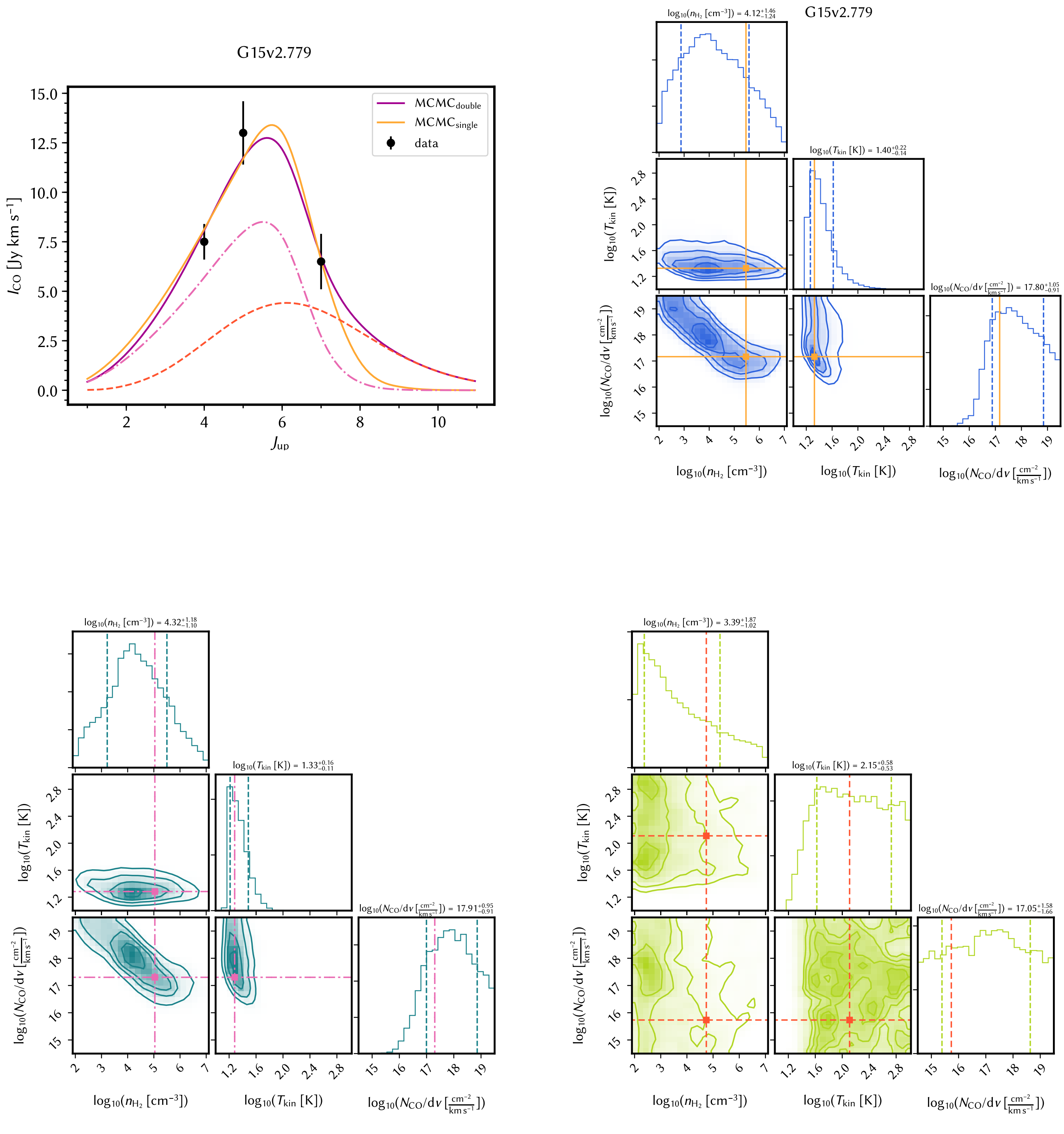

**Fig. C.1.** Continued.